\documentclass[aps,pra,superscriptaddress]{revtex4}

\usepackage[dvips]{graphicx}
\usepackage{indentfirst}
\usepackage{bm}
\usepackage{enumerate}

\def\be{\begin{equation}}
\def\ee{\end{equation}}
\def\bea{\begin{eqnarray}}
\def\eea{\end{eqnarray}}
\newcommand{\ket}[1]{\mbox{$|#1\rangle$}}

\newcommand{\avg}[1]{\mbox{$\langle#1\rangle$}}
\newcommand{\bfEdu}[2]{\mbox{${\bf E}_{#1}^{#2}$}}
\newcommand{\bfHdu}[2]{\mbox{${\bf H}_{#1}^{#2}$}}
\newcommand{\epsilondu}[2]{\mbox{$\epsilon_{#1}^{#2}$}}
\def\curl{\nabla\times}
\def\bfr{{\bf r}}
\def\bfrp{{\bf {r^\prime}}}
\def\bfE{{\bf E}}
\def\bfH{{\bf H}}
\def\bfD{{\bf D}}
\def\bfv{{\bf v}}
\def\bfw{{\bf w}}
\def\bff{{\bf f}}
\def\bfx{{\bf x}}
\def\bfs{{\bf s}}
\def\bfrho{\mbox{\boldmath{$\rho$}}}
\def\bfp0{{\bf{p_0}}}
\def\koneperp{k_{1\perp}}
\def\ktwoperp{k_{2\perp}}
\def\kiperp{k_{i\perp}}
\def\kp{k_{\parallel}}
\def\tildekp{\tilde{k}_{\parallel}}
\def\kappaoneperp{\kappa_{1\perp}}
\def\honeperp{h_{1\perp}}
\def\htwoperp{h_{2\perp}}
\def\hiperp{h_{i\perp}}
\def\hp{h_{\parallel}}
\def\Rcutoff{R_{\footnotesize\textrm{cutoff}}}

\def\Veff{V_{\footnotesize\textrm{eff}}}
\def\Aeff{A_{\footnotesize\textrm{eff}}}

\begin{document}
\title{Strong coupling of single emitters to surface plasmons}

\author{D.E.\ Chang}
\affiliation{Physics Department, Harvard University, Cambridge, MA 02138}

\author{A. S.\ S\o rensen}
\affiliation{Niels Bohr Institute, DK-2100 Copenhagen \O, Denmark}

\author{P.R.\ Hemmer}
\affiliation{Physics Department, Harvard University, Cambridge, MA 02138} 
\affiliation{Electrical Engineering Department, Texas A\&M University, College 
Station, TX 77843}

\author{M.D.\ Lukin}
\affiliation{Physics Department, Harvard University, Cambridge, MA 
02138}\affiliation{ITAMP, Harvard-Smithsonian Center for Astrophysics, 
Cambridge, MA 02138}

\date{\today}

\begin{abstract}
We propose a method that enables strong, coherent coupling between individual 
optical emitters and electromagnetic excitations in conducting nano-structures.  
The excitations are optical plasmons that can be localized to sub-wavelength 
dimensions. Under realistic conditions, the tight confinement causes optical 
emission to be almost entirely directed into the propagating plasmon modes via a 
mechanism analogous to cavity quantum electrodynamics.  We first illustrate this 
result for the case of a nanowire, before considering the optimized geometry of 
a nanotip.  We describe an application of this technique involving efficient 
single-photon generation on demand, in which the plasmons are efficiently 
out-coupled to a dielectric waveguide.  Finally we analyze the effects of 
increased scattering due to surface roughness on these nano-structures.
\end{abstract}

\maketitle

\section{Introduction}
In recent years there has been substantial interest in nanoscale optical devices 
based on local electric field enhancements and electromagnetic surface modes 
(surface plasmons) associated with sub-wavelength metallic systems.  Surface 
plasmons~\cite{smolyaninov03} are electromagnetic excitations associated with 
charge density waves on the surface of a conducting object.  The unique 
properties of plasmons on nanoscale metallic systems have produced a number of 
dramatic observed effects, such as single molecule detection with 
surface-enhanced Raman scattering (SERS)~\cite{kneipp97,nie97}, enhanced 
transmission through sub-wavelength apertures~\cite{ebbesen98,thio01}, and 
enhanced photoluminescence from quantum wells~\cite{hecker99}.  There is also 
considerable interest in these systems in applications such as 
biosensing~\cite{oldenburg02}, sub-wavelength 
imaging~\cite{smolyaninov05,zayats05}, and waveguiding and switching devices 
below the diffraction limit~\cite{takahara97,quinten98,brongersma00}.  Such 
sub-wavelength waveguiding of plasmons in metallic nanowires has been observed 
in a number of recent experiments~\cite{dickson00,krenn02a,ditlbacher05}.

At the same time, spurred in part by rapid developments in the fields of quantum 
computation and quantum information science, there has been strong interest in 
exploring new physical mechanisms that enable coherent coupling between 
individual quantum systems and photon fields.  Such a mechanism would enable 
quantum information to be passed over long distances and long-range interactions 
between systems.  These features are not only essential for quantum 
communication~\cite{ekert91,briegel98} but would also facilitate the scalability 
of quantum computers~\cite{svore04}. The required coupling between emitters and 
photons is difficult but has been achieved in a number of systems that reach the 
so-called ``strong-coupling" regime of cavity quantum electrodynamics 
(QED)~\cite{thompson92,brune96,wallraff04}. Recently several approaches to reach 
this regime on a chip at microwave frequencies have been 
suggested~\cite{childress04,sorensen04,blais04} and experimentally 
observed~\cite{wallraff04}, which utilize coupling between emitters and modes of 
superconducting transmission lines.  A key feature of these transmission lines 
is the reduction of the effective mode volume $\Veff$ for the photons, which in 
turn results in a substantial increase of the emitter-field coupling constant 
$g\propto 1/\sqrt{\Veff}$. Realization of analogous techniques with optical 
photons would open the door to many potential applications in quantum 
information science, and in addition lead to smaller mode volumes and faster 
interaction times.

In this paper we describe a method that enables strong, coherent coupling 
between individual emitters and electromagnetic excitations in conducting 
nano-structures on a chip at optical frequencies, via excitation of guided 
optical plasmons localized to nanoscale dimensions.  The strong coupling occurs 
due to the sub-wavelength confinement and small mode volumes associated with the 
surface plasmon modes.  We show that under realistic conditions optical emission 
can be almost entirely directed into these modes due to their large Purcell 
factors, in a manner analogous to cavity QED.  We first examine the case of a 
cylindrical nanowire, a simple geometry where the relevant physics can be 
understood analytically, before considering the more optimized geometry of a 
conducting nanotip.  We show that effective Purcell factors exceeding 
${\sim}10^3$ are possible in these systems, limited only by metal losses at 
optical frequencies.  Because of these losses the plasmon modes themselves are 
not suitable as carriers of information over long distances. However, we show 
that the plasmon excitation can be efficiently out-coupled to a propagating 
photon by evanescently coupling to a nearby co-propagating dielectric waveguide, 
as illustrated schematically in Fig.~\ref{fig:singlephotondevice}. This can be 
used, \textit{e.g.}, to create an efficient single photon source, or as part of 
an architecture to perform controlled interactions between distant qubits.  The 
achievable coupling between the plasmon and waveguide systems can be much 
stronger than the plasmon dissipation rates, and we find that single-photon 
generation efficiencies exceeding 95\% are possible for the simple geometries 
considered here.

This paper is organized as follows.  In Sec.~\ref{sec:cylinder} we calculate the 
mode structure of a conducting nanowire surrounded by some positive dielectric 
medium.  We show that the nanowire supports one fundamental plasmon mode with 
significantly reduced phase velocity, which is tightly localized on a scale 
${\sim}R$ around the wire surface.  We also calculate the dissipation rate of 
the fundamental mode as it propagates along the nanowire, due to metallic 
losses.  In Sec.~\ref{sec:dipoleemission} we calculate the emission properties 
of a dipole emitter near the nanowire as a function of emitter position and wire 
radius.  We show that under certain circumstances, emission into the guided 
plasmon modes is greatly enhanced over decay into radiative and non-radiative 
channels.  In fact, when optimized, the probability of emission into the plasmon 
mode approaches almost unity for small $R$ and is limited only by dissipative 
loss of the metal.  Because of its simple geometry, the nanowire is a system 
where the relevant physics can be understood and derived analytically, and from 
which we can proceed to design and understand better-optimized systems.  In 
Sec.~\ref{sec:nanotip}, we consider one such system, a conducting nanotip.  It 
will be seen that the enhancement of emission into plasmon modes found earlier 
is not exclusive to nanowires but arises quite generally as a feature of 
conducting nano-structures.  However, we will show that the nanotip is an 
optimized geometry that can significantly reduce the effects of propagative 
losses even while preserving this enhancement.  In Sec.~\ref{sec:opticalfiber} 
we consider the problem of out-coupling the plasmon modes, and study in detail 
the interaction between the plasmon modes of our nano-structures and the guided 
modes of a nearby dielectric waveguide.  We show that the plasmon modes can be 
efficiently out-coupled to the waveguide, and we propose an architecture for 
efficient single-photon generation on demand based on a tiered 
emitter/nano-structure/waveguide system.  We calculate the expected efficiencies 
for single photon generation, taking fully into account the propagative losses 
of the plasmons, the finite Purcell factors governing the interactions with the 
dipole emitter, and the non-unity coupling efficiency between the plasmon and 
waveguide modes.  In Sec.~\ref{sec:roughness}, we consider the effects of 
possible imperfections to the system, in particular, the adverse effect of 
surface roughness on our nano-structures.  In general, surface roughness can 
lead to radiative scattering of plasmons as well as increased non-radiative 
dissipation, which results in larger losses as the plasmons propagate along the 
structure.  We calculate the effects of these two processes and find only 
moderate increases in the total loss under reasonable parameters.  Finally, in 
Sec.~\ref{sec:conclusion} we summarize our results, while outlining possible 
physical realizations and discussing possible future directions of research in 
this area.

\section{Plasmon modes on a nanowire}\label{sec:cylinder}

The method for calculating the electromagnetic modes of a nanowire is briefly 
outlined here, with details of the calculation given in 
Appendix~\ref{sec:generaltheory}.  We consider a cylinder of radius $R$ of 
dimensionless electric permittivity $\epsilon_{2}$, which is centered along the 
$z$-axis and surrounded by a second dielectric medium $\epsilon_1$.  While we 
are particularly interested in the case of a conducting nanowire surrounded by 
some lossless positive dielectric~($\textrm{Re}\,\epsilon_2<0,\epsilon_1>0$), we 
note that at this point the discussion is quite general.  Like any other simple 
geometry with a high degree of symmetry, one can use separation of variables and 
find field solutions $\bfE,\bfH$ to Maxwell's Equations in each dielectric 
region~\cite{jackson99,stratton41}.  In cylindrical coordinates, the electric 
field is given by 
$\bfE_{i}(\bfr)=\mathcal{E}_{i,m}\bfE_{i,m}(k_{i\perp}\rho)e^{im\phi}e^{i\kp 
z}$, where $i=1,2$ denotes the regions outside and inside the cylinder, 
respectively.  Here $\kp$ is the longitudinal component of the wavevector, which 
is related the vacuum wavevector $k_0=\omega/c$, electric permittivity 
$\epsilon_i$, and transverse wavevector $\kiperp$ by 
$\epsilon_{i}k_0^2=\kp^2+\kiperp^2$, and $m$ is an integer characterizing the 
winding of the mode.  A similar expression holds for the magnetic field $\bfH$.  
For future reference, we also define the vacuum wavelength $\lambda_0=2\pi/k_0$, 
and $k_i=\sqrt{\epsilon_i}k_0$ as the wavevector in medium $i$.  The 
coefficients $\mathcal{E}_{i,m}$ and $\mathcal{H}_{i,m}$ multiplying the fields 
are not arbitrary but instead must satisfy a set of equations that enforces the 
necessary boundary conditions at the dielectric interface $\rho=R$.  The 
existence of a non-trivial solution requires that the matrix corresponding to 
this linear system have zero determinant~$(\textrm{det}\,M=0$), which upon 
simplifying yields the mode equation~\cite{stratton41,jackson99},
\be 
\frac{m^{2}k_{\|}^{2}}{R^2}\left(\frac{1}{k_{2\perp}^2}-\frac{1}{k_{1\perp}^2}\right)^{2} 
=\left(\frac{1}{k_{2\perp}}\frac{J_{m}^\prime(k_{2\perp}R)}{J_{m}(k_{2\perp}R)}-\frac{1}{k_{1\perp}}\frac{H_{m}^\prime(k_{1\perp}R)}{H_{m}(k_{1\perp}R)}\right) 
\left(\frac{k_2^2}{k_{2\perp}}\frac{J_{m}^\prime(k_{2\perp}R)}{J_{m}(k_{2\perp}R)}-\frac{k_1^2}{k_{1\perp}}\frac{H_{m}^\prime(k_{1\perp}R)}{H_{m}(k_{1\perp}R)}\right).\label{eq:modecondition} 
\ee
One can use the above equation, for example, to determine the allowed values of 
$\kp$ as functions of $m,R$, and $\epsilon_{i}$.

\begin{figure*}[p]
\begin{center}
\includegraphics[width=7cm]{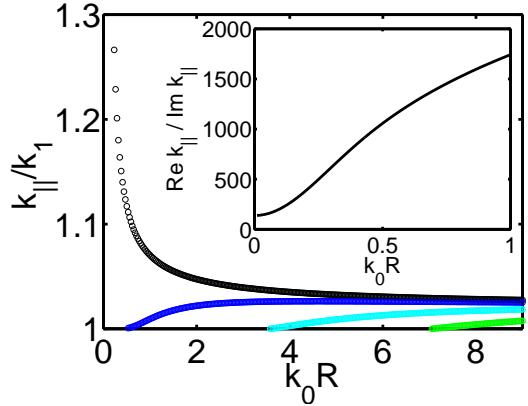}
\end{center}
\caption{Allowed plasmon modes $\kp$ as a function of $R$ for a silver nanowire 
embedded in a surrounding dielectric $\epsilon_{1}=2$, for frequency 
corresponding to a vacuum wavelength $\lambda_0=1\;\mu$m and room temperature. 
The fundamental ($m=0$) mode, in black, exhibits a $1/R$ dependence, while all 
other modes are effectively cut off as $R{\rightarrow}0$. Inset: the propagative 
losses for the fundamental mode, characterized by the ratio 
$\textrm{Re}\;\kp/\textrm{Im}\;\kp$, for the same 
parameters.\label{fig:wiremodes}}
\end{figure*}

We now focus on the case of a sub-wavelength, conducting metal wire surrounded 
by a normal, positive dielectric.  In Fig.~\ref{fig:wiremodes} we plot the 
allowed wavevectors $k_{\parallel}$, as determined through 
Eq.~(\ref{eq:modecondition}), for such a system as a function of $R$ for a few 
lowest-order modes in $m$.  For concreteness, all numerical results presented in 
this paper are for a silver nanowire~(or later, nanotip) at room temperature, 
$\lambda_0=1\;{\mu}\textrm{m}$, and with a surrounding dielectric 
$\epsilon_1=2$, although the physical processes described are not specific to 
silver or to some narrow frequency range.  The electric permittivity of silver 
at this frequency is $\epsilon_2{\approx}-50+0.6i$, as given 
in~\cite{johnson72}.  In plotting Fig.~\ref{fig:wiremodes} we have temporarily 
ignored the dissipative imaginary part of $\epsilon_2$, although we will address 
its effect later.  Ignoring $\textrm{Im}\,\epsilon_2$ results in purely real 
values of $\kp$.

We first qualitatively discuss the important features of the plasmon modes 
illustrated in Fig.~\ref{fig:wiremodes}, before deriving them more carefully.  
It is clear from the figure that the longitudinal component of the wavevector 
exceeds the wavevector in uniform dielectric, $k_{\parallel}>k_{1}$, which in 
turn causes the perpendicular component $k_{i\perp}=\sqrt{k_i^2-\kp^2}\equiv 
i\kappa_{i\perp}$ to be purely imaginary. Physically these relationships imply 
that the plasmon modes are non-radiative and are confined near the 
metal/dielectric interface, with the length scale of confinement determined by 
${\sim}1/\kappa_{i\perp}$.  Furthermore, these plasmon modes cannot couple 
directly to radiative fields, which have wavevectors $\kp{\leq}k_1$.  Of 
particular interest is the behavior of the plasmon modes in the nanowire limit 
$|k_i|R{\ll}1$.  In this limit, all higher-order modes~($|m|{\geq}1$) exhibit a 
cutoff as $R{\rightarrow}0$, as derived in Appendix~\ref{sec:cutoff}, while the 
$m=0$ fundamental plasmon mode exhibits a unique $\kp{\propto}1/R$ behavior. 
Physically, in this limit, the $m=0$ mode can be interpreted approximately as a 
quasi-static configuration of field and associated charge density wave on the 
wire.  As such, $R$ becomes the only relevant length scale, as the length scales 
$1/|k_i|$ associated with electrodynamic behavior become unimportant.  From the 
$1/R$ scaling of $\kp$ it follows that $\kappa_{1\perp}{\propto}1/R$, which 
states that the field outside the wire is tightly localized on a scale 
${\propto}R$ around the metal surface.  The corresponding small effective 
transverse mode area $\Aeff{\propto}R^2$ is responsible for the strong 
interaction strength of the fundamental mode with nearby emitters, as will be 
discussed in following sections.  We note that this behavior contrasts sharply 
with that of, \textit{e.g.}, a sub-wavelength normal dielectric waveguide or 
optical fiber, which runs into a ``confinement problem'' where the evanescent 
tails outside the device become exponentially large as 
$R{\rightarrow}0$~\cite{tong04}.

In practice $\epsilon_2$ is not purely real but has a small imaginary part 
corresponding to metal losses~(heating) at optical frequencies. Its effect is to 
add a small imaginary component to $k_{\parallel}$ corresponding to dissipation 
as the plasmon propagates along the wire.  In the inset of 
Fig.~\ref{fig:wiremodes} we plot $\textrm{Re}\;\kp/\textrm{Im}\;\kp$ for the 
fundamental mode as a function of $R$.  This quantity is proportional to the 
decay length in units of the plasmon wavelength 
$\lambda_{\footnotesize\textrm{pl}}{\equiv}2\pi/{\textrm{Re}}\,\kp$.  As $R$ 
decreases, it can be seen that this ratio decreases monotonically but approaches 
a nonzero constant, as will be shown below.  For silver at 
$\lambda_0=1\;\mu\textrm{m}$ and room temperature and $\epsilon_1=2$ this 
constant is approximately $140$.  The fact that this ratio does not approach 
zero even as $R{\rightarrow}0$ is important for potential applications involving 
plasmons on nanowires, as it implies that the plasmons can still travel multiple 
$\lambda_{\footnotesize\textrm{pl}}$ for devices of any size.  We also note that 
while all numbers and figures presented here are for room temperature, operating 
at lower temperatures might somewhat reduce the value of 
$\textrm{Im}\,\epsilon_2$ due to decreased losses from phonon-assisted 
absorption~\cite{mckay76}.

We now analyze the fundamental mode more carefully.  For $m=0$, one sees in 
Eq.~(\ref{eq:modecondition}) that one of the two terms on the right-hand side 
must equal zero. It can be shown that setting the first term to zero corresponds 
to a $TE$ mode, while the other case corresponds to a $TM$ mode~(see 
Appendix~\ref{sec:generaltheory}). The $TE$ mode equation does not have any 
solutions, and thus the fundamental mode is a $TM$ mode that satisfies the 
simplified equation~\cite{takahara97,cao05}
\be 
\frac{k_2^2}{\ktwoperp}\frac{J^{\prime}_{0}({\ktwoperp}R)}{J_{0}({\ktwoperp}R)}- 
\frac{k_1^2}{\koneperp}\frac{H^{\prime}_{0}({\koneperp}R)}{H_{0}({\koneperp}R)}=0.\label{eq:TMmodecondition} 
\ee
The fields themselves are given by~(see Appendix~\ref{sec:generaltheory})
\bea \bfE_{1} & = & 
b_{1}\left(\frac{i{\kp}{\koneperp}}{k_1^2}H_{0}^{\prime}\left({\koneperp}\rho\right)\hat{\rho}+\frac{{\koneperp}^2}{k_1^2}H_{0}\left({\koneperp}\rho\right)\hat{z}\right)e^{i\kp{z}}, 
\nonumber
\\ \bfE_{2} & = &
b_{2}\left(\frac{i{\kp}{\ktwoperp}}{k_2^2}J_{0}^{\prime}\left({\ktwoperp}\rho\right)\hat{\rho}+\frac{{\ktwoperp}^2}{k_2^2}J_{0}\left({\ktwoperp}\rho\right)\hat{z}\right)e^{i\kp{z}}, 
\nonumber \\ \bfH_{1} & = & 
\frac{i}{\omega\mu_{0}}{\koneperp}b_{1}H_{0}^{\prime}\left({\koneperp}\rho\right)e^{i\kp{z}}\hat{\phi}, 
\nonumber \\ \bfH_{2} & = & 
\frac{i}{\omega\mu_0}{\ktwoperp}b_{2}J_{0}^{\prime}\left({\ktwoperp}\rho\right)e^{i\kp{z}}\hat{\phi}, 
\label{eq:tm0mode} \eea
while the boundary conditions between the two dielectrics require that
\be 
\frac{b_{1}}{b_{2}}=\frac{{\ktwoperp}}{{\koneperp}}\frac{J_{0}^{\prime}\left({\ktwoperp}R\right)}{H_{0}^{\prime}\left({\koneperp}R\right)}.\label{eq:b1b2} 
\ee

The $1/R$ dependence of $\kp,\kiperp$ in the nanowire limit can be confirmed 
mathematically by considering Eq.~(\ref{eq:modecondition}) in the non-retarded 
limit~($c\rightarrow\infty$).  In this case, $k_{i\perp}=ik_{\parallel}$ and the 
mode equation~(\ref{eq:TMmodecondition}) reduces to
\be \frac{\epsilon_2}{\epsilon_1} = 
\frac{K^{\prime}_{0}(k_{\parallel}R)I_{0}(k_{\parallel}R)}{K_{0}(k_{\parallel}R)I^{\prime}_{0}({\kp}R)},\label{eq:staticmodecondition} 
\ee
where $I_{m},K_{m}$ are modified Bessel functions.  The solution to 
Eq.~(\ref{eq:staticmodecondition}) requires $k_{\parallel}R=C_{-1}$ to be 
constant and proves the aforementioned scaling law for $\kp$.  It is also 
straightforward to see that when $\epsilon_2$ acquires a small imaginary 
component, the constant $C_{-1}$ becomes complex as well, and that 
$\textrm{Re}\;\kp/\textrm{Im}\;\kp$ takes on some fixed, non-zero value. No 
closed-form solution exists for the equation above, although when 
$|k_{\parallel}R|{\ll}1$~(corresponding to large $|\epsilon_{2}/\epsilon_{1}|$) 
the equation asymptotically approaches
\be 
\frac{\epsilon_2}{\epsilon_1}=\frac{2}{(\gamma-\log{2}+\log{C_{-1}})(C_{-1})^2}, 
\ee
where $\gamma{\approx}0.577$ is Euler's constant.

Finally, it should be noted that the components of $\bfE_i$ in 
Eq.~(\ref{eq:tm0mode}) are proportional to $\kp\kiperp$ or $\kiperp^2$, while 
$\bfH_i$ is proportional to $\kiperp$.  Thus, in the nanowire limit when 
$\kp,|\kiperp|{\propto}1/R$, the magnetic fields are a factor of $R$ smaller 
than the electric fields, which is consistent with this mode being roughly a 
quasi-static configuration.

\section{Spontaneous emission near a metal nanowire}\label{sec:dipoleemission}
The small mode volume associated with the fundamental plasmon mode of a nanowire 
offers a possible mechanism to achieve strong coupling with nearby optical 
emitters, in analogy to the methods 
of~\cite{childress04,sorensen04,blais04,wallraff04}.  In this section we derive 
more rigorously the interaction between an emitter and nanowire, and show that 
under certain circumstances the small mode volume indeed leads to strongly 
preferential spontaneous emission into the guided plasmon modes via a mechanism 
equivalent to the Purcell effect~\cite{purcell46} in cavity QED.

The spontaneous emission rate of a dipole emitter in general becomes altered 
from its free-space value in the presence of some dielectric body.  In our 
system of interest, the dipole can possibly lose power radiatively to 
propagating photon modes, through excitation of the guided plasmon modes, or 
through non-radiative loss~(heating) in the wire.  The dipole in consideration 
can physically be formed by a single atom, a defect in a solid-state system, or 
any other system with a dipole-allowed transition.  In 
Sec.~\ref{subsec:radnonraddecay} we calculate the radiative and non-radiative 
rates using a quasi-static approach, while waiting until 
Sec.~\ref{subsec:plasmondecay} to treat the plasmon decay rate more thoroughly.  
In Sec.~\ref{subsec:purcellwire}, we show how the efficiency of emission into 
the plasmon modes can be optimized to yield Purcell factors in excess of 
${\sim}5{\times}10^2$, and discuss the physical origins of this limit.

\subsection{Radiative and non-radiative
decay rates}\label{subsec:radnonraddecay}

In this subsection we derive formulas for the decay rates of a dipole near a 
metal nanowire into radiative and non-radiative channels. This calculation 
closely follows that of~\cite{klimov04}, but is briefly presented here for 
completeness.

It is well-known that spontaneous emission rates can be obtained via classical 
calculations of the fields due to an oscillating dipole near the dielectric 
body~\cite{wylie84}, and this method will be employed here.  Specifically, we 
consider a (classical) oscillating dipole ${\bfp0}e^{-i{\omega}t}$ oriented 
along $\hat{\rho}$ and positioned a distance $d$ from the center of the wire, 
and wish to calculate the total fields of the system.  For nano-structures one 
can make a considerable simplification and consider the fields in the 
quasi-static limit~($\bfH{\approx}0$)~\cite{klimov04}, which satisfy
\bea \nabla\cdot\bfD & = & \rho_{\footnotesize{\textrm{ext}}}, \\
\nabla\times\bfE & = & 0. \eea
Here $\rho_{\footnotesize\textrm{ext}}(\bfr)$ is the external charge density.  
In the system of interest the external source is a dipole located at position 
$\bfrp$ outside the wire~(with radial coordinate $\rho'=d$), which has a 
corresponding charge configuration
\be 
\rho_{\footnotesize{\textrm{ext}}}(\bfr,\bfrp)=\left(\bf{p_0}\cdot\nabla^{\prime}\right)\delta\left(\bfr-\bfrp\right). 
\ee
For simplicity we omit the harmonic time dependence from our expressions for the 
source and all fields.  Note that the $\delta(\bfr-\bfrp)$ term above 
corresponds to a (unitless) point charge source, while the operator 
$(\bfp0\cdot\nabla')$ generally converts the point charge solution to that of a 
dipole.  It is therefore convenient to write $\bfE_{i}$ in similar form,
\be 
\bfE_{i}(\bfr,\bfrp)=-\nabla\left(\bfp0\cdot\nabla^{\prime}\right)\Phi_{i}\left(\bfr,\bfrp\right),\label{eq:pseudopotential} 
\ee
where $\Phi_{i}(\bfr,\bfrp)$ are ``pseudopotentials'' that satisfy 
$\nabla^{2}\Phi_{1}=-\delta(\bfr-\bfrp)/\epsilon_0\epsilon_1$ and 
$\nabla^{2}\Phi_{2}=0$.  Here the indices $1,2$ again denote the regions outside 
and inside the cylinder, respectively.  Clearly, $\Phi$ physically corresponds 
to the potential due to a point charge at $\bfrp$, while the dipole potential 
follows from $\Phi_{dip}\equiv(\bfp0\cdot\nabla')\Phi$.

To solve for the fields, it is convenient to further separate $\Phi_{1}$ into 
``free'' and ``reflected'' components $\Phi_{0}$ and $\Phi_{r}$, respectively, 
where $\Phi_{r}$ represents a source-free contribution that ensures that 
boundary conditions are satisfied and $\Phi_{0}$ is the solution for a point 
charge in a medium of uniform electric permittivity $\epsilon_1$.  We will 
expand the known source term $\Phi_0$ in a basis appropriate for the cylindrical 
geometry, and expand the source-free terms $\Phi_{r,2}$ in a similar basis that 
satisfies Laplace's Equation~($\nabla^{2}\Phi_{r,2}=0$).  The unknown 
coefficients multiplying the basis functions of $\Phi_{r,2}$ will then be 
determined by enforcing the proper boundary conditions at the dielectric 
interface.  Mathematically, the proper expansions are given by
\bea \Phi_{0}(\bfr,\bfrp) & = &
\frac{1}{4\pi\epsilon_{0}\epsilon_1}\frac{1}{\left|\bfr-\bfrp\right|}\nonumber \\
& = &
\frac{1}{2\pi^2\epsilon_0\epsilon_1}\sum_{m=0}^{\infty}\left(2-\delta_{m,0}\right)\cos\left(m(\phi-\phi^\prime)\right)\int_{0}^{\infty}dh\cos\left(h(z-z^\prime)\right)K_{m}(h\rho^\prime)I_{m}(h\rho)\;\;\;(\rho<\rho^\prime)\label{eq:Phi0}, \\
\Phi_{r}(\bfr,\bfrp) & = & 
\frac{1}{2\pi^2\epsilon_0}\sum_{m=0}^{\infty}\left(2-\delta_{m,0}\right)\cos\left(m(\phi-\phi^\prime)\right)\int_{0}^{\infty}dh\;\alpha_{m}(h)\cos\left(h(z-z^\prime)\right)K_{m}(h\rho^\prime)K_{m}(h\rho),\label{eq:Phir}
\\ \Phi_{2}(\bfr,\bfrp) & = &
\frac{1}{2\pi^2\epsilon_0}\sum_{m=0}^{\infty}\left(2-\delta_{m,0}\right)\cos\left(m(\phi-\phi^\prime)\right)\int_{0}^{\infty}dh\;\beta_{m}(h)\cos\left(h(z-z^\prime)\right)K_{m}(h\rho^\prime)I_{m}(h\rho),\label{eq:Phi1} 
\eea
where $\alpha_{m}(h),\beta_m(h)$ thus far are unknown amplitude coefficients.  
We obtain a set of two coupled equations for $\alpha_{m}(h),\beta_m(h)$ by 
requiring continuity of $\Phi$ and ${\bf D}_\perp$ at the boundary, $\rho=R$.  
Because of the translational symmetry of the system, these equations are 
uncoupled in $h$ and can easily be solved~(this is in contrast to the case where 
translational symmetry is broken due to surface roughness, as discussed in 
Sec.~\ref{sec:roughness}).  The solutions are given by~\cite{klimov04}
\bea \alpha_{m}(h) & = & 
\frac{\left(\epsilon-1\right)I_{m}^{\prime}(hR)I_{m}(hR)}{\epsilon_{1}I_{m}(hR)K_m^\prime(hR)-{\epsilon_2}K_{m}(hR)I_m^\prime(hR)}, 
\nonumber
\\ \beta_{m}(h) & = &
\frac{I_{m}(hR)K_{m}^\prime(hR)-K_m(hR)I_m^\prime(hR)}{\epsilon_{1}I_{m}(hR)K_m^\prime(hR)-{\epsilon_2}K_{m}(hR)I_m^\prime(hR)},\label{eq:alphabeta} 
\eea
where we have defined $\epsilon\equiv\epsilon_{2}/\epsilon_{1}$.  Note that 
Eq.~(\ref{eq:alphabeta}) along with Eqs.~(\ref{eq:Phi0})-(\ref{eq:Phi1}) give 
the total electric field of the dipole and wire system.

To calculate the radiative emission into free space, we consider the far-field 
properties of the system.  Physically, the presence of the emitter induces some 
dipole moment $\delta{\bf p}$ in the nanowire, which results in a total radiated 
power proportional to the square of the total dipole moment of the system, 
$P_{\footnotesize\textrm{rad}}\propto\Gamma_{\footnotesize\textrm{rad}}\propto|\bfp0+\delta{\bf 
p}|^2$.  We can determine $\delta{\bf p}$ by finding the dipole-like 
contribution to the ``reflected'' potential 
$\Phi_{dip,r}=(\bfp0\cdot\nabla')\Phi_{r}(\bfr,\bfrp)$ far away from the source, 
which on physical grounds must behave like $\rho^{-2}$ for large $\rho$.  It is 
straightforward to show that the $m=1$ term in Eq.~(\ref{eq:Phir}) is 
responsible for this contribution, with all other $m$ terms yielding faster 
decays in $\rho$.  Because of the asymptotic behavior of 
$K_{m}(x){\approx}\sqrt{\pi/2x}e^{-x}$ when $x{\gg}1$, it can be seen that for 
large $\rho$ the integrand appearing in~(\ref{eq:Phir}) is significant only over 
a small region $h\stackrel{<}{\sim}\rho^{-1}$.  As a result, we can safely 
replace $K_{1}(h\rho^{\prime})$ and $\alpha_{1}(h)$ by their expansions around 
$h=0$.  After this simplification the integral can in fact be evaluated exactly 
and yields
\bea \Phi_{r}^{(m=1)} & {\approx} & 
-\frac{1}{2\pi^2\epsilon_0\epsilon_1}\cos\left(\phi-\phi^\prime\right)\int_{0}^{\infty}dh\;{\cos}\;h\left(z-z^\prime\right)\frac{1}{h\rho^{\prime}}K_{1}(h\rho)\frac{\epsilon-1}{\epsilon+1}h^{2}R^{2}
\nonumber \\ & = &
-\frac{1}{4\pi\epsilon_0\epsilon_1}\frac{\epsilon-1}{\epsilon+1}\cos\left(\phi-\phi^\prime\right)\frac{R^2}{\rho'}\frac{\rho}{(\rho^2+(z-z')^2)^{3/2}}, 
\eea
with a corresponding reflected potential
\bea \Phi_{dip,r}^{(m=1)}(\bfr,\bfrp) & = & 
\left(\bfp0\cdot\nabla^\prime\right)\Phi_{r}^{(m=1)}(\bfr,\bfrp) \nonumber
\\ & = & 
\frac{p_0}{4\pi\epsilon_{0}\epsilon_1}\frac{\epsilon-1}{\epsilon+1}\cos\left(\phi-\phi^\prime\right)\frac{R^2}{d^2}\frac{\rho}{(\rho^2+z^2)^{3/2}}.\label{eq:Vwire} 
\eea
In evaluating the above expression we have chosen $\rho'=d$ and $z'=0$ as the 
dipole coordinates, and $\bfp0=p_{0}\hat{\rho}$ as the dipole orientation.  
Comparing Eq.~(\ref{eq:Vwire}) to the potential due to a dipole $\delta{\bf p}$ 
in uniform dielectric $\epsilon_1$, $V_{\delta{\bf p}}=\frac{\delta{\bf 
p}{\cdot}\bfr}{4{\pi}\epsilon_{0}\epsilon_{1}r^3}$, we can readily identify 
\be \delta{\bf p}=p_{0}\frac{\epsilon-1}{\epsilon+1}\frac{R^2}{d^2}\hat{\rho} \ee
as the induced dipole moment in the wire, from which it follows that the 
radiative spontaneous emission rate is given by~\cite{klimov04}
\be 
\frac{\Gamma_{\footnotesize\textrm{rad}}}{\Gamma_0}=\left|1+\frac{\epsilon-1}{\epsilon+1}\frac{R^2}{d^2}\right|^2.\label{eq:gammarad}\;\;\;\;\;(d{\geq}R) 
\ee
Here $\Gamma_0$ is defined to be the spontaneous emission rate in uniform 
dielectric $\epsilon_1$~\cite{note2}.  Away from the plasmon 
resonance~($\epsilon{\approx}-1$), the radiative decay rate changes slightly 
from $\Gamma_0$ and reflects some moderate change in the radiative density of 
states in the vicinity of the nanowire.  We note that this decay rate is 
well-behaved in either limit $R,(d-R){\rightarrow}0$.

To calculate the other decay rates, one utilizes the fact that the total power 
loss of an oscillating dipole is proportional to the electric field in 
quadrature at the dipole's location, specifically, 
$\Gamma_{\footnotesize\textrm{total}}{\propto}\textrm{Im}\left(\bfp0\cdot\bfE_{1}(\bfrp,\bfrp)\right)$.  
Having divided up $\bfE_1$ into free and reflected components, we note that the 
contribution to $\bfE_{1}$ from the free field simply is associated with the 
decay rate in uniform dielectric $\epsilon_1$, and thus we concentrate on the 
contribution from $\Phi_{r}(\bfr,\bfrp)$.  First we note that the coefficient 
$\alpha_{0}(h)$ derived in Eq.~(\ref{eq:alphabeta}) contains a pole at the point 
where the denominator vanishes.  This pole corresponds to an excitation of a 
natural mode~(the fundamental plasmon mode) of the system.  This can immediately 
be seen by comparing the denominator of $\alpha_{0}$ to 
Eq.~(\ref{eq:staticmodecondition}), which gives the plasmon mode in the nanowire 
limit.  The pole lies at $hR=C_{-1}$ and agrees with the plasmon wavevector 
derived in Sec.~\ref{sec:cylinder}, as expected.  Evaluating the contribution of 
this pole to $\bfE_{r}(\bfrp,\bfrp)$ gives the decay rate into the fundamental 
plasmon mode, and is discussed more carefully in the next subsection.  At the 
same time, in the limit $d{\rightarrow}R$ one expects some type of divergence to 
occur in the non-radiative decay rate.  Physically, such a divergence results 
from the large currents in the wire generated by the near-field of the dipole 
and their resulting dissipation.  We can find the leading-order term to this 
divergent decay rate by carefully evaluating the leading-order divergence in the 
reflected field.

In particular, while the pole associated with the $m=0$ term in $\Phi_{r}$ 
yields the spontaneous emission rate into the plasmon modes, we will show that 
the mathematical origin of the divergence is the significant contribution to the 
field of an infinite number of terms with $m>0$ as $d{\rightarrow}R$.  
Specifically, in this limit, for a dipole oriented along $\hat{\rho}$,
\bea \frac{\Gamma_{\footnotesize\textrm{non-rad}}}{\Gamma_0} & \approx & 
\frac{6\pi\epsilon_0}{k_0^3\sqrt{\epsilon_1}}\frac{\textrm{Im}\,\hat{\rho}\cdot\bfE_{r}(\bfrp,\bfrp)}{p_0} 
\nonumber
\\ & = & 
-\frac{6\pi\epsilon_0}{k_0^3\sqrt{\epsilon_1}}\textrm{Im}\,\hat{\rho}\cdot\nabla\left(\hat{\rho}\cdot\nabla'\right)\Phi_{r}(\bfr,\bfrp)\Big|_{\bfr=\bfrp} 
\nonumber
\\ & \approx & 
-\frac{6}{{\pi}k_0^{3}\sqrt{\epsilon_1}}\sum_{m=1}^{\infty}\int_{0}^{\infty}dh\;h^{2}K_{m}^\prime(hd)^2\textrm{Im}\;\alpha_{m}(h) 
\nonumber
\\ & \equiv &
\frac{6}{{\pi}k_0^{3}\sqrt{\epsilon_1}}\sum_{m=1}^{\infty}\int_{0}^{\infty}dh\;f_{m}(h,d,R).\label{eq:gammanonradformula} 
\eea
The divergent nature of the above expression can be shown by examining the 
asymptotic behavior of the functions $f_{m}$,
\bea f_{m}(h,d,R)\approx\left\{
\begin{array}{r@{\quad\quad}l@{\quad}}
\displaystyle\frac{m}{2d^2\epsilon_1}\textrm{Im}\left(\frac{\epsilon-1}{\epsilon+1}\right)\left(\frac{R}{d}\right)^{2m}
& h{\rightarrow}0 \\
\displaystyle\frac{h}{2d\epsilon_1}\textrm{Im}\left(\frac{\epsilon-1}{\epsilon+1}\right)e^{-2h(d-R)} 
& h{\rightarrow}\infty \end{array}. \right. \eea
From the above expressions, we see that $f_m$ as a function of $h$ has a 
characteristic width of about $\left[2(d-R)\right]^{-1}$, yet at the same time 
the quantity $m(R/d)^{2m}$ reaches a maximum around 
$\tilde{m}\approx\frac{d}{2(R-d)}$ as $d{\rightarrow}R$.  This confirms the 
non-vanishing contribution of an infinite number of terms with $m>0$ to the 
decay rate.  The exact behavior of the functions $f_m$ at small $h$, including 
the peak around $\tilde{m}$, and the tails at large $h$ is well-modelled by a 
Lorentzian approximation,
\be 
f_{m}(h,d,R){\approx}\frac{\frac{m}{2d^2\epsilon_1}\textrm{Im}\left(\frac{\epsilon-1}{\epsilon+1}\right)\left(\frac{R}{d}\right)^{2m}}{1+h^2\left(d-R\right)^2}, 
\ee
which allows the integration and sum in Eq.~(\ref{eq:gammanonradformula}) to be 
performed exactly.  The resulting decay rate is given by
\be 
\frac{\Gamma_{\footnotesize\textrm{non-rad}}}{\Gamma_0}{\approx}\frac{3}{16k_{0}^{3}(d-R)^3\epsilon_1^{3/2}}\textrm{Im}\left(\frac{\epsilon-1}{\epsilon+1}\right).\label{eq:gammanonrad} 
\ee
Note that for $|\epsilon|{\gg}1$ and small $\textrm{Im}\;\epsilon$, 
$\textrm{Im}\left(\frac{\epsilon-1}{\epsilon+1}\right){\approx}2\textrm{Im}\;\epsilon/\left(\textrm{Re}\;\epsilon\right)^2$, 
which makes it clear that the non-radiative spontaneous emission rate is 
proportional to the dissipative part of the electric permittivity.

\subsection{Decay rate into plasmon
modes}\label{subsec:plasmondecay}

In this subsection we quantify the spontaneous emission rate 
$\Gamma_{\footnotesize\textrm{pl}}$ of a dipole into the surface plasmon modes 
on a nanowire.  As shown in the previous subsection, the coefficient 
$\alpha_{0}(h)$ characterizing the reflected field contains a pole at 
$h=C_{-1}/R$ that corresponds to excitation of the natural surface plasmon mode 
of the system.  The contribution of this pole to the quantity 
$\textrm{Im}\left(\bfp0\cdot\bfE_{1}(\bfrp,\bfrp)\right)$ yields the spontaneous 
emission rate into the plasmon modes and can readily be evaluated.  Before 
proceeding further, we first note that in the presence of metal losses, the 
distinction between $\Gamma_{\footnotesize\textrm{pl}}$ and 
$\Gamma_{\footnotesize\textrm{non-rad}}$ is not perfectly well-defined, since 
the plasmons eventually dissipate due to heating as well.  Thus, for 
concreteness, we will define $\Gamma_{\footnotesize\textrm{pl}}$ to be the decay 
rate resulting from the pole in the limit that $\textrm{Im}\,\epsilon_{2}=0$, 
and take the plasmon wavevector $\kp$ and $C_{-1}$ to be purely real in this 
subsection.  In particular, for a dipole oriented along $\hat{\rho}$, 
\bea \frac{\Gamma_{\footnotesize\textrm{pl}}}{\Gamma_0} & = & 
\frac{6\pi\epsilon_{0}}{k_{0}^3\sqrt{\epsilon_1}}\left(\frac{\textrm{Im}\,\hat{\rho}\cdot\bfE_{r}(\bfrp,\bfrp)}{p_0}\right)_{pole} 
\nonumber \\ & = & 
-\frac{6\pi\epsilon_{0}}{k_{0}^3\sqrt{\epsilon_1}}\textrm{Im}\left(\hat{\rho}\cdot\nabla\left(\hat{\rho}\cdot\nabla'\right)\Phi_{r}(\bfr,\bfrp)\big|_{\bfr=\bfrp}\right)_{pole} 
\nonumber \\ & = & 
-\frac{3}{{\pi}k_0^{3}\sqrt{\epsilon_{1}}}\textrm{Im}\left(\int_{0}^{\infty}dh\,h^{2}K_{1}^{2}(hd)\alpha_{0}(h)\right)_{pole},\label{eq:poleeqn} 
\eea
where we have explicitly indicated that we are interested in the pole 
contribution to the expressions above.  It is convenient to explicitly separate 
out the pole of $\alpha_{0}$, and approximately describe the behavior around the 
pole's vicinity by
\bea \alpha_{0}(h) & {\approx} & 
\frac{1}{\epsilon_1}\frac{(\epsilon_2-\epsilon_1)I_{1}(C_{-1})I_{0}(C_{-1})}{(h-C_{-1}/R)R\frac{d\chi(C_{-1})}{dx}}, 
\eea
where
\be 
\chi(x)=\epsilon_{1}I_{0}(x)K_{0}'(x)-\epsilon_{2}K_{0}(x)I_{0}'(x).\label{eq:chiqv} 
\ee
This separation allows us to easily evaluate Eq.~(\ref{eq:poleeqn}) and yields 
the decay rate
\bea \Gamma_{\footnotesize\textrm{pl}} & = & 
\alpha_{\footnotesize\textrm{pl}}{\Gamma_0}\frac{K_{1}^2(C_{-1}d/R)}{(k_{0}R)^3} 
\nonumber \\ & \approx & 
\alpha_{\footnotesize\textrm{pl}}{\Gamma_0}\frac{K_{1}^2(\kappaoneperp d 
)}{(k_{0}R)^3},\label{eq:gammaplasmon} \eea
where we have identified $\kappaoneperp{\approx}C_{-1}/R$ in the nanowire 
limit.  The coefficient $\alpha_{\footnotesize\textrm{pl}}$ is given by 
\be 
\alpha_{\footnotesize\textrm{pl}}=\frac{3(\epsilon_1-\epsilon_2)}{\epsilon_{1}^{3/2}}\frac{C_{-1}^{2}I_{1}(C_{-1})I_{0}(C_{-1})}{d\chi(C_{-1})/dx} 
\ee
and most importantly depends only on $\epsilon_{1,2}$.

While the derivation above is straightforward, one can gain some physical 
understanding of the result and its relation to the Purcell effect by using 
Fermi's Golden Rule.  This rule states that, once the plasmon modes are 
quantized, the decay rate is given by
\be 
\Gamma_{\footnotesize\textrm{pl}}=2{\pi}g^{2}(\bfr,\omega)D(\omega),\label{eq:goldenrule} 
\ee
where $g(\bfr,\omega)$ is the position-dependent coupling strength between the 
quantized field and emitter, and $D(\omega)$ is the plasmon density of states on 
the nanowire.

Canonical quantization of a dispersive medium is a difficult and subtle 
problem~\cite{kennedy88,blow90,milonni95,dung98,garrison04}, and thus here we 
present a simple~\textit{ad hoc} quantization scheme that captures the relevant 
physics.  To quantize the plasmon modes, we take the field solution in 
Eq.~(\ref{eq:tm0mode}) and normalize the energy~(again, ignoring 
$\textrm{Im}\,\epsilon_2$) to
\be 
\hbar\omega={\int}d^{3}\bfr\left(\epsilon_{0}\frac{d}{d\omega}\left(\omega\epsilon(\bfr,\omega)\right)\left|\hat{\bfE}(\bfr)\right|^{2}+\mu_{0}\left|\hat{\bfH}(\bfr)\right|^2\right).\label{eq:normalization} 
\ee
The electric field term in Eq.~(\ref{eq:normalization}) gives the correct 
expression for the classical energy density in a dispersive 
medium~\cite{jackson99}, and the coupling parameter at position $\bfr$ then 
simply follows through the relation 
$g(\bfr)={\bf{p_0}}\cdot\hat{\bfE}(\bfr)/\hbar$.  To evaluate the dispersive 
term, we assume that the conductor forming the wire exhibits Drude-like behavior 
with plasma frequency $\omega_p$, and that we operate well below the plasma 
frequency so that the permittivity is given by 
$\epsilon_2(\omega)=1-\omega_p^2/{\omega^2}{\approx}-{\omega_p^2}/{\omega^2}$.
 For such a metal this dispersive term is positive and given by 
$\frac{d}{d\omega}\left(\omega\epsilon_{2}(\omega)\right){\approx}|\epsilon_{2}(\omega)|$. 
Furthermore, we recall from Sec.~\ref{sec:cylinder} that in the nanowire limit 
the magnetic fields are smaller than the electric fields by a factor $R$.  
Combining these results, we find that the field energy is primarily electric, and
\be 
\hbar\omega{\approx}\epsilon_{0}{\int}d^{3}\bfr|\epsilon(\bfr,\omega)|\left|\hat{\bfE}(\bfr)\right|^{2}.\label{eq:Drudenormalization} 
\ee
Evaluating this equation readily leads to a normalization coefficient 
$b_{1}{\approx}\sqrt{{\hbar\omega k_0^4 \epsilon_1^2 R^2}/{\epsilon_0 \tilde{V} 
C_{-1}^4 L}}$, where $\tilde{V}$ is a dimensionless parameter that depends only 
on the permittivities $\epsilon_{1,2}$,
\be 
\tilde{V}=\frac{8\epsilon_1^2}{{\pi}C_{-1}^2}\left(\frac{1}{|\epsilon_2|}\frac{K_{1}^2(C_{-1})}{I_{1}^2(C_{-1})}\int_{0}^{C_{-1}}dx\,x(I_1^2(x)+I_0^2(x))+\frac{1}{\epsilon_1}\int_{C_{-1}}^{\infty}dx\,x(K_1^2(x)+K_0^2(x))\right). 
\ee
For a dipole oriented in the radial direction at position $\rho'=d$, the 
position-dependent coupling strength immediately follows,
\bea g(d) & = & \frac{p_0}{\hbar}
b_{1}\left|\frac{{\kp}{\koneperp}}{k_1^2}H_{0}^{\prime}\left({\koneperp}d\right)\right| 
\nonumber
\\ & = & \frac{2}{\pi}p_{0}\sqrt{\frac{\omega}{\hbar\epsilon_{0}V_{\footnotesize\textrm{eff}}}}K_{1}({\kappaoneperp}d).\label{eq:gscaling} 
\eea
The effective mode volume defined above is given by 
$V_{\footnotesize\textrm{eff}}=\tilde{V}R^{2}L$ and is proportional to the 
cross-sectional area of the wire and the quantization length $L$.  This result 
reflects the transverse confinement of the plasmon on a scale comparable to 
$R$.  Note that the presence of the $1/\sqrt{\Veff}$ term in $g$ is responsible 
for the strong coupling between plasmon modes and emitter as $R{\rightarrow}0$.

Assuming a Drude model, a scaling law for the density of states 
$D(\omega)=2(L/2\pi)(d\kp/d\omega)$ can also be derived~(the factor of $2$ 
accounts for forward- and backward-propagating plasmons):
\bea D(\omega) & = & \frac{L}{\pi}\frac{d\kp}{d\omega}\nonumber \\
& {\approx} & 
\frac{L}{\pi}\frac{d}{d\omega}\left(\frac{C_{-1}\left(\epsilon_1,\epsilon_2(\omega)\right)}{R}\right)\nonumber
\\
 & \approx & \frac{L}{\pi{R}}\frac{{\partial}C_{-1}}{\partial\epsilon_2}\frac{2|\epsilon_2|}{\omega}.\label{eq:densityofstates} \eea
The important feature of Eq.~(\ref{eq:densityofstates}) is the $1/R$ dependence 
due to the reduced group velocity $d\omega/d\kp{\propto}{\omega}R$ of plasmons 
on the nanowire.

Combining the results of Eqs.~(\ref{eq:gscaling}) and~(\ref{eq:densityofstates}) 
into Eq.~(\ref{eq:goldenrule}), one finds that the decay rate into plasmons in 
the nanowire limit behaves like
\be \Gamma_{\footnotesize\textrm{pl}}\propto
\Gamma_{0}\frac{K_{1}({\kappaoneperp}d)^2}{(k_{0}R)^3}, \ee
which agrees with the results derived previously.  Again, the proportionality 
constant depends only on $\epsilon_{1,2}$.  Physically, the spontaneous emission 
rate into the plasmon modes increases like $1/R^3$ as $R{\rightarrow}0$ due to 
the simultaneous reduction in group velocity~$(v_g{\propto}R)$ of these modes 
and an increase in coupling strength~($g^2{\propto}1/R^2$) due to the 
localization of the field energy to a region whose size is proportional to the 
cross-sectional area of the wire.  

\subsection{Purcell factor of a nanowire}\label{subsec:purcellwire}

Comparing the spontaneous emission rates given by 
Eqs.~(\ref{eq:gammarad}),~(\ref{eq:gammanonrad}), and~(\ref{eq:gammaplasmon}), 
we now qualitatively discuss the behavior one should expect as the position of 
the emitter is varied.  In the limit that $d/R{\gg}1$ clearly the spontaneous 
emission rate is dominated by radiative decay and is equal to the spontaneous 
emission rate $\Gamma_0$ in a uniform dielectric.  As one brings the emitter 
closer to the wire surface, the change in the electromagnetic mode structure 
near the wire results in some modified radiative decay rate 
$\Gamma_{\footnotesize\textrm{rad}}$ which never exceeds approximately 
$4\Gamma_0$ for large $|\epsilon|$.  When the emitter position $d$ approaches 
$d{\sim}1/|\koneperp|{\sim}R/|C_{-1}|$, the emitter starts to be able to decay 
into the localized plasmon fields, with the rate scaling with wire size like 
$1/R^3$.  The spontaneous emission rate into plasmons continues to grow as the 
emitter is brought even closer to the wire edge, $d{\rightarrow}R$.  However, 
the efficiency or probability of plasmon excitation eventually decreases due to 
the large non-radiative decay rate experienced by the dipole very near the wire, 
which diverges like $1/(d-R)^3$.  We thus expect some optimal efficiency of 
spontaneous emission into the plasmon modes to occur when the emitter is 
positioned at a distance $\mathcal{O}(R)$ away from the wire edge, and for this 
optimal efficiency to improve as $R{\rightarrow}0$.

This result is illustrated in Fig.~\ref{fig:fractionalpower}a, where we have 
numerically evaluated the spontaneous emission rates derived previously.  
Specifically, we plot as a function of $R$ the ``error'' probability 
$P_E=1-\Gamma_{\footnotesize\textrm{pl}}/(\Gamma_{\footnotesize\textrm{pl}}+\Gamma')$ 
that a single, excited quantum emitter fails to decay into the plasmon mode.  
Here, 
$\Gamma'=\Gamma_{\footnotesize\textrm{rad}}+\Gamma_{\footnotesize\textrm{non-rad}}$ 
denotes the total emission rate into channels other than the fundamental plasmon 
mode, and the error probability has been optimized over the emitter position 
$d$.  It can be seen that as $R{\rightarrow}0$, the probability of emission into 
the plasmons approaches almost unity.  Examining this limit more carefully, the 
error in fact approaches a small factor 
$P_E{\propto}\textrm{Im}\;\epsilon/(\textrm{Re}\;\epsilon)^2$, explicitly 
indicating that the efficiency is limited by dissipative losses, as will be more 
carefully shown below.  For the chosen parameters the probability of emission 
into the plasmons is well over $99\%$ as $R{\rightarrow}0$, with a corresponding 
effective Purcell factor $\Gamma_{\footnotesize\textrm{pl}}/\Gamma'\approx 
5.2\times 10^2$.  Again, we emphasize that these properties are specifically a 
result of the conducting properties of the nanowire.  This can be contrasted 
with emission into the guided modes of a sub-wavelength optical fiber, which 
drops exponentially as $R{\rightarrow}0$ due to the weak confinement of these 
guided modes~\cite{klimov04}.  In Fig.~\ref{fig:fractionalpower}b, we plot 
$\log_{10}P_E$ as functions of $R$ and $d/R$.  It can be seen that achieving a 
large Purcell factor does not depend too sensitively on the emitter position $d$.

We now prove that the maximum efficiency as $R{\rightarrow}0$ is indeed limited 
by a small factor related to the dissipative losses of the metal. We consider 
the quantity
\be \frac{\Gamma'}{\Gamma_{\footnotesize\textrm{pl}}}{\approx} 
\frac{\left(1+R^{2}/d^{2}\right)^2+\alpha_{\footnotesize\textrm{non-rad}}(k_{0}(d-R))^{-3}}{\alpha_{\footnotesize\textrm{pl}}{(k_{0}R)^{-3}}K_{1}(\kappa_{1\perp}d)^2},\label{eq:defchi} 
\ee
where 
$\alpha_{\footnotesize\textrm{non-rad}}\approx(3/8\epsilon_1^{3/2})\textrm{Im}\,\epsilon/(\textrm{Re}\,\epsilon)^2$ 
is a small parameter explicitly characterizing the losses in the metal.  
Defining $y{\equiv}(d-R)/R$, and using $\kappa_{1\perp}{\approx}C_{-1}/R$ along 
with the asymptotic expression $K_{1}(x)\approx\sqrt{\pi/2x}e^{-x}$ for large 
$x$, we can re-write $\Gamma'/\Gamma_{\footnotesize\textrm{pl}}$ as
\be \frac{\Gamma'}{\Gamma_{\footnotesize\textrm{pl}}}{\approx} 
\frac{2C_{-1}}{\pi\alpha_{\footnotesize\textrm{pl}}}\left(1+y\right)\left[(k_{0}R)^3\left(1+\frac{1}{(1+y)^2}\right)^2+\frac{\alpha_{\footnotesize\textrm{non-rad}}}{y^3}\right]e^{2C_{-1}(1+y)}.\label{eq:emissionfraction} 
\ee
Note that the first term in the brackets corresponds to radiative decay and 
vanishes in the limit that $R{\rightarrow}0$, and thus the ultimate limit to the 
efficiency of plasmon generation is due to a balance between the plasmon and 
non-radiative decay rates.  In this limit, a straightforward calculation yields 
a minimum in the expression above at 
$y_0=(1-C_{-1}+\sqrt{1+C_{-1}(4+C_{-1})})/2C_{-1}$, which confirms that the 
optimum position of the emitter is on the order of a few radii away from the 
wire edge, while the corresponding value of the minimum is proportional to 
$\alpha_{\footnotesize\textrm{non-rad}}$.

\section{Spontaneous emission near a nanotip}\label{sec:nanotip}

In Sections~\ref{sec:cylinder} and~\ref{sec:dipoleemission} we derived and 
discussed the physics of plasmon modes on a nanowire and spontaneous emission of 
a nearby dipole emitter.  For this simple geometry it was possible to find 
analytical solutions and understand the relevant physics of emitter/plasmon 
coupling in conducting nano-structures.  In particular, it was seen that for 
such structures, the tight transverse confinement of the plasmon modes leads to 
a large effective Purcell factor for an optimally positioned dipole emitter as 
the relevant size scale decreases, with the maximum enhancement limited by 
non-radiative decay.  At the same time, however, it is evident that the 
$R{\rightarrow}0$ limit is accompanied by enhanced losses as the plasmon 
propagates, due to the tighter confinement of fields in the metal, and a 
reduction in the plasmon wavelength $\lambda_{\footnotesize\textrm{pl}}$ that 
could make out-coupling more difficult.  Such factors could clearly impose 
limits for applications such as quantum information, but can be circumvented 
with simple design improvements.  In this section, we investigate one specific 
design, a metallic nanotip.  As in the nanowire case, one expects a 
sub-wavelength plasmon mode volume, determined here by the tip curvature, and an 
associated enhancement of emission into the plasmon modes.  At the same time, 
though, the tip can rapidly expand to larger sizes where the propagative losses 
of the plasmons are less severe, and where $\lambda_{\footnotesize\textrm{pl}}$ 
is not as small.  In the nanotip case we are not able to obtain full 
electrodynamic solutions for the plasmon modes.  However, in a manner similar to 
that described in Sec.~\ref{sec:dipoleemission}, we will calculate all of the 
relevant decay rates in the quasistatic limit and describe an approximative 
method to calculate the effects of propagative losses along the nanotip.  We 
will also compare these results to those obtained via fully electrodynamic 
numerical simulations, and we find that these two approaches agree closely.

In the following we will consider a nanotip whose surface can be parameterized 
as a paraboloid of revolution with symmetry along the $z$-axis.  Specifically, 
we suppose that the surface of the nanotip is described by
\be z=\frac{1}{2}\left(\frac{x^2+y^2}{v_0^2}-v_0^2\right),\label{eq:tipequation} 
\ee
a paraboloid of revolution with apex at $z=-v_0^2/2$~(the reason for the offset 
of the apex will become apparent below).  We now introduce a change of 
coordinates,
\bea x & = & uv\cos\phi, \\ y & = & uv\sin{\phi}, \\ z & = & 
\frac{1}{2}\left(u^2-v^2\right). \eea
While these coordinates may seem awkward~(note, for example, that $u,v$ have 
units of $\sqrt{length}$), they are convenient for deriving expressions for the 
fields and spontaneous emission rates, which we will then express in more 
``natural'' coordinates at the end of the calculation.  In these parabolic 
coordinates, the nanotip profile of Eq.~(\ref{eq:tipequation}) is defined by a 
surface of constant $v=v_0$.  More generally, any constant $v$ defines some 
paraboloid of revolution in this system, while the unit vectors $\hat{u}$ and 
$\hat{v}$ run normally and tangentially to these surfaces, respectively.

Now, as in the nanowire case, we are interested in seeking the quasistatic field 
solution for a point charge source in the vicinity of the nanotip, from which we 
can obtain the field due to a dipole $\bfp0$.  In particular, we seek solutions 
of the total field of the form~(\ref{eq:pseudopotential}) with appropriate 
boundary conditions.  Like before, we separate the pseudopotential $\Phi_1$ 
outside the nanotip into its free and reflected components $\Phi_{0,r}$, and use 
an integral representation of the free pseudopotential suitable for parabolic 
coordinates,
\be 
\Phi_{0}(\bfr,\bfrp)=\frac{1}{2\pi\epsilon_0\epsilon_1}\sum_{m=0}^{\infty}(2-\delta_{m,0})\cos\,m(\phi-\phi')\int_{0}^{\infty}dq\;qJ_{m}(qu)J_{m}(qu')I_{m}(qv)K_{m}(qv').\;\;\;\;(v<v') 
\ee
Because $\Phi_{0}$ fully accounts for the point source, $\Phi_{r,2}$ then 
satisfy Laplace's Equation.  Using separation of variables, it is 
straightforward to show that the solutions to Laplace's Equation are given in 
parabolic coordinates by ${\sim}J_{m}(qu)G_{i,m}(qv)e^{im\phi}$, where 
$G_{1,m}(qv)=K_{m}(qv)$ and $G_{2,m}(qv)=I_{m}(qv)$ are non-divergent functions 
in their regions of applicability.  We then define the following expansions,
\bea \Phi_{r}(\bfr,\bfrp) & = & 
\frac{1}{2\pi\epsilon_0\epsilon_1}\sum_{m=0}^{\infty}(2-\delta_{m,0})\cos\,m(\phi-\phi')\int_{0}^{\infty}dq\;q\alpha_{m}(q)J_{m}(qu')K_{m}(qv')J_{m}(qu)K_{m}(qv),\label{eq:Phirnanotip} 
\\ \Phi_{2}(\bfr,\bfrp) & = & 
\frac{1}{2\pi\epsilon_0\epsilon_1}\sum_{m=0}^{\infty}(2-\delta_{m,0})\cos\,m(\phi-\phi')\int_{0}^{\infty}dq\;q\beta_{m}(q)J_{m}(qu')K_{m}(qv')J_{m}(qu)I_{m}(qv), 
\eea
where the coefficients $\alpha,\beta$ will be determined by imposing boundary 
conditions at the nanotip surface $v=v_0$.  It can be easily shown that the 
continuity of $\Phi$ and ${\bf D}_{\perp}$ imply that
\bea \alpha_{m}(q) & = & 
\frac{(\epsilon_1-\epsilon_2)I_{m}'(qv_0)I_{m}(qv_0)}{\epsilon_{2}I_{m}'(qv_0)K_{m}(qv_0)-\epsilon_{1}I_{m}(qv_0)K_{m}'(qv_0)},\label{eq:alphatip} 
\\ \beta_{m}(q) & = & 
\frac{-\epsilon_{1}\left(I_{m}(qv_0)K_{m}'(qv_0)-I_{m}'(qv_0)K_{m}(qv_0)\right)}{\epsilon_{2}I_{m}'(qv_0)K_{m}(qv_0)-\epsilon_{1}I_{m}(qv_0)K_{m}'(qv_0)}. 
\eea
Note that the coefficients $\alpha_{m}(q)$, along with 
Eq.~(\ref{eq:Phirnanotip}), completely determine the reflected field.

The calculation of the radiative and non-radiative spontaneous emission rates 
proceeds in the same manner as the nanowire case.  To calculate 
$\Gamma_{\footnotesize\textrm{rad}}$, we again look for a dipole term in the 
far-field~(large $v$) that corresponds to an induced dipole moment $\delta{\bf 
p}$ in the nanotip, and then use the relationship 
$\Gamma_{\footnotesize\textrm{rad}}{\propto}|\bfp0+\delta{\bf p}|^2$.  At the 
same time, we look for a divergent contribution to the reflected field at the 
dipole location as its position $v'$ approaches $v_0$, which yields the leading 
term of the non-radiative decay rate through 
$\Gamma_{\footnotesize\textrm{non-rad}}{\propto}\textrm{Im}(\bfp0\cdot\bfE_r(\bfrp,\bfrp))$.  
This divergence is physically due to the dissipation of divergent currents 
induced in the metal by the dipole.  For a dipole positioned along the 
$z$-axis~($u'=0$), 
\bea \bfp0\cdot\bfE_{r}(\bfrp,\bfrp) & = & 
-\frac{p_0^2}{4\pi\epsilon_0\epsilon_1}\int_{0}^{\infty}dq\;\frac{q^3}{v'^2}\alpha_{1}(q)K_1^2(qv'),\;\;\;\;\;\;(\bfp0\perp\hat{z}) 
\nonumber \\ \bfp0\cdot\bfE_{r}(\bfrp,\bfrp) & = & 
-\frac{p_0^2}{2\pi\epsilon_0\epsilon_1}\int_{0}^{\infty}dq\;\frac{q^3}{v'^2}\alpha_{0}(q)K_1^2(qv'),\;\;\;\;\;\;(\bfp0\parallel\hat{z})\label{eq:pEtip} 
\eea
as shown more carefully in Appendix~\ref{sec:tiprates}.  Mathematically, the 
divergence as $v'{\rightarrow}v_0$ occurs due to the presence of a long tail 
${\sim}e^{-2q(v'-v_0)}$ in the integrand for large $q$.  Because of the 
similarity of the decay rate calculations with those in 
Sec.~\ref{sec:dipoleemission}, we simply state the results here, while providing 
more details in Appendix~\ref{sec:tiprates}.  For a dipole positioned along the 
$z$-axis at $v=v'$, the radiative and non-radiative spontaneous emission rates 
are given by
\bea \frac{\Gamma_{\footnotesize\textrm{rad}}}{\Gamma_0} & = & 
\left|1-\frac{v_0^2}{v'^2}\left(1-\frac{\epsilon_2}{\epsilon_1}\right)\right|^2,\;\;\;\;\;(\bfp0\parallel\hat{z}) 
\nonumber
\\ \frac{\Gamma_{\footnotesize\textrm{rad}}}{\Gamma_0} & = & 
\left|1+\frac{\epsilon_1-\epsilon_2}{\epsilon_1+\epsilon_2}\frac{v_0^2}{v'^2}\right|^2,\;\;\;\;\;(\bfp0\perp\hat{z}) 
\label{eq:tiprad} \eea
and
\bea \Gamma_{\footnotesize\textrm{non-rad}}/\Gamma_{0} & {\approx} & 
\frac{3}{16k_0^3\epsilon_1^{3/2}}\frac{1}{v^{{\prime}3}(v'-v_0)^3}\textrm{Im}\left(\frac{\epsilon_2-\epsilon_1}{\epsilon_2+\epsilon_1}\right),\;\;\;\;\;\;(\bfp0\perp\hat{z}) 
\nonumber \\ \Gamma_{\footnotesize\textrm{non-rad}}/\Gamma_{0} & {\approx} & 
\frac{3}{8k_0^3\epsilon_1^{3/2}}\frac{1}{v^{\prime{3}}(v'-v_0)^3}\textrm{Im}\left(\frac{\epsilon_2-\epsilon_1}{\epsilon_2+\epsilon_1}\right).\;\;\;\;\;\;(\bfp0\parallel\hat{z}) 
\label{eq:tipnonrad} \eea

Finally we consider the decay rate into the fundamental plasmon mode of the 
nanotip, which is associated with the contribution of the poles in the integrand 
of Eq.~(\ref{eq:pEtip}) to 
$\textrm{Im}\left(\bfp0\cdot\bfE_r(\bfrp,\bfrp)\right)$.  The presence of a pole 
indicates the excitation of a natural mode of the system.  Examining the 
solutions to $\alpha_{0,1}$ given in Eq.~(\ref{eq:alphatip}), one finds that 
$\alpha_1$ has no pole in the range $0\leq q \leq\infty$.  Physically, the 
absence of a pole means that a dipole simultaneously oriented perpendicular to 
$\hat{z}$ and located along the $z$-axis does not excite the fundamental plasmon 
mode of the nanotip.  This is easily understood since a dipole oriented this way 
is anti-symmetric with respect to $180^{\circ}$ rotations about $\hat{z}$, while 
the plasmon mode is symmetric.  On the other hand, $\alpha_0$ does have a pole 
corresponding to plasmon excitation.  This pole is located at $q_0=C_{-1}/v_0$, 
where $C_{-1}$ is the solution to Eq.~(\ref{eq:staticmodecondition}).  
Evaluating the contribution of this pole to the field is straightforward and 
yields a plasmon decay rate
\be 
\frac{\Gamma_{\footnotesize\textrm{pl}}}{\Gamma_0}=\frac{3\pi}{k_0^3\epsilon_1^{3/2}}\frac{C_{-1}^3}{v_{0}^{4}v'^2}K_1^2(q_{0}v')\frac{(\epsilon_1-\epsilon_2)I_{1}(C_{-1})I_0(C_{-1})}{d\chi(C_{-1})/dx},\label{eq:tipplasmon} 
\ee
where $\chi(x)$ is defined in Eq.~(\ref{eq:chiqv}).  As in the nanowire case, 
the decay rate $\Gamma_{\footnotesize\textrm{pl}}$ into the plasmon mode given 
by Eqs.~(\ref{eq:tipplasmon}) and~(\ref{eq:chiqv}) is evaluated in the limit 
that $\textrm{Im}\,\epsilon_2=0$, such that $q_0$ and $C_{-1}$ are purely real.

Having derived the decay rates in parabolic coordinates, we now define a more 
natural set of parameters to describe the system.  Let us introduce a length 
scale $w$ that characterizes the nanotip via $\rho(z)=\sqrt{wz}\;(z \geq 0)$, 
where $\rho$ is the radius of the nanotip at position $z$~(note also the 
corresponding shift in the apex of the tip from $z=-v_0^2/2$ to $z=0$).  
Furthermore, let $z=-d<0$ be the position of the emitter~($d$ is the distance 
between the emitter and end of the nanotip).  In terms of these parameters, the 
spontaneous emission rates derived above can be re-written as
\bea \frac{\Gamma_{\footnotesize\textrm{rad}}}{\Gamma_0} & = & 
\left|1+\left(1+4d/w\right)^{-1}\left(\frac{\epsilon_2}{\epsilon_1}-1\right)\right|^2, 
\\ \frac{\Gamma_{\footnotesize\textrm{non-rad}}}{\Gamma_0} & = & 
\frac{3}{8\epsilon_1^{3/2}}\frac{1}{(k_{0}d)^3}\textrm{Im}\left(\frac{\epsilon_2-\epsilon_1}{\epsilon_2+\epsilon_1}\right), 
\\ \frac{\Gamma_{\footnotesize\textrm{pl}}}{\Gamma_0} & = & 
\tilde{\alpha}_{\footnotesize\textrm{pl}}\frac{1}{(k_{0}w)^3(1+4d/w)}K_{1}^{2}(C_{-1}\sqrt{1+4d/w}), 
\eea
where $\tilde{\alpha}_{\footnotesize\textrm{pl}}$ only depends on 
$\epsilon_{1,2}$ and is given by
\be 
\tilde{\alpha}_{\footnotesize\textrm{pl}}=\frac{24\pi}{\epsilon_{1}^{3/2}}C_{-1}^{3}\frac{(\epsilon_1-\epsilon_2)I_{1}(C_{-1})I_{0}(C_{-1})}{d\chi(C_{-1})/dx}. 
\ee

\begin{figure*}[p]
\begin{center}
\includegraphics[width=17cm]{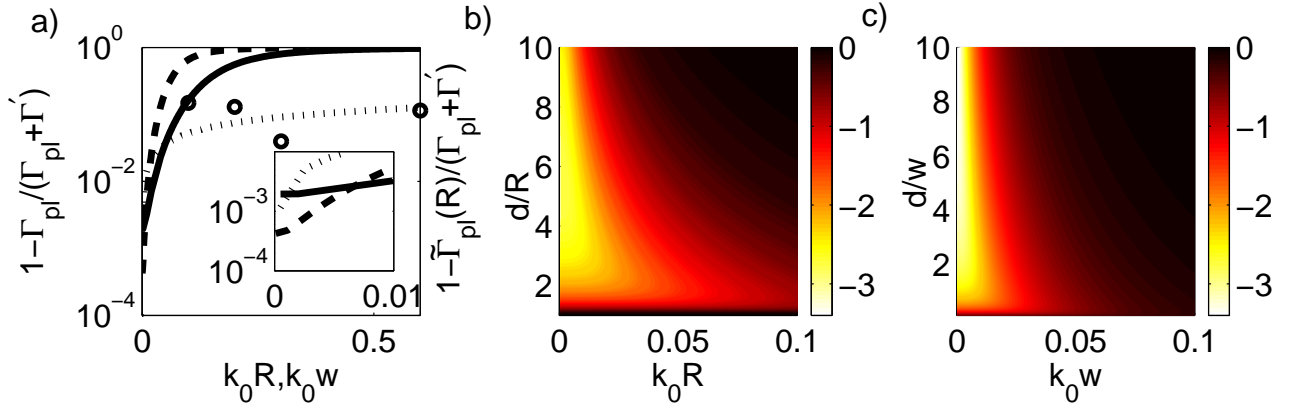}
\end{center}
\caption{a.  Solid line: Probability of error, 
$P_E=1-\Gamma_{\footnotesize\textrm{pl}}/(\Gamma'+\Gamma_{\footnotesize\textrm{pl}})$, 
in which an emitter fails to emit into the fundamental plasmon mode for a 
nanowire, plotted as a function of $R$ and optimized over the emitter position.  
Dashed line: optimized $P_E$ vs. curvature parameter $w$ for a nanotip. Dotted 
line: effective probability of error, 
$\tilde{P}_E=1-\tilde{\Gamma}_{\footnotesize\textrm{pl}}(R)/(\Gamma'+\Gamma_{\footnotesize\textrm{pl}})$ 
for emission into a nanotip and successful propagation to final radius $R$.  
Solid points: effective error probability $\tilde{P}_E$ for a nanotip, 
calculated numerically through boundary element method.  Inset: same plot, 
zoomed in near $R,w=0$.  b.  Contour plot of $\log_{10}P_{E}$ for a nanowire, as 
functions of $R$ and $d/R$.  c.  Contour plot of $\log_{10}P_E$ for a nanotip, 
as functions of $w$ and $d/w$.\label{fig:fractionalpower}}
\end{figure*}

Having obtained the spontaneous emission rates near a nanotip into the different 
possible channels, it is once again possible to optimize the efficiency of 
emission into the plasmon modes for given $w$ by varying the emitter position 
$d$.  The corresponding optimized error probability, 
$P_E=1-\Gamma_{\footnotesize\textrm{pl}}/(\Gamma_{\footnotesize\textrm{pl}}+\Gamma')$, 
is plotted as a function of $w$ in Fig.~\ref{fig:fractionalpower}a.  In analogy 
to the nanowire system, a large effective Purcell enhancement of 
$\Gamma_{\footnotesize\textrm{pl}}/\Gamma'{\approx}2.5\times 10^3$ arises as 
$w{\rightarrow}0$, due to a balance between the small mode volumes associated 
with the plasmons and the non-radiative decay rate.  In 
Fig.~\ref{fig:fractionalpower}c, we plot $\log_{10}P_E$ as functions of $w$ and 
$d/w$.  Once again, it can be seen that the error is not too sensitive to the 
emitter position.

Because the plasmon modes here were obtained through a quasistatic 
approximation, this calculation yields no information about dissipative losses 
as the plasmon propagates along the nanotip.  For example, in this limit 
$\bfH{\approx}0$ so one cannot obtain the Poynting vector for the system.  To 
estimate the effect of propagative losses, however, we can make an eikonal 
approximation~\cite{stockman04}, assuming that the plasmons are emitted 
completely into the end of the tip~($z=0$), and that the propagative losses 
thereafter at any position $z$ are described locally by the nanowire solution at 
radius $\rho(z)$.  This motivates us to define an effective decay rate
\be 
\tilde{\Gamma}_{\footnotesize\textrm{pl}}(R)=\Gamma_{\footnotesize\textrm{pl}}\textrm{exp}\left(-2\int_{0}^{z(R)}\textrm{Im}\,\kp(\rho(z))dz\right),\label{eq:eikonal} 
\ee
which equals the rate of decay into the plasmons multiplied by the probability 
that an emitted plasmon will propagate without dissipation to some final nanotip 
radius $R$.  Here $\Gamma_{\footnotesize\textrm{pl}}$ is the decay rate for the 
nanotip obtained earlier, while $\kp(\rho)$ is the plasmon wavevector for a 
nanowire of radius $\rho$.  One can also define a corresponding effective error 
probability 
$\tilde{P}_E(R)=1-\tilde{\Gamma}_{\footnotesize\textrm{pl}}(R)/(\Gamma_{\footnotesize\textrm{pl}}+\Gamma')$ 
for the nanotip.  Physically, $\tilde{P}_E$ is the probability that the plasmon 
mode is either not excited by the emitter, or is excited but fails to 
successfully propagate to final radius $R$.  In Fig.~\ref{fig:fractionalpower}a 
we plot this quantity as a function of $R$, when optimized over possible tip 
parameters $w$ and $d$.  It can be seen that the effective error probability for 
a nanotip compares favorably to that of a nanowire when $k_0{R}\gtrsim 0.05$.  
In other words, the nanotip configuration is able to simultaneously exhibit a 
strong Purcell effect and reduce propagative losses.  We note that the nanotip 
system also has the added benefit of generating guided plasmons along a single 
direction of propagation.

Finally we discuss the limits of validity of the equations derived above for the 
nanotip.  The quasistatic decay rates are valid in the regime 
$|k_i|w,|k_i|d{\ll}1$, which implies that propagative phases associated with the 
electrodynamic solution can be ignored over the length scales of interest.  At 
the same time, Eq.~(\ref{eq:eikonal}) assumes that the plasmon mode at some tip 
radius $\rho$ adiabatically follows the nanowire solution of corresponding 
radius.  One can define an adiabatic parameter 
$\beta=d(1/\textrm{Re}\,\kp(\rho))/dz$ associated with the propagation, which 
must be small for such an assumption to be valid.  Physically, a small $\beta$ 
corresponds to a small correction to the propagative phase due to the variation 
of $\textrm{Re}\,\kp(\rho(z))$, compared to the ${\sim}2\pi$ phase acquired over 
a distance of the plasmon wavelength.  Assuming, for example, that we are 
considering sufficiently small length scales that 
$\textrm{Re}\,\kp{\approx}|\kp|{\approx}|C_{-1}|/\rho(z){\approx}|C_{-1}|/\sqrt{wz}$, 
the condition that $\beta{\ll}1$ implies that the eikonal approximation is valid 
only in regions where $z{\gg}w/|C_{-1}^2|$.  It can be seen that 
$z_c{\equiv}w/|C_{-1}^2|$ represents some cross-over value, below which 
Eq.~(\ref{eq:eikonal}) clearly is invalid.  On the other hand, 
$|1/q_0^2|=v_0^2/|C_{-1}^2|{\sim}w/|C_{-1}^2|$ sets the relevant length scale 
for the plasmons on the nanotip, and one expects dissipation to occur on length 
scales much longer than this.  Thus, as long as the losses predicted by 
Eq.~(\ref{eq:eikonal}) for $z<z_c$ remain small, one can effectively use this 
equation for all $z$ even if it is not strictly valid for $z<z_c$.  A 
straightforward calculation confirms that the predicted loss, 
$1-\tilde{\Gamma}_{\footnotesize\textrm{pl}}(R_c)/{\Gamma}_{\footnotesize\textrm{pl}}{\sim}1-\textrm{exp}(-\textrm{Im}\,C_{-1}/|C_{-1}|)$, 
is indeed negligible, where $R_c{\equiv}R(z_c)$.  Finally, in practice, for the 
applications of interest we will be primarily interested in nanotip devices 
whose radii do not grow indefinitely, but rather expand until they reach some 
final, constant radius $R$.  Such devices, for example, are more likely to be 
easily out-coupled, as discussed further in the next section.  For 
Eq.~(\ref{eq:eikonal}) to remain valid for such a system, the spontaneous 
emission rate into plasmons for this device must be close to the rate calculated 
for an infinite, perfectly paraboloidal tip.  This imposes the additional 
requirement that the final radius $R$ be much larger than $R_c$.

\begin{figure*}[p]
\begin{center}
\includegraphics[width=15cm]{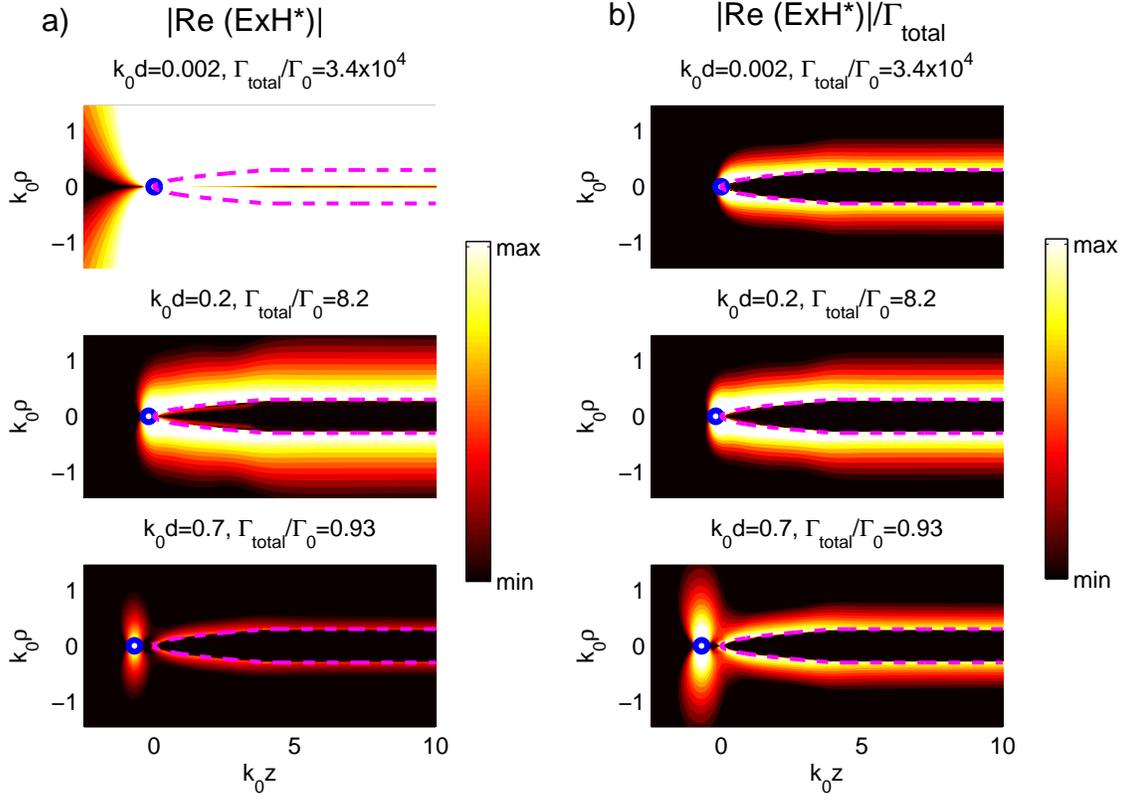}
\end{center}
\caption{Numerically calculated fields due to a dipole emitter near a conducting 
nanotip, obtained by boundary element method.  a.  The energy flux $|\textrm{Re} 
(\bfE\times\bfH^{\ast})|$, in arbitrary units.  The position of the emitter is 
denoted by the blue circles, while the boundary of the nanotip is given by the 
dotted lines .  The plots shown are for a final nanotip radius of $k_{0}R=0.3$, 
curvature parameter $k_{0}w=0.022$, and emitter positions 
$k_{0}d=0.002,0.2,0.7$.  It can be seen that both the total spontaneous emission 
rate $\Gamma_{\footnotesize\textrm{total}}$ and the emission rate into plasmons 
increase as the emitter is brought closer to the nanotip.  b.  The quantity 
$|\textrm{Re} (\bfE\times\bfH^{\ast})|/\Gamma_{\footnotesize\textrm{total}}$, 
for the same parameters.  This quantity is proportional to the energy flux 
normalized by the total power output of the emitter.  The $k_{0}d=0.002$ plot is 
mostly dark, indicating that most of the decay is into non-radiative channels.  
The $k_{0}d=0.2$ case is characterized by bright spots along the entire edge of 
the nanotip, which indicates efficient plasmon excitation.  The $k_{0}d=0.7$ 
case exhibits the typical lobe pattern associated with radiative decay of a 
dipole.\label{fig:BEMplot}}
\end{figure*}

To check the analytical results derived above for the nanotip, we have also 
performed detailed numerical simulations using boundary element 
method~(BEM)~\cite{abajo02}.  Details of our implementation are given in 
Appendix~\ref{sec:BEM}.  BEM simulations are fully electrodynamic solvers of 
Maxwell's Equations, and they were used to obtain the classical electromagnetic 
field solutions of an oscillating dipole emitter $\bfp0 e^{-i\omega t}$ near a 
nanotip.  The results of a few sample simulations are shown in 
Fig.~\ref{fig:BEMplot}, for a tip curvature parameter $k_{0}w=0.022$, final 
radius $k_{0}R=0.3$, and varying emitter positions $k_{0}d=0.002,0.2,0.7$.  In 
Fig.~\ref{fig:BEMplot}a, we plot the quantity $|\textrm{Re} 
(\bfE\times\bfH^{\ast})|$, which is proportional to the Poynting vector and 
corresponds to the total energy flux of the system.  The nanotip is assumed to 
be composed of silver in a surrounding dielectric $\epsilon_1=2$, and its 
boundary is given by the dotted line, while the emitter positions are denoted by 
the circles.  The total spontaneous emission rate is given via 
$\Gamma_{\footnotesize\textrm{total}}=(\Gamma_{\footnotesize\textrm{pl}}+\Gamma'){\propto}\textrm{Im}\left(\bfp0\cdot\bfE_{1}(\bfrp,\bfrp)\right)$ 
and is determined numerically for each configuration by finding the total field 
at the dipole location.  On the other hand, the effective emission rate 
$\tilde{\Gamma}_{\footnotesize\textrm{pl}}(R)$ into the plasmons is determined 
by a best fit of the fields in the region of constant $R$ to the known plasmon 
solution on a nanowire given in Eq.~(\ref{eq:tm0mode}), and then calculating the 
total power transport of this best-fit mode through the integrated Poynting 
vector.  This total power is directly proportional to 
$\tilde{\Gamma}_{\footnotesize\textrm{pl}}(R)$.  The figure confirms the 
qualitative behavior that we expect and have described previously.  In 
particular, the generated plasmon field and total spontaneous emission rate are 
largest for very small separations and decrease as the emitter is placed further 
away from the end of the nanotip.  In Fig.~\ref{fig:BEMplot}b, we plot 
$|\textrm{Re} (\bfE\times\bfH^{\ast})|/\Gamma_{\footnotesize\textrm{total}}$, 
which is proportional to the energy flux normalized by the total power output of 
the emitter.  This quantity yields information about the efficiency of decay 
into the various channels.  For small separations~($k_{0}d=0.002$), the plot is 
mostly dark, which indicates that the decay of the dipole is predominantly 
non-radiative.  For $k_{0}d=0.2$, the maximum~(corresponding to bright spots in 
the plot) is located along the entire surface of the nanotip, which indicates 
highly efficient plasmon excitation.  Here, although the total emission rate 
into plasmons decreases from the $k_{0}d=0.002$ case~(as seen in 
Fig.~\ref{fig:BEMplot}a), the efficiency increases dramatically due to less 
competition from non-radiative decay. Finally, for $k_{0}d=0.7$, the maximum 
appears as the typical lobe pattern associated with radiative decay.

In Fig.~\ref{fig:fractionalpower}a, we have plotted the numerically optimized 
values of 
$\tilde{P}_{E}(R)=1-\tilde{\Gamma}_{\footnotesize\textrm{pl}}(R)/(\Gamma_{\footnotesize\textrm{pl}}+\Gamma')$ 
for a few values of $R$.  It can be seen that the values of $\tilde{P}_{E}$ 
obtained through analytical approximations and numerical BEM closely agree.  
Unlike the theoretical predictions, however, the numerically calculated error 
probability does not increase monotonically with $R$.  We believe that the 
origin of this is that for the numerically optimized parameters, the condition 
$R{\gg}R_c$ under which the theoretical predictions hold is only weakly 
satisfied, and the excitation region for the plasmons cannot strictly be thought 
of as a single point at the end of the tip~($z=0$).

\section{Single photon generation via coupling to dielectric waveguide}\label{sec:opticalfiber}
We have shown in previous sections that a single emitter can spontaneously emit 
into the guided plasmon modes of a nearby nano-structure with high probability.  
This prospect of efficient conversion between excitation of the emitter and a 
single photon has a number of applications in the fields of quantum computing 
and quantum information. In this section, we consider one particular 
application, involving the use of such a system as an efficient single-photon 
source.  The concepts behind single-photon generation on demand with an 
individual emitter in a cavity have been discussed 
elsewhere~\cite{michler00,pelton02,mckeever04} and will not be presented in 
detail here. We note also that the ideas behind single-photon sources can be 
extended to create long-distance entanglement between emitters, as detailed, 
\textit{e.g.}, in~\cite{vanenk97}.

\begin{figure*}[p]
\begin{center}
\includegraphics[width=8cm]{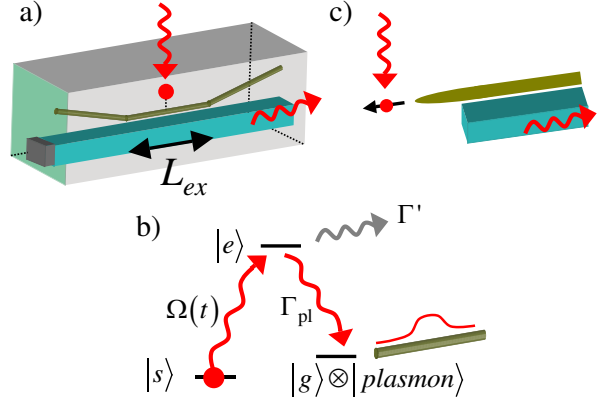}
\end{center}
\caption{a.  An emitter coupled to a nanowire is optically excited and decays 
with high probability into the plasmon modes of the nanowire.  A single photon 
source is created by evanescently coupling the nanowire to a nearby dielectric 
waveguide over a length $L_{ex}$.  The single photon source can potentially be 
uni-directional, \textit{e.g.}, by capping one end of the waveguide with a 
reflective surface.  b.  An internal-level scheme that allows for shaping of the 
outgoing single photon pulses.  An emitter that starts in state $\ket{s}$ is 
coupled to excited state $\ket{e}$ via a time-dependent external field 
$\Omega(t)$.  We assume that the excited state $\ket{e}$ is coupled to state 
$\ket{g}$ via the plasmon modes, causing $\ket{e}$ to decay into $\ket{g}$ with 
high probability, while simultaneously generating a single photon in the plasmon 
modes.  The shape of the photon wavepacket is determined by $\Omega(t)$.  c.  A 
similar scheme for single photon generation using an emitter coupled to a 
nanotip instead of a nanowire.  Note that this scheme is naturally 
uni-directional, as the generated plasmons propagate in a single direction.  
\label{fig:singlephotondevice}}
\end{figure*}

Because of dissipative losses in metals, the plasmon modes are not directly 
suitable as carriers of information over long distances.  We show, however, that 
plasmonic devices can serve as an effective intermediate step, and in particular 
can be efficiently out-coupled to the modes of a co-propagating dielectric 
waveguide.  The single photon device is illustrated schematically in 
Fig.~\ref{fig:singlephotondevice}.  In Fig.~\ref{fig:singlephotondevice}a, an 
optically addressable emitter with multiple internal levels sits in the vicinity 
of a conducting nanowire.  The emitter is strongly coupled to the nanowire, such 
that single photons on demand can be generated with high efficiency in the 
plasmon modes by external manipulation of the emitter.  The addressability of 
the emitter along with the internal levels allows for shaping of this 
single-photon pulse~\cite{cirac97}, as illustrated in 
Fig.~\ref{fig:singlephotondevice}b.  Here, a three-level emitter is shown with 
two ground or metastable states $\ket{s},\ket{g}$, which both have 
dipole-allowed transitions to the excited state $\ket{e}$.  We assume that the 
system is prepared initially in the state $\ket{s}$, and that the 
$\ket{e}-\ket{g}$ transition is coupled via the plasmon modes of the nanowire, 
\textit{i.e.}, the state $\ket{e}$ can decay at a rate 
$\Gamma_{\footnotesize\textrm{pl}}$ into $\ket{g}$ by emitting a photon into the 
plasmon modes.  In addition, there is a small rate $\Gamma'$ at which the 
excited state can decay without emitting a plasmon.  A single photon in the 
plasmon modes of the nanowire is generated with high probability by exciting the 
$\ket{s}-\ket{e}$ transition with some external pulse $\Omega(t)$ and the 
subsequent decay into $\ket{g}$, with the shape of the single photon wavepacket 
controlled by the shape of $\Omega(t)$.  We further assume that the plasmon is 
then evanescently coupled to a nearby dielectric waveguide, as shown in 
Fig.~\ref{fig:singlephotondevice}a, which co-propagates with the nanowire over 
some distance $L_{ex}$ over which this coupling is non-negligible.  The coupling 
is a reversible process, and the distance $L_{ex}$ is optimized to maximize 
efficiency of ending up with a single photon in the waveguide~(\textit{i.e.}, to 
prevent further Rabi oscillations back into the nanowire).  A similar setup with 
a nanotip is illustrated in Fig.~\ref{fig:singlephotondevice}c.  Here the 
nanotip radius $\rho(z)$ expands to some final radius $R$ at which point 
coupling with the waveguide starts to occur.  Initiating the coupling once the 
nanotip has reached a constant radius allows the two systems to be easily 
coupled, as discussed below.  When optimized, we estimate that single-photon 
generation efficiencies exceeding ${\sim}90\%$ are possible in this tiered 
configuration.

\begin{figure*}[p]
\begin{center}
\includegraphics[width=7cm]{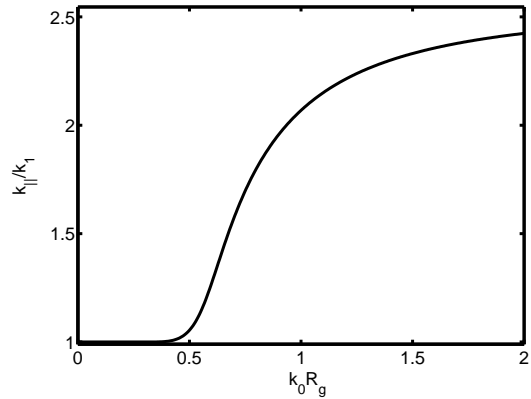}
\end{center}
\caption{Wavevector $\kp$ of the fundamental guided modes of a cylindrical 
dielectric waveguide with core permittivity $\epsilon_{c}=13$ and surrounding 
permittivity $\epsilon_1=2$, plotted as a function of core radius 
$R_{g}$.\label{fig:fundamentalfibermodes}}
\end{figure*}

To treat the problem analytically, we consider the simple situation of our 
nano-structure coupled to a cylindrical dielectric waveguide~(\textit{e.g.}, an 
optical fiber) of radius $R_{g}$, such that the modes can be calculated 
analytically using the methods described in Appendix~\ref{sec:generaltheory}. It 
can be shown that the fundamental modes of the waveguide are degenerate 
$m={\pm}1$ modes that are not cut off as $R_{g}{\rightarrow}0$. The dependence 
of their wavevector $\kp$ on $R_{g}$ is shown in 
Fig.~\ref{fig:fundamentalfibermodes}, for a core permittivity $\epsilon_g=13$ 
and surrounding permittivity $\epsilon_1=2$. These parameters correspond closely 
to that of a Si/Si$\textrm{O}_2$ guide at $\lambda_0=1\;\mu$m.  To simplify the 
calculation, we also assume that coupling between the wire and higher-order 
waveguide modes is negligible.  This can be achieved, for example, by operating 
below the cutoff radius of higher-order modes or by operating with sufficiently 
large wavevector mismatch between the plasmon and higher-order guide modes.

We make the ansatz that the total field of the system is given by a 
superposition of the unperturbed modes of the nano-structure and waveguide.  
While this cannot strictly be correct, as such a solution violates boundary 
conditions at each interface, we rely on such an assumption to give us the 
correct qualitative behavior without resorting to more
complex numerical calculations.  Specifically, we assume that the total electric field for the system takes the form %
\be 
\bfE_{T}(\bfr)=\sum_{\mu=w,g}\sum_{i=1}^{N_{\mu}}C_{\mu,i}(z)\bfE_{\mu,i}(\bfr),\label{eq:Etotalansatz} 
\ee
where $\mu$ indexes the nano-structure~($w$) and waveguide~($g$) systems, and 
$i=1,{\cdots},N_{\mu}$ runs over the modes of system $\mu$.  In the following we 
will explicitly treat the nanotip case, where the plasmons propagate in a single 
direction, although this argument can easily be extended to the nanowire.  We 
emphasize that we are considering coupling of the plasmon mode to the waveguide 
once the nanotip has already expanded to its final radius $R$, at which point 
the plasmon mode solution becomes identical to that of a nanowire.  In our case, 
$N_w=1$ as we only consider the fundamental plasmon mode of the nanotip, while 
$N_{g}=2$ as we take into account the two degenerate, co-propagating fundamental 
modes of the waveguide.  $\bfE_{\mu,i}(\bfr)$ here represents the unperturbed 
solution of mode $i$ in system $\mu$~(without the presence of the other 
system).  A similar expression holds for the total magnetic field.

With the ansatz of Eq.~(\ref{eq:Etotalansatz}) for the total field of the 
combined waveguide and nanotip system, one can derive exact equations of 
evolution~\cite{barclay03} based on Lorentz reciprocity for the coefficients 
$C_{\mu,i}$.  Explicitly separating out the plane-wave dependence of the 
unperturbed fields, 
$\bfE_{\mu,i}(\bfr)=\bfE_{\mu,i}(\bfrho)e^{i\kp{}_{\mu,i}z}$, the $N_w+N_g$ 
coupled-mode equations take the form
\be 
\sum_{\nu=w,g}\sum_{j=1}^{N_{\nu}}P_{\mu,i;\nu,j}(z)\frac{dC_{\nu,j}}{dz}=i\omega\epsilon_{0}\sum_{\nu=w,g}\sum_{j=1}^{N_{\nu}}K_{\mu,i;\nu,j}(z)C_{\nu,j}(z),\label{eq:wgcoupling} 
\ee
as derived in detail in Appendix~\ref{sec:coupledmodeeqns}.  The coefficients to 
the system of equations above are given by
\bea P_{\mu,i;\nu,j}(z) & = & 
e^{i(\kp{}_{\nu,j}-\kp{}_{\mu,i}^{\ast})z}\int\,d\bfrho\,\left(\bfE_{\nu,j}(\bfrho)\times\bfH_{\mu,i}^{\ast}(\bfrho)+\bfE_{\mu,i}^{\ast}(\bfrho)\times\bfH_{\nu,j}(\bfrho)\right)\cdot\hat{z}, 
\\ K_{\mu,i;\nu,j}(z) & = & 
e^{i(\kp{}_{\nu,j}-\kp{}_{\mu,i}^{\ast})z}\int\,d\bfrho\,\bfE_{\nu,j}(\bfrho)\cdot\bfE_{\mu,i}^{\ast}(\bfrho)\left(\epsilon_{T}(\bfrho)-\epsilon_{\nu}(\bfrho)\right), 
\eea
where $\epsilon_{T}(\bfrho)$ is the electric permittivity of the combined 
system.  Clearly, the presence of the phase factors 
$e^{i(\kp{}_{\nu,j}-\kp{}_{\mu,i}^{\ast})z}$ in the equations above indicate 
that, at least under weak coupling, significant power transfer between the two 
systems will not take place unless the two systems are approximately 
mode-matched with respect to $\kp$.  In practice, this implies that for a final 
tip radius $R$, there is some ideal waveguide size $R_g$ that allows for maximum 
transfer efficiency between the two systems.  A similar optimization of the 
waveguide parameters exists in the case of arbitrary coupling strength between 
the two systems, although this problem is more complex because one must account 
for factors such as the phase shift of one system due to the other.  We 
emphasize that the coupled-mode equations above are exact within the ansatz of 
Eq.~(\ref{eq:Etotalansatz}).  For example, for two lossless systems these 
equations conserve power, and for a lossy system~(such as a nanotip) the effects 
of losses are treated exactly.  By convention, the integrals appearing in the 
diagonal matrix elements $P_{\mu,i;\mu,i}$ are typically set to $1$.

For the waveguide and nanotip systems coupled over a length $L_{ex}$, the exact 
single-photon generation efficiency will depend on the details of how the two 
systems are brought together and separated apart.  In practice, for example, the 
two systems should be brought together slowly enough that the introduction of 
the waveguide does not cause significant back-scattering of the plasmon, yet 
quickly enough that this introduction length is small compared to the plasmon 
decay length.  Furthermore, in reality the coupling region will not be a step of 
length $L_{ex}$ but will be characterized by some smooth transition.  To avoid 
the many details associated with this introduction and separation and to 
approximately calculate the efficiency, we will consider an idealized system and 
make three assumptions:

\begin{enumerate}[(i)]
\item  The decay rates of the emitter are not affected by the presence of the 
nearby dielectric waveguide.  In particular, the Purcell factors and error 
probabilities calculated earlier for the nanotip are unchanged. \item  The 
radius of the nanotip is given by $\rho(z)=\sqrt{wz}$ for $z<z_0$ and becomes 
constant, $R\equiv\rho(z_0)=\sqrt{wz_0}$, for $z{\geq}z_0$.  For $z{\geq}z_0$ 
the plasmon mode solution becomes identical to the nanowire solution, and in 
particular has well-defined $\kp$ which allows it to be easily mode-matched with 
the waveguide.  It is assumed that coupling between the nanotip and waveguide 
begins at $z=z_0$, with the initial field amplitudes of the coupled system given 
by
\bea C_w(z_0) & = & \left(1-\tilde{P}_{E}(R)\right)^{1/2}, 
\\ C_{g,i}(z_0) & = & 0, \eea
where the effective plasmon excitation probability $1-\tilde{P}_{E}(R)$ is 
already optimized for a given $R$ over the nanotip curvature and emitter 
position.
\item  Eq.~(\ref{eq:wgcoupling}) exactly describes the coupling between the two 
systems in the region $z_0{\leq}z{\leq}z_0+L_{ex}$.  To estimate the probability 
of transfer from nanowire to waveguide after distance $L_{ex}$ when the two 
systems are once again separated, we project the total field of 
Eq.~(\ref{eq:Etotalansatz}) at $z=z_0+L_{ex}$ into the waveguide mode.  
Specifically, the projected field amplitude in the waveguide in either of the 
degenerate modes $i$ is given by
\be 
C_{\footnotesize\textrm{proj},i}(z_0+L_{ex})=2\int\,d{\bfrho}\,\left(\bfE_{T}(\bfr)\times\bfH_{g,i}^{\ast}(\bfr)\right)\cdot\hat{z}, 
\ee
where the factor of $2$ arises due to the normalization convention adopted here 
for the unperturbed modes, $P_{g,i;g,i}=\int d\bfrho 
\left(\bfE_{g,i}\times\bfH_{g,i}^{\ast}+\bfE_{g,i}^{\ast}\times\bfH_{g,i}\right)=1$.
\end{enumerate}

Because of the symmetry, the projected field strengths 
$|C_{\footnotesize\textrm{proj},i}|^2$ calculated above are equal for the two
degenerate waveguide modes, and the quantity 
$2|C_{\footnotesize\textrm{proj},i}|^2$ then corresponds to the efficiency of 
single photon generation.  Here the additional factor of $2$ accounts for the 
mode degeneracy.  This quantity takes completely into account the propagative 
losses of the plasmons, imperfect coupling between the nanotip and waveguide, 
and the Purcell factor of the nanotip.  

\begin{figure}
\begin{center}
\includegraphics[width=12cm]{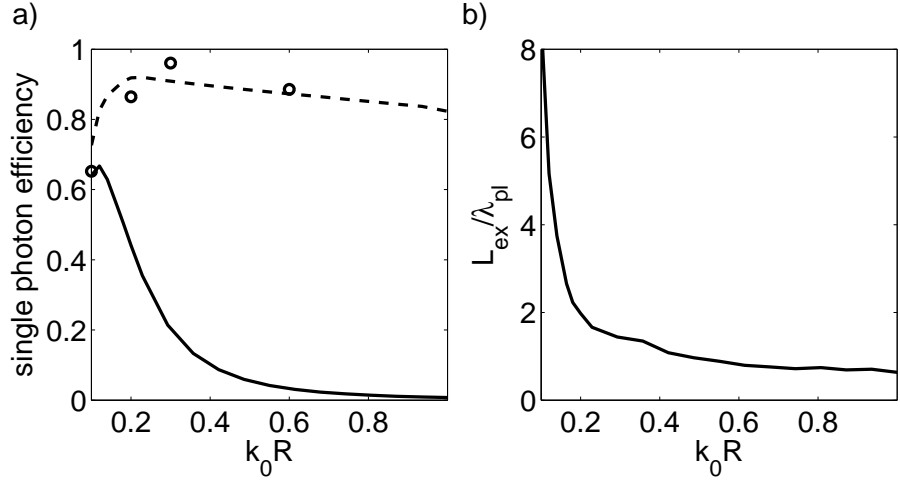}
\end{center}
\caption{a.  Optimized efficiencies of single photon generation vs. $R$. We have 
assumed that coupling to waveguide modes other than the fundamental mode is 
negligible, \textit{i.e.}, the waveguide is effectively in the single-mode 
regime.  Solid line: theoretical efficiency using a nanowire.  Dotted line: 
theoretical efficiency using a nanotip. Solid points: nanotip efficiency based 
on boundary element method simulations, combined with coupled-mode equations.  
b.  Optimal coupling length $L_{ex}$ for a nanotip as a function of $R$.  Here 
$L_{ex}$ is given in units of the plasmon wavelength 
$\lambda_{\footnotesize\textrm{pl}}$ at that particular 
$R$.\label{fig:photonefficiency}}
\end{figure}

In Fig.~\ref{fig:photonefficiency}a we plot the efficiency of single photon 
generation as a function of $R$, for both the nanowire and nanotip systems.  For 
each $R$ the plotted efficiencies have been optimized over all other possible 
parameters of the system.  For the nanowire configuration, we have assumed that 
the resulting forward- and backward-propagating waves in the waveguide can be 
perfectly combined.  In the figure we have also included points obtained by our 
BEM simulations of a nanotip.  Here, we have taken the numerically optimized 
values of $\tilde{P}_{E}$ and plugged them in as initial values for the 
coupled-mode theory above.  It can be seen that the numerical simulations agree 
well with our theoretical predictions.  We find that photon efficiencies of 
approximately $70\%$ are possible for the nanowire, while efficiencies exceeding 
$95\%$ are possible for the nanotip.  In Fig.~\ref{fig:photonefficiency}b we 
plot the optimal coupling length $L_{ex}$, in units of 
$\lambda_{\footnotesize\textrm{pl}}$, as a function of $R$ for the 
nanotip~($L_{ex}$ for the nanowire should be twice that of the nanotip, to 
account for the transfer of the forward- and backward-propagating components of 
the emitted plasmon).  It can be seen that the out-coupling to the waveguide can 
in principle occur quite rapidly, over length scales of a few 
$\lambda_{\footnotesize\textrm{pl}}$.

The existence of an optimum $R$ for photon generation can be intuitively 
understood.  For smaller $R$ the coupling between the emitter and plasmon modes 
can be quite large.  However, these tightly-confined plasmon modes are 
accompanied by higher propagative losses and cannot be as efficiently coupled to 
the waveguide system.  The coupling efficiency between plasmons and waveguide 
modes improves for larger $R$.  For the nanowire, however, the larger radius 
results in weaker coupling between the plamson and emitter, while for the 
nanotip the accumulated propagative loss increases as the final radius $R$ 
grows.  

\section{Effects of surface roughness}\label{sec:roughness}
In previous sections we have treated the problem of plasmon propagation on 
smooth nanowires and nanotips, taking into account inherent dissipative losses 
characterized by $\textrm{Im}\;\epsilon_2$.  In practice, however, these 
structures are not perfectly smooth, and the surface roughness can give rise to 
new scattering mechanisms for the plasmons.  While the general solution for the 
fields in the presence of arbitrary roughness is a complicated problem, we 
calculate the effects in two limits.  In Sec.~\ref{subsec:radiativelosses} we 
calculate the losses on a nanowire due to radiative scattering in the limit of 
small roughness and zero heating~($\textrm{Im}\,\epsilon_2=0$). Here the 
plasmons experience no inherent loss due to the metal but can receive momentum 
kicks from the roughness that cause them to scatter radiatively. In 
Sec.~\ref{subsec:nonradiativelosses} we calculate the effects of small roughness 
for a nanowire in the non-retarded limit, where radiative effects are ignored 
but the effects of increased dissipative losses are treated.  While we 
explicitly treat only the nanowire case here, we note that the results obtained 
can also be incorporated into our model for nanotip losses via 
Eq.~(\ref{eq:eikonal}).

\subsection{Radiative losses}\label{subsec:radiativelosses}
For simplicity we consider a wire with axial symmetry, but with a surface 
profile given by $\rho_0(z)=R+p\zeta(z)$, where $R$ is the average radius of the 
wire, $\zeta(z)$ is some random function describing the roughness, and $p$ is an 
expansion parameter that will be taken to equal $1$ at the end.  We will 
calculate in perturbation theory the radiated field scattered from the 
roughness, from an initial field corresponding to the fundamental plasmon mode 
for a perfectly smooth wire.  Because of the symmetry, the only non-zero 
components of the fields remain $E_{\rho}$, $E_z$, and $H_{\phi}$, which will 
also have axial symmetry.  As will be seen later, it suffices for now to 
consider only $E_z$, as the other components depend on $E_z$ in a simple way 
through Maxwell's Equations.  We proceed by breaking up the total field along 
$z$ in region $i$ into incident and scattered fields
\be 
E^{\footnotesize\textrm{total}}_{i,z}=E^{0}_{i,z}+E^{s}_{i,z},\label{eq:totalfield} 
\ee
where $E^{0}_{i,z}$ is the $z$-component of the fundamental plasmon mode given 
by Eqs.~(\ref{eq:tm0mode}) and~(\ref{eq:b1b2}), and further assume that the 
scattered field can be expanded in a power series
\be 
E^{s}_{i,z}=\sum_{n=1}^{\infty}p^{n}E^{(n)}_{i,z}.\label{eq:scatteredpowerseries} 
\ee
In the following we will calculate the first-order scattered field 
$E^{(1)}_{i,z}$.  We make the ansatz that $E^{(1)}_{i,z}$ can be expanded in the 
form~\cite{agassi86}
\bea E^{(1)}_{1,z} & = & 
\int_{-\infty}^{\infty}d\hp\,H_{0}(\honeperp\rho)\frac{\honeperp^2}{k_1^2}A(\hp)e^{i{\hp}z} 
\nonumber \\ E^{(1)}_{2,z} & = & 
\int_{-\infty}^{\infty}d\hp\,J_{0}(\htwoperp\rho)\frac{\htwoperp^2}{k_2^2}B(\hp)e^{i{\hp}z},\label{eq:Eexpansion} 
\eea
where each Fourier component is an outgoing solution of the wave equation with 
appropriate boundary conditions at $\rho=0$ and $\rho=\infty$, as derived in 
Eq.~(\ref{eq:cylindricalfields}).  From Eq.~(\ref{eq:cylindricalfields}) one 
also sees that $E_{\rho},H_{\phi}$ are determined completely once $E_z$ is 
known.  Using these relations, the total~(incident plus scattered) fields 
$\bfE^{\footnotesize\textrm{total}}$ and $\bfH^{\footnotesize\textrm{total}}$ 
are straightforward but lengthy to write down, and are given to order $p$ in 
Eq.~(\ref{eq:fields}) in Appendix~\ref{sec:radiativelossesappendix}.

The coefficients $A(\hp),B(\hp)$ are determined by enforcing continuity of the 
tangential fields at the boundary $\rho_{0}(z)$. Specifically, we require that
\bea 
(\hat{t}\cdot\bfE_{1}^{\footnotesize\textrm{total}})\big|_{\rho=R+p\zeta(z)} & = 
& (\hat{t}\cdot\bfE_{2}^{\footnotesize\textrm{total}})\big|_{\rho=R+p\zeta(z)}, 
\hat{t}=\frac{\hat{z}+p\frac{d\zeta}{dz}\hat{\rho}}{\sqrt{1+p^2\left(\frac{d\zeta}{dz}\right)^2}} 
\nonumber
\\ H_{\phi,1}^{\footnotesize\textrm{total}}\big|_{\rho=R+p\zeta(z)} & = & H_{\phi,2}^{\footnotesize\textrm{total}}\big|_{\rho=R+p\zeta(z)},\label{eq:boundarycond} \eea
where $\hat{t}(z)$ is the unit tangent vector to the interface.  These equations 
can be solved perturbatively by expanding them in $p$ and solving at each 
order.  It should be noted that the expansion should be done carefully, as 
dependence in $p$ is contained not only in the fields given in
Eq.~(\ref{eq:fields}) but also in the surface profile $\rho_{0}(z)=R+p\zeta(z)$ 
and tangent vector $\hat{t}$.  The $\mathcal{O}(p^{0})$ equation is trivially 
satisfied by the fundamental plasmon mode for a smooth nanowire.  To solve to 
$\mathcal{O}(p)$, it is useful to first introduce the Fourier transform of the 
surface roughness,
\be \zeta(z)=\int_{-\infty}^{\infty}\frac{d\hp}{2\pi}e^{i\hp 
z}\tilde{\zeta}(\hp).\label{eq:roughnesstransform} \ee
Using the Fourier transform $\tilde{\zeta}(\hp)$, the $\mathcal{O}(p)$ equations 
become algebraic in Fourier space and have solutions~(see 
Appendix~\ref{sec:radiativelossesappendix})
\bea A(\hp) & = & \frac{\tilde\zeta(\hp-\kp)}{2\pi}\frac{k_1^2}{\honeperp}f(\hp) 
\nonumber
\\ B(\hp) & = &
\frac{\tilde\zeta(\hp-\kp)}{2\pi}\frac{k_2^2}{\htwoperp}g(\hp),\label{eq:AB} \eea
where $\kp$ denotes the unperturbed plasmon wavevector~(in this section we take 
$\textrm{Im}\,\epsilon_2=0$ so that $\kp$ and $C_{-1}$ are purely real) .  The 
scattering coefficients $f(\hp),g(\hp)$ are complicated functions of $\hp$ and 
$R$ and are given in Appendix~\ref{sec:radiativelossesappendix}.  Physically, 
the equations above state that, to first order, the surface roughness 
contributes single momentum kicks to the unperturbed plasmon fields with a 
strength determined by the Fourier components of the roughness.  From this point 
forward we set $p=1$.

We now consider some random surface profile such that
\bea \langle\zeta(z)\rangle & = & 0 \nonumber \\
\langle\zeta(z)\zeta(z^{\prime})\rangle & = & 
\delta^{2}e^{-(z-z^{\prime})^2/a^2},\label{eq:spacecorrelations} \eea
with corresponding correlations
\bea \langle\tilde{\zeta}(k)\rangle & = & 0 \nonumber
\\ \langle\tilde{\zeta}(k)\tilde{\zeta}^{\ast}(k^{\prime})\rangle & = &
2{\pi}^{3/2}\delta^{2}ae^{-\frac{1}{4}a^{2}k^2}\delta(k-k^{\prime})\label{eq:correlations} 
\eea
for the Fourier components.  Physically $\delta$ and $a$ correspond respectively 
to the typical amplitude and length of a rough patch on the surface of the 
wire.  It is also useful to define $s=\delta/a$ as a typical ``slope'' to the 
roughness.  To calculate the power radiated due to the surface roughness we will 
find the ensemble-averaged Poynting vector far from the wire.  It is sufficient 
to consider just the component of $\avg{{\bf S}}$ oriented along $\hat\rho$, 
given outside the wire by
\be 
S_{\rho}=-\frac{1}{2}\avg{E_{1,z}^{\footnotesize\textrm{total}}H^{\ast\,{\footnotesize\textrm{total}}}_{1,\phi}}, 
\ee
where the fields 
$E_{1,z}^{\footnotesize\textrm{total}},H_{1,\phi}^{\footnotesize\textrm{total}}$ 
are given to first order by Eq.~(\ref{eq:fields}).  The calculation of 
$S_{\rho}$ simplifies further because the incident plasmon field decays 
exponentially away from the wire, and thus to lowest order only the first-order 
scattered fields will contribute to the Poynting vector at large $\rho$, which 
physically corresponds to the power radiated away to infinity.  Specifically, 
the radiated power per unit area is given by
\bea S_{\rho} & = & 
-\frac{1}{2}\avg{E^{(1)}_{1,z}H^{(1)\ast}_{1,\phi}}\;\;\;\;\;\left(\rho\rightarrow\infty\right)
\\ & = &
\frac{1}{2\omega\mu_0}\int_{-\infty}^{\infty}d{\hp}d{\hp^{\prime}}\frac{i{\honeperp}^2\honeperp^{\prime\ast}}{k_1^2}H_{0}(\honeperp\rho)H^{\prime\ast}_{0}(\honeperp^{\prime}\rho)\avg{A(\hp)A^{\ast}(\hp^{\prime})}e^{i(\hp-\hp^{\prime})z}.\label{eq:Sfull} 
\eea
Substituting the solution for $A(\hp)$ derived in Eq.~(\ref{eq:AB}) and using 
the correlations in Eq.~(\ref{eq:correlations}), it is straightforward to 
evaluate the integral over $\hp'$ and arrive at
\be 
S_{\rho}=\frac{i\epsilon_{0}\epsilon_{1}\omega}{4\sqrt{\pi}}s^{2}a^{3}\int_{-k_1}^{k_1}d{\hp}e^{-\frac{1}{4}a^2(\hp-\kp)^2}{\honeperp}H_{0}(\honeperp\rho)H^{\prime\ast}_{0}(\honeperp\rho)\left|f(\hp)\right|^2.\label{eq:S} 
\ee
In the expression above we have truncated the bounds of the integral to 
${\pm}k_{1}$ because we are interested in the Poynting vector far away from the 
wire, where only radiative fields $|\hp|\leq k_{1}$ should contribute.  With 
knowledge of the Poynting vector it is then possible to find the dissipation 
rate of the plasmons due to radiative scattering, given by
\be 
\Gamma_{\footnotesize\textrm{rad,rough}}=\lim_{\rho\to\infty}\frac{2\pi{\rho}S_{\rho}}{\frac{1}{4}{\int}d\bfrho\,\epsilon_{0}\frac{d}{d\omega}\left(\epsilon(\bfrho,\omega)\omega\right)\left|\bfE(\bfrho)\right|^2+\mu_{0}\left|\bfH(\bfrho)\right|^2}.\label{eq:dissipationratio} 
\ee
The denominator on the right-hand side of the equation above can be identified 
with the plasmon energy per unit length.

\begin{figure*}[p]
\begin{center}
\includegraphics[width=7cm]{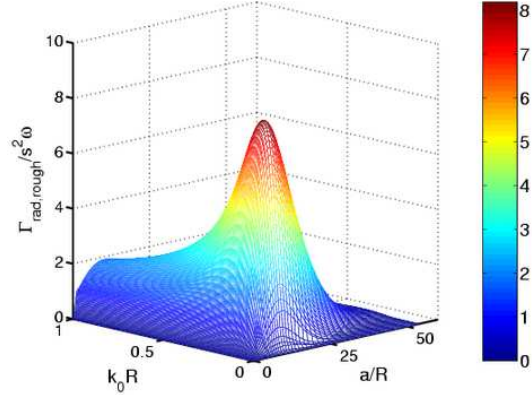}
\end{center}
\caption{The plasmon dissipation rate due to radiative scattering off of surface 
roughness, $\Gamma_{\footnotesize{\textrm{rad,rough}}}/s^2\omega$, as functions 
of wire radius $R$ and correlation length $a/R$.  The numbers are calculated for 
a silver nanowire at $\lambda_0=1\,\mu$m and $\epsilon_1=2$. 
\label{fig:radroughness}}
\end{figure*}

We first qualitatively discuss the behavior of 
$\Gamma_{\footnotesize\textrm{rad,rough}}$ before deriving various limits more 
quantitatively.  From Eq.~(\ref{eq:S}) it is clear that 
$\Gamma_{\footnotesize\textrm{rad,rough}}$ scales explicitly like $\delta^2$ or 
$s^2$.  Physically, this occurs because the lowest-order contribution to the 
Poynting vector far away from the wire is due to the combination of a 
first-order scattered electric field and first-order scattered magnetic field.  
In Fig.~\ref{fig:radroughness} the quantity  
$\Gamma_{\footnotesize{\textrm{rad,rough}}}/s^2\omega$ is evaluated numerically 
as a function of wire radius $R$ and correlation length $a/R$, for a silver 
nanowire at $\lambda_0=1\,\mu$m and $\epsilon_1=2$.  We are particularly 
interested in the nanowire limit, when the plasmon wavevector 
$\kp{\approx}C_{-1}/R$.  We see that for fixed $R$, the scattering reaches a 
peak for some particular value of $a/R$.  More careful inspection reveals that 
the maximum occurs when 
$a{\propto}R/C_{-1}\propto\lambda_{\footnotesize\textrm{pl}}$.  This result 
makes intuitive sense, since the characteristic momentum kick ${\sim}1/a$ that 
the plasmon wavevector $\kp$ receives due to roughness must be on the order of 
$C_{-1}/R$ in order for the resulting wavevector to lie in the radiative range 
between $-k_1$ and $k_1$.  In the limit $a/R{\gg}C_{-1}$, one observes an 
exponential suppression of scattering, due to the fact that the roughness has a 
very narrow momentum distribution and cannot possibly contribute a large kick to 
$\kp$.  In fact, in this regime one physically expects for the plasmon 
wavevector to adiabatically vary with the changing wire radius.  In the other 
limit $a/R{\ll}C_{-1}$, the scattering also decreases, but with a polynomial 
dependence on $R$, as will be proven below.  Here, the momentum distribution of 
the roughness becomes very wide, and thus the probability of receiving a kick 
that results in a final momentum between $\pm k_{1}$ becomes quite small.  
Finally, for fixed slope $s$, it can be seen that the scattering decreases as 
$R{\rightarrow}0$ at any correlation length $a$.  This result is also easily 
understood, as the plasmon wavevector $\kp$ becomes increasingly far-removed 
from the range of radiative wavevectors.  In Table~\ref{table:roughness}, we 
calculate the scattering rates for wire sizes $k_{0}R=0.1,0.2,0.3$~(or 
$R{\approx}16,32,48$~nm), for a few chosen roughness parameters.  The scattering 
rates are given as a percentage increase in $\textrm{Im}\;\kp$ over the values 
for a smooth nanowire.  It can be seen that strong suppression of radiative 
scattering occurs both for smaller $R$ and when $a$ is either much larger or 
much smaller than $R$, which confirms our earlier observations.  Furthermore, it 
is evident that under reasonable parameters, the losses in the system are 
increased only slightly due to radiative scattering, around an amount of $10\%$ 
or less.

We now analyze more carefully the behavior of the radiative scattering in the 
nanowire regime.  For concreteness, we will consider a field normalized by 
Eq.~(\ref{eq:normalization}), in which case the denominator of 
Eq.~(\ref{eq:dissipationratio}) becomes $\hbar\omega/4L$.  To simplify the 
expression further, we first note that since we are interested in the far 
field~($\rho\rightarrow\infty$), we can take the asymptotic limits of the Hankel 
functions in Eq.~(\ref{eq:S}), 
$H_{0}(\honeperp\rho)H_{0}^{\prime\ast}(\honeperp\rho){\approx}-2i/(\pi\honeperp\rho)$.  
One can also derive an asymptotic relationship of $f(\hp)$ as 
$R{\rightarrow}0$~(see Appendix~\ref{sec:radiativelossesappendix}), which upon 
substitution yields
\bea \Gamma_{\footnotesize\textrm{rad,rough}} & {\approx} & 
\pi^{3/2}\frac{|\phi|^2}{\tilde{V}}\omega\epsilon_{1}s^{2}a^{3}\int_{-k_1}^{k_1}d\hp\; 
\honeperp^2 
e^{-(1/4)a^2(\kp-\hp)^2},\label{eq:nanowirerad}\;\;\;\;(R{\rightarrow}0) \\ \phi 
& \equiv & 
\frac{H_{0}^{\prime}\left(iC_{-1}\right)}{J_{0}^{\prime}\left(iC_{-1}\right)}J_{0}''(iC_{-1})-H_{0}''(iC_{-1}).\label{eq:phidef}
\eea

From the equation above, it is clear that there are three distinct regimes of 
interest defined by the quantity $\alpha{\equiv}{\kp}a=2\pi 
a/\lambda_{\footnotesize\textrm{pl}}{\approx}C_{-1}a/R$, which characterizes the 
typical extent of a rough patch compared to the plasmon wavelength.  In the 
limit $\alpha{\ll}1$, one can approximate the exponential in the integrand of 
Eq.~(\ref{eq:nanowirerad}) as a constant, which leads to straightforward 
evaluation of the integral,
\be 
\Gamma_{\footnotesize\textrm{rad,rough}}{\approx}\frac{4}{3}\pi^{3/2}\frac{|\phi|^2}{\tilde{V}}\omega\epsilon_{1}^{5/2}s^{2}\left(\frac{k_{0}R}{C_{-1}}\right)^3\alpha^3.\;\;\;\;\;\;(\alpha{\ll}1) 
\ee
Here, the noise spectrum of Eq.~(\ref{eq:correlations}) becomes very wide and 
leads to an $\alpha^3$ scaling of the dissipation rate.  In the opposite limit 
$\alpha{\gg}1$, the value of the exponential term becomes exponentially small, 
with a corresponding exponential suppression of the scattering rate.  A more 
careful evaluation of the integrand yields
\be 
\Gamma_{\footnotesize\textrm{rad,rough}}{\approx}8\pi^{3/2}\frac{|\phi|^2}{\tilde{V}}\omega\epsilon_{1}^{3/2}s^2\frac{k_{0}R}{C_{-1}\alpha}e^{-(1/4)a^2(\kp-k_1)^2}.\;\;\;\;\;\;(\alpha{\gg}1) 
\ee
Finally, one can show that for fixed, sub-wavelength $R$ the radiative 
scattering is most significant when $\alpha{\sim}\mathcal{O}(1)$.  In this case, 
the exponential appearing in Eq.~(\ref{eq:nanowirerad}) is neither exponentially 
small nor constant.  However, one can make the rough approximation 
$e^{-(1/4)a^2(\kp-\hp)^2}{\approx}1-(1/4)a^2(\kp-\hp)^2$ to get an idea of the 
scaling in this regime.  It is straightforward to show that the scattering rate 
has a maximum with respect to $\alpha$ at $\alpha{\approx}(12/5)^{1/2}$, with a 
corresponding maximum decay rate
\be \max_{a} \left\{ 
\Gamma_{\footnotesize\textrm{rad,rough}}\right\}{\propto}\frac{|\phi|^2}{\tilde{V}}\omega\epsilon_{1}^{5/2}s^{2}\left(\frac{k_{0}R}{C_{-1}}\right)^3. 
\ee
Again, the radiative scattering is most significant when the length scale $a$ of 
the roughness is on the order of the plasmon wavelength, and the maximum 
scattering~(for fixed $s$) decreases as $R{\rightarrow}0$ due to the increasing 
mismatch between $\kp$ and radiative wavevectors.

We now consider the limits of validity of the derivations above, specifically 
considering the expansions made in Eq.~(\ref{eq:pexpansions}) that are necessary 
for the perturbative method used here.  The first of these expansions requires 
that $|d\zeta/dz|{\ll}1$, which can be re-written as a condition on the slope, 
$s{\ll}1$.  Physically this requirement states that the typical length of a 
rough patch be much larger than its typical height.  The second line of 
Eq.~(\ref{eq:pexpansions}) requires that $|k_{i\perp}\zeta|{\ll}1$.  In the 
nanowire regime this requirement is equivalent to 
$\delta{\ll}\lambda_{\footnotesize\textrm{pl}}$, which states that the height of 
a rough patch must be much smaller than the plasmon wavelength.  Finally, the 
third line of Eq.~(\ref{eq:pexpansions}) requires $|\hiperp\zeta|{\ll}1$, within 
the range of $\hiperp$ that are appreciably scattered into.  From 
Eqs.~(\ref{eq:AB}) and~(\ref{eq:correlations}), we see that the relevant range 
for the parallel component of the wavevector is given by 
$\kp-1/a\stackrel{<}{\sim}\hp\stackrel{<}{\sim}\kp+1/a$, and thus the largest 
relevant transverse wavevector is 
$|\hiperp{}_{,\footnotesize\textrm{max}}|{\sim}\max_{-1\leq\theta\leq 
1}\left|\sqrt{\epsilon_{i}(\omega/c)^2-(\kp+\theta/a)^2}\right|$.  In the 
nanowire regime, $\kp{\approx}C_{-1}/R$, there are two limiting cases.  The 
first is when the correlation length $a$ is much larger than $R$, $a{\gg}R$, in 
which case $|\hiperp{}_{,\footnotesize\textrm{max}}|{\sim}C_{-1}/R$ and 
$|\hiperp\zeta|{\ll}1$ reduces to 
$\delta{\ll}\lambda_{\footnotesize\textrm{pl}}$.  In the other limiting case, 
$R{\gg}a$, one finds that $|\hiperp{}_{,\footnotesize\textrm{max}}|{\sim}1/a$ 
and the corresponding requirement is given by $s \ll 1$.  

We finally note that while the radiative scattering goes like $\delta^2$ or 
$s^2$, the relevant quantity for dissipative~(heating) losses due to roughness 
becomes ${\bf S}$ inside the wire.  For this quantity the lowest-order 
correction to the smooth wire solution will come from a combination of a 
first-order and zeroth-order field.  Thus one expects roughness-induced 
dissipative losses to contribute a decay term proportional to $\delta$ or $s$, 
which for small roughness will dominate over radiative scattering.  This 
correction will be treated in the next subsection.

\subsection{Non-radiative losses}\label{subsec:nonradiativelosses}

To study the effects of surface roughness on non-radiative losses, we will make 
one simplifying assumption and calculate these losses in the quasi-static 
limit.  To do this we will proceed in a manner similar to that in 
Sec.~\ref{subsec:radnonraddecay}, where we found the quasi-static fields 
associated with a smooth nanowire.  Here the calculations for the fields yielded 
the presence of poles whose positions and widths give the real and imaginary 
parts of the wavevector $\kp$.  The case of a smooth nanowire was particularly 
easy to treat because of the translational symmetry of the system.  A system 
containing surface roughness lacks such translational symmetry and therefore 
must be considered more carefully, but the calculation proceeds in much the same 
way.   In particular, we will find expressions for the pseudopotentials 
$\Phi_1=\Phi_0+\Phi_r$ and $\Phi_2$ that satisfy the necessary boundary 
conditions in the presence of surface roughness.  We can once again find the 
positions and widths of the poles associated with the system, which are now 
altered by the roughness.

We first write down appropriate expansions for $\Phi_{0,r,2}(\bfr,\bfrp)$.   The 
expansion for the incident component $\Phi_0$, given originally in 
Eq.~(\ref{eq:Phi0}), is slightly re-written here,
\bea \Phi_{0}(\bfr,\bfrp) & = &
\frac{1}{4\pi\epsilon_{0}\epsilon_1}\frac{1}{\left|\bfr-\bfrp\right|} \nonumber \\
& = & 
\frac{1}{2\pi^2\epsilon_0\epsilon_1}\sum_{m=0}^{\infty}\left(2-\delta_{m,0}\right)\cos\left(m(\phi-\phi^\prime)\right)\int_{0}^{\infty}dh\cos\left(h(z-z^\prime)\right)K_{m}(h\rho^\prime)I_{m}(h\rho)\;\;\;(\rho<\rho^\prime) 
\nonumber
\\ & = & 
\frac{1}{4\pi^2\epsilon_0\epsilon_1}\sum_{m=-\infty}^{\infty}e^{im(\phi-\phi^\prime)}\int_{-\infty}^{\infty}dh\,e^{ih(z-z^\prime)}\tilde{K}_{m}(h\rho^\prime)\tilde{I}_{m}(h\rho)\;\;\;(\rho<\rho^\prime).\label{eq:Phi0expansion} 
\eea
The functions $\tilde{K}_{m}(x),\tilde{I}_{m}(x)$ are defined by
\bea \tilde{K}_{m},\tilde{I}_{m}(x) & = & K_{m},I_{m}(|x|). \eea
We also break up $\Phi_{r,2}$ into Fourier components that satisfy Laplace's 
equation, and assume that these expressions hold up to the interface:
\bea \Phi_{r}(\bfr,\bfrp) & = & 
\frac{1}{4\pi^2\epsilon_0\epsilon_1}\sum_{m=-\infty}^{\infty}e^{im(\phi-\phi^\prime)}\int_{-\infty}^{\infty}dh\,e^{ih(z-z')}\tilde{K}_{m}(h\rho)\alpha_{m}(h),\label{eq:Phirexpansion}
\\ \Phi_{2}(\bfr,\bfrp) & = &
\frac{1}{4\pi^2\epsilon_0\epsilon_1}\sum_{m=-\infty}^{\infty}e^{im(\phi-\phi^\prime)}\int_{-\infty}^{\infty}dh\,e^{ih(z-z')}\tilde{I}_{m}(h\rho)\beta_{m}(h).\label{eq:Phi2expansion} 
\eea
To describe the surface roughness, we assume an interface with axial symmetry as 
before, $\rho_{0}(z)=R+p\zeta(z)$, where the roughness profile $\zeta$ satisfies 
the correlations given in Eqs.~(\ref{eq:spacecorrelations}) 
and~(\ref{eq:correlations}).  The coefficients $\alpha_{m}$, $\beta_{m}$ are 
determined by the boundary conditions, namely that $\Phi$ and $\bfD_{\perp}$ 
must be continuous at the interface:
\bea \Phi_{1}(\bfr,\bfrp)\big|_{\rho=R+p\zeta(z)} & = & \Phi_{2}(\bfr,\bfrp)\big|_{\rho=R+p\zeta(z)}, \nonumber \\
\epsilon_{1}\hat{n}\cdot\nabla\Phi_{1}(\bfr,\bfrp)\big|_{\rho=R+p\zeta(z)} & = & 
\epsilon_{2}\hat{n}\cdot\nabla\Phi_{2}(\bfr,\bfrp)\big|_{\rho=R+p\zeta(z)},\;\;\;\hat{n}=\frac{\hat{\rho}-p\frac{d\zeta}{dz}\hat{z}}{\sqrt{1+p^{2}\left(\frac{d\zeta}{dz}\right)^2}}.\label{eq:phiboundaryeqns} 
\eea
Plugging in the expressions for $\Phi_i$ given by 
Eqs.~(\ref{eq:Phi0expansion}),~(\ref{eq:Phirexpansion}), 
and~(\ref{eq:Phi2expansion}), one can expand both boundary condition equations 
to $\mathcal{O}(p^2)$.  Then, replacing $\zeta(z)$ in these equations with its 
Fourier transform given in Eq.~(\ref{eq:roughnesstransform}), one obtains a set 
of coupled equations for $\alpha_{m},\beta_{m}(h)$ completely in Fourier space, 
which, unlike the case of a smooth nanowire, is not de-coupled in $h$.  It is 
tedious but straightforward to show that, to $\mathcal{O}(p^2)$, this set of 
equations is given by the matrix integral equation
\bea & & M_{0}(h)\left(\begin{array}{l} \alpha_{m}(h) \\
\beta_{m}(h)
\end{array}\right)+p\int\frac{dq}{2\pi}\tilde\zeta(h-q)M_{1}(h,q)\left(\begin{array}{l} \alpha_{m}(q) \\ \beta_{m}(q)
\end{array}\right) \nonumber \\ & & +p^{2}\int\frac{dq\,dq^{\prime}}{(2\pi)^2}\tilde\zeta(h-q-q^{\prime})\tilde\zeta(q^{\prime})M_{2}(h,q,q^{\prime})\left(\begin{array}{l} \alpha_{m}(q) \\ \beta_{m}(q)
\end{array}\right)+\mathcal{O}(p^3)=\nonumber \\ & & \;\;\;\;\;\bfv_{0}(h)+p\int\frac{dq}{2\pi}\tilde\zeta(h-q)\bfv_{1}(h,q)+p^{2}\int\frac{dq\,dq^{\prime}}{(2\pi)^2}\tilde\zeta(h-q-q^{\prime})\tilde\zeta(q^{\prime})\bfv_{2}(h,q,q^{\prime})+\mathcal{O}(p^3).\label{eq:roughnessmatrix} \eea
The matrices $M_i$ and vectors $\bfv_i$ are complicated expressions and are 
given in Appendix~\ref{sec:matvec}.  We note, however, that in the case of no 
surface roughness~($\tilde{\zeta}=0$), the solution to the resulting equation 
$M_{0}(h)\cdot\left(\alpha_{m}(h) \; \beta_{m}(h)\right)^{T}=\bfv_{0}(h)$ 
reduces to that of a smooth nanowire.

We now discuss how to solve Eq.~(\ref{eq:roughnessmatrix}) in the presence of 
surface roughness, using the methods detailed in~\cite{rahman80}.  One might 
first consider expanding $\alpha_{m},\beta_{m}$ in a power series of $p$, in a 
manner similar to the field expansion in Eq.~(\ref{eq:scatteredpowerseries}) for 
the case of radiative scattering, and then solving the $\mathcal{O}(p^{n+1})$ 
equations based on the $\mathcal{O}(p^n)$ solutions.  However, one expects that 
such a perturbative solution would simply yield poles for each higher-order 
correction with the same location as that of the unperturbed solutions 
$\alpha^{(0)}_{m},\beta^{(0)}_{m}$.  Mathematically, this occurs because each 
calculation of the next correction involves an inversion $M_{0}^{-1}(h)$.  On 
the other hand, physically we expect for the surface roughness to result in some 
shift of the pole that is not predicted by such a perturbative 
method~\cite{rahman80}.  We thus consider an alternate approach, in which we 
symbolically sum the perturbation series in Eq.~(\ref{eq:roughnessmatrix}) to 
all orders and then only keep the lowest order result in $p$.  Let us 
symbolically write Eq.~(\ref{eq:roughnessmatrix}) in the form
\be 
(\mathcal{M}_{0}+\delta\mathcal{M})\bfx=\bff_0+\delta\bff,\label{eq:symboliceq} 
\ee
where $\mathcal{M}_{0}$ and $\bff_0$ are non-random matrices and vectors, 
respectively, $\delta\mathcal{M}$ is a random $2\times 2$ matrix integral 
operator, $\delta\bff$ is a random vector, and $\bfx$ is a column vector with 
components $\alpha_{m},\beta_{m}$.  We now define the averaging operator
\be Px=\avg{x}, \ee
and the operator $Q=1-P$.  We can apply $P,Q$ to Eq.~(\ref{eq:symboliceq}) to get
\bea P\mathcal{M}_{0}\bfx+P\delta\mathcal{M}\bfx & = & P(\bff_0+\delta\bff), \\
 Q\mathcal{M}_{0}\bfx+Q\delta\mathcal{M}\bfx & = & Q(\bff_0+\delta\bff), \eea
which after some manipulation results in the set of equations
\bea & & 
\left(\mathcal{M}_{0}+P\delta\mathcal{M}\right)\avg{\bfx}+P\delta\mathcal{M}Q\bfx=P(\bff_0+\delta\bff),\label{eq:avgbff}
\\ & &
Q\bfx=(1+\mathcal{M}^{-1}_{0}Q\delta\mathcal{M})^{-1}\mathcal{M}^{-1}_{0}Q(\bff_0+\delta\bff)-(1+\mathcal{M}^{-1}_{0}Q\delta\mathcal{M})^{-1}\mathcal{M}^{-1}_{0}Q\delta\mathcal{M}\avg{\bfx}.\label{eq:Qx} 
\eea
One can then substitute Eq.~(\ref{eq:Qx}) into Eq.~(\ref{eq:avgbff}) and solve 
for $\avg{\bfx}$, in which case one obtains
\be 
\left(\mathcal{M}_{0}+\avg{(1+\delta\mathcal{M}\mathcal{M}^{-1}_{0}Q)^{-1}\delta\mathcal{M}}\right)\avg{\bfx} 
=\avg{(1+\delta\mathcal{M}\mathcal{M}^{-1}_{0}Q)^{-1}(\bff_0+\delta\bff)}.\label{eq:symbolicsum} 
\ee
We note that, unlike a perturbative expansion and solution for 
$\alpha_{m},\beta_{m}$, the equation above is thus far exact.  Now, we assume 
that $\delta\mathcal{M}$ and $\delta\bff$ can be expanded in powers of $p$ in 
the form
\bea \delta\mathcal{M} & = & 
p\delta\mathcal{M}_{1}+p^2\delta\mathcal{M}_{2}+\cdots, \nonumber \\ \delta\bff 
& = & p\delta\bff_{1}+p^2\delta\bff_{2}+\cdots.\label{eq:Mfexpansion} \eea
Comparing Eqs.~(\ref{eq:roughnessmatrix}) and~(\ref{eq:Mfexpansion}), we see 
that $\avg{\delta\mathcal{M}_{1}}=\avg{\delta\bff_1}=0$ since 
$\avg{\tilde{\zeta}}=0$.  With this result, and utilizing the definition 
$Q=1-P$, one can proceed to expand Eq.~(\ref{eq:symbolicsum}) up to 
$\mathcal{O}(p^2)$, which yields~(after setting $p=1$)
\be 
\left(\mathcal{M}_{0}+\avg{\delta\mathcal{M}_{2}}-\avg{\delta\mathcal{M}_{1}\mathcal{M}^{-1}_{0}\delta\mathcal{M}_{1}}\right)\avg{\bfx}= 
\avg{\bff_{0}}+\avg{\delta\bff_{2}}-\avg{\delta\mathcal{M}_{1}\mathcal{M}^{-1}_{0}\delta\bff_{1}}.\label{eq:symbolicsecorder} 
\ee
Substituting the corresponding terms of Eq.~(\ref{eq:roughnessmatrix}) into the 
equation above and using the second-order correlations given by 
Eq.~(\ref{eq:correlations}), we find after simplifying that
\bea
& & \left[M_{0}(h)+\frac{s^{2}a^3}{2\sqrt{\pi}}\int\;dq\left(e^{-a^{2}q^{2}/4}M_{2}(h,h,q)-e^{-a^{2}(h-q)^{2}/4}M_{1}(h,q)M^{-1}_{0}(q)M_{1}(q,h)\right)\right]\left\langle\begin{array}{l} \alpha_{m}(h) \\
\beta_{m}(h)
\end{array}\right\rangle= \nonumber \\ & &
\;\;\;\;\;\left[\bfv_{0}(h)+\frac{s^{2}a^3}{2\sqrt{\pi}}\int\;dq\left(e^{-a^{2}q^{2}/4}\bfv_{2}(h,h,q)-e^{-a^{2}(h-q)^2/4}M_{1}(h,q)M^{-1}_{0}(q)\bfv_{1}(q,h)\right)\right],\label{eq:nonperturbativesolution} 
\eea
where $s=\delta/a$.

We now discuss the solution to $\alpha_{0}(h)$, which contains a pole 
corresponding to the fundamental plasmon mode $m=0$.  When $\epsilon_2$ is a 
negative real number, $\alpha_{0}$ has a pole on the real $h$-axis whose 
position gives the new, shifted plasmon wavevector $\tildekp$.  When 
$\epsilon_2$ has a non-zero imaginary component, $\alpha_{0}$ will have a 
resonance feature along this axis whose peak corresponds $\textrm{Re}\,\tildekp$ 
and whose width corresponds to $\textrm{Im}\,\tildekp$.  A quick inspection of 
the equation above reveals that $\tildekp{R}=\tilde{C}_{-1}(\epsilon_{i},s,a/R)$ 
is a constant that depends only on the quantities $\epsilon_i$,$s$, and $a/R$.  
Unfortunately, because of the complexity of 
Eq.~(\ref{eq:nonperturbativesolution}) it is difficult to derive other scaling 
results for $\tildekp$ even in limiting cases.  However, 
Eq.~(\ref{eq:nonperturbativesolution}) can be solved numerically.  In practice, 
for known parameters, the matrices $M$ and vectors ${\bf v}$ can be readily 
evaluated over some range of $h$, from which the solutions to the system 
$\alpha_{0},\beta_{0}(h)$ in that range immediately follow.  The resulting 
resonance in $\textrm{Im}\,\alpha_{0}(h)$ as a function of $h$ is then fitted to 
a Lorentzian, with its peak giving the shifted wavevector 
$\textrm{Re}\,\tildekp$ and its half-width giving $\textrm{Im}\,\tildekp$.  In 
Table~\ref{table:roughness2}, we give the resulting losses and wavevector shifts 
for a few roughness parameters, as calculated through 
Eq.~(\ref{eq:nonperturbativesolution}).  Again the numbers that we have used are 
for a silver nanowire at $\lambda_0=1\,\mu$m and $\epsilon_1=2$.  The shifts in 
$\textrm{Re}\,\tildekp$~(or equivalently, $\textrm{Re}\,\tilde{C}_{-1}$) and 
increases in the loss parameter 
$\textrm{Im}\,\tildekp/\textrm{Re}\,\tildekp$~(or 
$\textrm{Im}\,\tilde{C}_{-1}/\textrm{Re}\,\tilde{C}_{-1}$) are given in terms of 
the percentage increase over their values for a smooth nanowire.  Again it can 
be seen that for reasonable parameters, surface roughness adds only a moderate 
amount of loss to the system.

\section{Conclusions and outlook}\label{sec:conclusion}
We have demonstrated that the subwavelength confinement of guided plasmon modes 
on conducting nano-structures leads to strong coupling between these modes and 
nearby emitters in the optical domain.  This strong coupling leads to large 
effective Purcell factors for emission into the plasmon modes, which are limited 
only by heating losses in the conductor.  While losses prevent the plasmon modes 
from being useful photonic carriers of information, we have shown that they can 
be efficiently out-coupled, \textit{e.g.}, to a dielectric waveguide. We 
estimate that single photon generation efficiencies exceeding $95\%$ are 
possible in such a tiered system.  Finally we have analyzed the effects of 
plasmon scattering due to moderate surface roughness on these nano-structures.

Rapid advances in recent years in fabrication techniques for 
nanowires~\cite{xia02,schultz02}, nanotips~\cite{libioulle95}, and 
sub-wavelength dielectric waveguides~\cite{tong03,vlasov04} puts such a system 
in experimental reach.  Quantum dots or single color centers might serve as 
physical realizations of solid-state emitters, which could be used to achieve 
strong-coupling cavity QED and quantum information devices on a chip at optical 
frequencies.  It is also interesting to consider real, individual atoms 
interacting with nanowires and the challenges associated with constructing 
nanoscale traps.  These traps might in part be formed by the plasmon fields 
themselves.

We emphasize that the physical mechanisms that lead to strong coupling are not 
restricted to the nano-structures considered here but can be quite a general 
feature of the plasmon modes associated with sub-wavelength conducting devices. 
It is thus likely that the efficiencies calculated here are not fundamentally 
limited but can be further improved by proper design.  Photonic crystal-like 
structures for plasmons~\cite{maier04}, for example, may be a promising approach 
to achieve tight confinement while simultaneously reducing losses.  Similar 
schemes may also help to improve coupling between the plasmons and dielectric 
waveguide modes.  Such approaches are likely to improve the performance of 
plasmon cavity QED even further.

The authors thank Atac Imamoglu for useful discussions.  This work was supported 
by the ARO-MURI, ARDA, NSF, the Sloan and Packard Foundations, and by the Danish 
Natural Science Research Council.

\appendix
\section{General theory of electromagnetic modes of a cylinder}\label{sec:generaltheory}

The solution to the electromagnetic modes of a cylinder has been known for quite 
some time~\cite{stratton41,jackson99} and is briefly derived here.

We consider a cylinder of radius $R$ of dimensionless electric permittivity 
$\epsilon_{2}$, centered along the $z$-axis and surrounded by a second 
dielectric medium $\epsilon_{1}$.  For non-magnetic media the electric and 
magnetic fields in frequency space satisfy the wave equation
\be \nabla^{2}\left\{\begin{array}{c} \bfE(\bfr) \\ \bfH(\bfr)
\end{array} \right\}+\frac{\omega^2}{c^2}\epsilon(\bfr)\left\{\begin{array}{c} \bfE(\bfr) \\ \bfH(\bfr)
\end{array} \right\}=0.\label{eq:waveeqn}
\ee
The solutions to Eq.~(\ref{eq:waveeqn}) are perhaps most easily derived by first 
finding scalar solutions of the equation and then constructing vector solutions. 
Working in cylindrical coordinates, scalar solutions of Eq.~(\ref{eq:waveeqn}) 
satisfying the necessary boundary conditions take the form 
$\psi_{1}{\propto}H_{m}\left(k_{1\perp}\rho\right)e^{im\phi+ik_{\parallel}z}$ 
and 
$\psi_{2}{\propto}J_{m}\left(k_{2\perp}\rho\right)e^{im\phi+ik_{\parallel}z}$ 
outside and inside the cylinder, respectively.  Here $J_m$ and $H_m$ are Bessel 
functions and Hankel functions of the first kind, respectively, while 
$k_{i\perp}=\sqrt{k_i^2-k_{\parallel}^2}$ and 
$k_{i}=\omega\sqrt{\epsilon_{i}}/c$.  $J_m$ is well-behaved at $\rho=0$, while 
$H_m(x){\sim}e^{ix}$ for large $x$ satisfies outgoing-wave conditions. It is 
easy to verify that two independent vector solutions to Eq.~(\ref{eq:waveeqn}) 
are given by $\bfv_{i}=\frac{1}{k_i}\nabla\times\left(\hat{z}\psi_{i}\right)$ 
and $\bfw_{i}=\frac{1}{k_i}\nabla\times\bfv_{i}$.  The curl relations of 
Maxwell's Equations then imply that $\bfE$ and $\bfH$ must take the form
\bea \bfE_{i}(\bfr) & = & a_{i}\bfv_{i}(\bfr)+b_{i}\bfw_{i}(\bfr), \\
\bfH_{i}(\bfr) & = & 
-\frac{i}{\omega\mu_{0}}k_{i}\left(a_{i}\bfw_{i}(\bfr)+b_{i}\bfv_{i}(\bfr)\right), 
\eea
where $a_i,b_i$ are constant coefficients.  Expanding out these expressions in 
detail,
\bea \bfE_{i}(\bfr) & = & 
\left[\left(\frac{im}{k_{i}\rho}a_{i}F_{i,m}(\kiperp\rho)+\frac{i\kp\kiperp}{k_{i}^2}b_{i}F^{\prime}_{i,m}(\kiperp\rho)\right)\hat{\rho}+\left(-\frac{\kiperp}{k_{i}}a_{i}F^{\prime}_{i,m}(\kiperp\rho)-\frac{m\kp}{k_{i}^{2}\rho}b_{i}F_{i,m}(\kiperp\rho)\right)\hat{\phi}\right. 
\nonumber
\\ & &
\left.+\frac{\kiperp^2}{k_{i}^2}b_{i}F_{i,m}(\kiperp\rho)\hat{z}\right]e^{im\phi+i{\kp}z},\nonumber
\\ \bfH_{i}(\bfr) & = &
-\frac{i}{\omega\mu_{0}}k_{i}\left[\left(\frac{i\kp\kiperp}{k_{i}^2}a_{i}F^{\prime}_{i,m}(\kiperp\rho)+\frac{im}{k_{i}\rho}b_{i}F_{i,m}(\kiperp\rho)\right)\hat{\rho}-\left(\frac{m\kp}{k_{i}^{2}\rho}a_{i}F_{i,m}(\kiperp\rho)+\frac{\kiperp}{k_{i}}b_{i}F^{\prime}_{i,m}(\kiperp\rho)\right)\hat{\phi}\right. 
\nonumber \\ & & 
\left.+\frac{\kiperp^{2}}{k_{i}^2}a_{i}F_{i,m}(\kiperp\rho)\hat{z}\right]e^{im\phi+i{\kp}z},\label{eq:cylindricalfields} 
\eea
where $F_{1,m}(x)=H_{m}(x)$ and $F_{2,m}(x)=J_{m}(x)$.

Up to this point $a_{i},b_{i}$ are arbitrary coefficients, whose relationship 
becomes fixed by imposing boundary conditions between the cylinder and 
surrounding dielectric.  Requiring that the tangential components 
$E_{\phi},E_{z},H_{\phi},H_z$ of the fields be continuous at the boundary 
results in a linear system of four equations, which we write in abbreviated 
matrix form as $M(a_{1}\;a_{2}\;b_{1}\;b_{2})^{T}=0$~\cite{note1}.  A 
non-trivial solution for the fields requires that ${\det}\,M=0$, which after 
some work simplifies to the mode equation given in Eq.~(\ref{eq:modecondition}).

One special case of interest is that of a $TM$ mode with no winding~($m=0$).  
The component of $\bfH$ along $\hat{z}$ by definition vanishes, which implies 
that the coefficients $a_i$ in Eq.~(\ref{eq:cylindricalfields}) vanish.  The 
condition ${\det}\,M=0$ is then significantly easier to evaluate in this 
situation.  In particular, $a_i=0$ implies that the field components $E_{\phi}$ 
and $H_z$ vanish, and continuity of the remaining tangential field components 
$E_z$ and $H_{\phi}$ at the boundary requires that
\be \left(\begin{array}{cc} \frac{\koneperp^2}{k_1^2}H_{0}(\koneperp{R}) & 
-\frac{\ktwoperp^2}{k_2^2}J_{0}(\ktwoperp{R}) \\ 
\frac{i}{\omega\mu_0}{\koneperp}H_{0}'({\koneperp}R) & 
-\frac{i}{\omega\mu_0}{\ktwoperp}J_{0}'({\ktwoperp}R) \end{array}\right) 
\left(\begin{array}{c} b_{1} \\ b_{2} \end{array}\right) = 
\left(\begin{array}{c} 0 \\ 0 \end{array}\right). \ee
Setting the determinant of the above matrix equal to zero immediately yields the 
mode equation of Eq.~(\ref{eq:TMmodecondition}), and it is also immediately seen 
that the ratio of the coefficients $b_{1,2}$ must be given by 
Eq.~(\ref{eq:b1b2}).  

\section{Derivation of cutoff for higher-order
modes}\label{sec:cutoff} 

In this section we show that to a very good approximation, a nanowire 
essentially supports a single, fundamental $m=0$ plasmon mode.  In particular, 
for all higher-order plasmon modes $|m|{\geq}2$ a cutoff wire size $\Rcutoff$ 
exists below which such modes cannot exist, while the $|m|=1$ plasmon modes 
exhibit an exponential growth in their mode volumes as $R{\rightarrow}0$.  For 
simplicity, we will assume in this section that we are dealing with a lossless 
system~$(\textrm{Im}\,\epsilon_2=0)$.

\subsection{Behavior of $|m|{\geq}2$ modes}

We are interested here in the behavior of the $|m|{\geq}2$ modes near cutoff, 
which is characterized by a small deviation of the plasmon wavevector $\kp$ from 
$\sqrt{\epsilon_1}\omega/c$~(see Fig.~\ref{fig:wiremodes}).  To simplify algebra 
in the derivation of $\Rcutoff$, from this point forward we make the mode 
equation~(\ref{eq:modecondition}) dimensionless by setting $\omega/c=1$, and we 
will assume that $m$ is positive~(the case where $m$ is negative follows this 
derivation with a few minor modifications).  Furthermore, it is useful to define 
a small quantity $\delta=\kp-\sqrt{\epsilon_1}$, where we specifically consider 
the positive $\kp$ solution.  On physical grounds, any mode with positive $\kp$ 
must satisfy $\delta{\geq}0$, because if $\kp<\sqrt{\epsilon_1}$ the fields 
outside of the wire would be radiative in nature and implies that the system is 
continually radiating energy out to infinity without a source.  It follows that 
any value of $R$ where $\delta=0$ becomes a solution to 
Eq.~(\ref{eq:modecondition}) for some $m$ then corresponds to a critical point 
in behavior, and specifically is a cutoff beyond which modes cease to exist for 
that given $m$.  To find this $R=\Rcutoff$, it is useful to expand the two sides 
of Eq.~(\ref{eq:modecondition}) in $\delta$.  We will find that both sides have 
contributions to these expansions that are divergent at $\delta=0$~(terms that 
behave like $\delta^{-n}$, where $n>0$), and we will show that, for $m{\geq}2$, 
there exists one value of $R$ that equates these two divergent contributions; 
\textit{i.e.}, $\delta=0$ satisfies the mode equation at this particular value 
$R=R_{\footnotesize\textrm{cutoff}}$.

It is straightforward to show that the divergent contribution to the expansion 
of the left-hand side of Eq.~(\ref{eq:modecondition}) is given by
\be 
LHS=\frac{m^2}{4R^{2}\delta^{2}}-\frac{m^{2}(3\epsilon_1+\epsilon_2)}{4R^{2}\sqrt{\epsilon_1}(\epsilon_1-\epsilon_2)\delta} 
+\mathcal{O}(\delta^0).\label{eq:higherorderLHS} \ee
To expand the right-hand side, we first note that the quantity 
$(1/\ktwoperp)J_{m}^{\prime}({\ktwoperp}R)/J_{m}({\ktwoperp}R)=(1/\sqrt{\epsilon_2-\epsilon_1})\tilde{J}_{m}(\sqrt{\epsilon_2-\epsilon_1}R)+\mathcal{O}(\delta^1)$ 
is well-behaved near $\delta=0$.  Here we have defined 
$\tilde{J}_{m}(x)=J_{m}^{\prime}(x)/J_{m}(x)$.  Then, using the identity
\be H_{m}(ix)=\frac{2}{{\pi}i^{m+1}}K_{m}(x),\label{eq:KHrelation} \ee
where $K_{m}(x)$ is a modified Bessel function of the second kind, and the 
expansions
\bea K_{m}(x) & = & 
\frac{(m-1)!}{2}\left(\frac{2}{x}\right)^{m}-\frac{(m-2)!}{2}\left(\frac{2}{x}\right)^{m-2}+O(x^{4-m})\;\;\;(m{\geq}2),\label{eq:Kmexpansion}
\\ \koneperp & = &
i\left(\sqrt{2\delta\sqrt{\epsilon_1}}+\frac{\delta^{3/2}}{2^{3/2}\epsilon_{1}^{1/4}}+\mathcal{O}(\delta^{5/2})\right)\label{eq:koneperpexpansion}
\\ & \equiv & i\kappa_{1\perp},\label{eq:kappaoneperpexpansion} \eea
it is tedious but straightforward to expand the expression 
$(1/\koneperp)(H_{m}'({\koneperp}R)/H_{m}(\koneperp R))$ as well.  Performing 
these expansions and simplifying, one finds that
\be 
RHS=\frac{m^2}{4R^{2}\delta^2}-\frac{m^2}{4R^{2}\sqrt{\epsilon_1}\delta}+\frac{m\sqrt{\epsilon_1}}{2(m-1)\delta}+\frac{{i}m(\epsilon_1+\epsilon_2)\tilde{J}_{m}(\sqrt{\epsilon_2-\epsilon_1}R)}{2R\sqrt{\epsilon_{1}(\epsilon_1-\epsilon_2)}\delta}+\mathcal{O}(\delta^0). 
\label{eq:higherorderRHS} \ee
Comparing Eqs.~(\ref{eq:higherorderLHS}) and~(\ref{eq:higherorderRHS}), we see 
that $\delta=0$ is a solution provided that these terms are equal to 
$\mathcal{O}(\delta^{-1})$, \textit{i.e.},
\be 
\frac{m}{R}\frac{\epsilon_1+\epsilon_2}{\epsilon_2-\epsilon_1}=\frac{R\epsilon_1}{m-1}+\frac{i(\epsilon_1+\epsilon_2)\tilde{J}_{m}(\sqrt{\epsilon_2-\epsilon_1}R)}{\sqrt{\epsilon_1-\epsilon_2}}.\label{eq:Rcutoff} 
\ee
The solution $R=\Rcutoff$ to Eq.~(\ref{eq:Rcutoff}) gives the cutoff wire size 
below which the mode $m$ cannot exist.  In the regime of 
interest~($\epsilon_1>0$,$\epsilon_2<0$,$\epsilon_1+\epsilon_2<0$), the first 
and second terms are positive while the third term is a negative function~(for 
$R>0$) that behaves like $-1/R$ for small $R$ and approaches a constant for 
large $R$.  It can be seen then that a solution exists for any $m{\geq}2$, which 
establishes that these modes are indeed cut off in the nanowire limit.

\subsection{Behavior of $|m|=1$ mode}

For simplicity we will assume that $m=1$, as the case of $m=-1$ follows this 
derivation closely.  The case of $m=1$ must be studied separately because the 
expansion of $K_{m}(x)$ given in Eq.~(\ref{eq:Kmexpansion}) only holds for 
$m{\geq}2$.  The different asymptotic scaling of $K_{1}(x)$ leads to unique 
behavior of the $m=1$ mode in the nanowire limit.  In particular we will show 
that this mode does not strictly have a cutoff size, but that 
$\kp\rightarrow\sqrt{\epsilon_1}$ exponentially in the limit $R{\rightarrow}0$.  
In turn, the magnitude of $\koneperp$ becomes exponentially small, which 
corresponds to an exponential growth in the spatial extent or mode volume.

Again defining $\delta=\kp-\sqrt{\epsilon_1}$, we are interested in finding an 
approximate solution to Eq.~(\ref{eq:modecondition}) in the limit of small 
$\delta$.  We proceed by expanding both sides of the equation as a series in the 
small parameter.  The expression for the left-hand side given by 
Eq.~(\ref{eq:higherorderLHS}) remains valid for $m=1$.  For the right-hand side, 
we anticipate that both the quantities ${\koneperp}R$ and $\koneperp$ will be 
small as $R{\rightarrow}0$~(these assumptions can be checked for consistency at 
the end of the calculation), and we thus expand around ${\koneperp}R=0$ the term
\bea \frac{1}{\koneperp}\frac{H_{1}^{\prime}({\koneperp}R)}{H_{1}({\koneperp}R)} 
& = & 
\frac{1}{i\kappa_{1\perp}}\frac{H_{1}^{\prime}(i\kappa_{1\perp}R)}{H_{1}(i\kappa_{1\perp}R)}
\\ & = &
\frac{1}{i\kappa_{1\perp}}\left(\frac{i}{\kappa_{1\perp}R}-i\left(\gamma+\log\frac{\kappa_{1\perp}R}{2}\right)\kappa_{1\perp}R+O(\kappa_{1\perp}^{2}R^2)\right), 
\eea
where $\gamma{\approx}0.577$ is Euler's constant.  Here we have used 
Eq.~(\ref{eq:KHrelation}) to convert $H_{m}(ix)$ to $K_{m}(x)$ and the expansion
\be 
K_{1}(x)=\frac{1}{x}+\left(\frac{\gamma}{2}-\frac{\log{2}}{2}-\frac{1}{4}+\frac{\log{x}}{2}\right)x+O(x^3). 
\ee
We now assume that $\kappa_{1\perp}R$ is small enough that 
$\gamma{\ll}|\log\kappa_{1\perp}R|$, such that
\be 
\frac{1}{\koneperp}\frac{H_{1}^{\prime}({\koneperp}R)}{H_{1}({\koneperp}R)}{\approx} 
\frac{1}{i\kappa_{1\perp}}\left(\frac{i}{\kappa_{1\perp}R}-i\kappa_{1\perp}R\log\frac{\kappa_{1\perp}R}{2}\right). 
\ee
Furthermore, having assumed that $\koneperp$~(and by extension, 
$\kappa_{1\perp}$) is a small quantity, we can now expand the expression above 
in terms of $\delta$ using Eqs.~(\ref{eq:koneperpexpansion}) 
and~(\ref{eq:kappaoneperpexpansion}).  Making this substitution, and after a bit 
of algebra, one finds that the expansion of the right-hand side of 
Eq.~(\ref{eq:modecondition}) is given by
\be 
RHS{\approx}\frac{1}{4R^{2}\delta^2}+\frac{\epsilon_1+3\epsilon_2-2R^{2}\epsilon_1\left(\epsilon_{1}-\epsilon_2\right)\log\left({\delta}R^{2}\sqrt{\epsilon_1}/2\right)} 
{4R^{2}\sqrt{\epsilon_1}\left(\epsilon_1-\epsilon_2\right)\delta}. \ee
Finally, equating the left- and right-hand sides to $\mathcal{O}(\delta^{-1})$ 
gives the solution
\be 
\delta{\approx}\frac{2}{R^{2}\sqrt{\epsilon_1}}\exp\left(-\frac{2\left(\epsilon_1+\epsilon_2\right)}{R^{2}\epsilon_{1}\left(\epsilon_2-\epsilon_1\right)}\right).\label{eq:m1behavior} 
\ee
It follows that in the nanowire limit,
\bea \kappaoneperp & = & \left(\kp^2-\epsilon_1\right)^{1/2} \\ & \approx & 
\left(2\delta\sqrt{\epsilon_1}\right)^{1/2} \\ & \approx & 
\frac{2}{R}\exp\left(-\frac{\epsilon_1+\epsilon_2}{R^{2}\epsilon_{1}\left(\epsilon_2-\epsilon_1\right)}\right)\label{eq:kappabehavior}. 
\eea

Eqs.~(\ref{eq:m1behavior}) and~(\ref{eq:kappabehavior}) indicate that the $m=1$ 
plasmon mode does not have a cutoff in the nanowire limit, but instead that its 
longitudinal wavevector approaches $\sqrt{\epsilon_1}$ exponentially, with a 
corresponding exponential increase in its transverse 
extent~(${\sim}1/\kappaoneperp$) and mode volume.  It is therefore 
well-justified to say that this mode is effectively cut off, as the coupling 
strength to this mode becomes strongly suppressed as $R{\rightarrow}0$.

\section{Radiative and non-radiative decay rates near a 
nanotip}\label{sec:tiprates}

Here we derive more carefully the expressions given in Eqs.~(\ref{eq:tiprad}) 
and~(\ref{eq:tipnonrad}) for the radiative and non-radiative spontaneous 
emission rates near a nanotip.

To calculate the radiative rate, we should consider our expression for $\Phi_r$ 
in Eq.~(\ref{eq:Phirnanotip}) in the far-field~(large $v$) limit, where the 
$K_{m}(qv)$ terms in $\Phi_{r}$ decay exponentially with $v$.  Because of this 
exponential dependence at large $v$, to good approximation it suffices to expand 
the terms $\alpha_{m}(q),J_{m}(qu'),K_{m}(qv')$ around $q=0$.  The only 
non-trivial expansions occur for the terms $\alpha_{m}(q)$ and are given by
\bea \alpha_{0}(q) & = & 
\frac{1}{2}\left(1-\frac{\epsilon_2}{\epsilon_1}\right)q^2v_{0}^2+\mathcal{O}(q^4), 
\\ \alpha_{1}(q) & = & 
\frac{1}{2}\frac{\epsilon_1-\epsilon_2}{\epsilon_1+\epsilon_2}q^2v_{0}^2+\mathcal{O}(q^4), 
\;\;\cdots \eea
These expansions allow for exact evaluations of the integral.  It can be 
verified that the dipole contributions to $\Phi_r$ originate from the $m=0,1$ 
terms in the sum, which are readily found to be
\bea \Phi_{r}^{(m=0)}(\bfr,\bfrp) & \approx & \frac{1}{4\pi\epsilon_0\epsilon_1}v_0^2\left(1-\frac{\epsilon_2}{\epsilon_1}\right)\frac{4(v^2-u^2)}{(u^2+v^2)^3}\ln\frac{v}{v'}+\delta\Phi(\bfr), \\
\Phi_{r}^{(m=1)}(\bfr,\bfrp) & \approx & 
\frac{1}{4\pi\epsilon_0\epsilon_1}\cos(\phi-\phi')\frac{\epsilon_1-\epsilon_2}{\epsilon_1+\epsilon_2}\frac{v_0^{2}u'}{v'}\frac{8uv}{(u^2+v^2)^3}. 
\eea
Here $\delta\Phi$ is a complicated function, but most importantly contains no 
dependence on $\bfrp$.  Recalling that the pseudopotentials derived above 
correspond to a point charge source, we can immediately obtain the potentials 
due to a dipole $\bfp0 e^{-i\omega t}$ at $\bfrp$ by applying the operator 
$\left(\bfp0\cdot\nabla'\right)$ to these expressions.  In parabolic coordinates 
the gradient operator is given by
\be \nabla=\frac{1}{\sqrt{u^2+v^2}}\left(\hat{u}\frac{\partial}{\partial 
u}+\hat{v}\frac{\partial}{\partial 
v}\right)+\frac{1}{uv}\hat{\phi}\frac{\partial}{\partial \phi}, \ee
and for a dipole located on the $z$-axis~($u'=0$), we find that
\bea \Phi_{dip,r}^{(m=0)} & \approx & 
-\frac{1}{4\pi\epsilon_0\epsilon_1}\left(1-\frac{\epsilon_2}{\epsilon_1}\right)\frac{v_0^2}{v'^2}\frac{(\hat{v}\cdot\bfr)(\bfp0\cdot\hat{v})}{r^3}, 
\\ \Phi_{dip,r}^{(m=1)} & \approx & 
\frac{1}{4\pi\epsilon_0\epsilon_1}\frac{\epsilon_1-\epsilon_2}{\epsilon_1+\epsilon_2}\frac{v_0^2}{v'^2}\frac{\left(\bfp0-\hat{v}(\bfp0\cdot\hat{v})\right)\cdot\bfr}{r^3}. 
\eea
From these expressions one can immediately identify the induced dipole moments 
in the nanotip,
\bea \delta{\bf p} & = & 
-{\hat{v}}p_{0}\frac{v_0^2}{v'^2}\left(1-\frac{\epsilon_2}{\epsilon_1}\right),\;\;\;\;(\bfp0\parallel\hat{z}) 
\\ \delta{\bf p} & = & 
\hat{u}p_{0}\frac{\epsilon_1-\epsilon_2}{\epsilon_1+\epsilon_2}\frac{v_0^2}{v'^2},\;\;\;\;(\bfp0\perp\hat{z}) 
\eea
and arrive at the radiative decay rates given in Eq.~(\ref{eq:tiprad}).

The leading term for the non-radiative decay rate is found by calculating the 
divergence in the reflected field $\bfE_{r}(\bfrp,\bfrp)$ as 
$v{\rightarrow}v_0$.  The reflected field 
$\bfE_r=-\nabla(\bfp0\cdot\nabla')\Phi_{r}$ is in general difficult to evaluate, 
but simplifies considerably for a dipole located on-axis~($u'=0$) due to the
presence of the $J_{m}(qu')$ term in $\Phi_{r}$, given in 
Eq.~(\ref{eq:Phirnanotip}).  The operation $\nabla'$ causes terms like 
$J_{m}(0)$ and $J_{m}'(0)$ to appear in $\bfE_r$, which are non-zero only when 
$m=0$ and $m=1$, respectively.  This immediately leads to the expressions
\bea \bfp0\cdot\bfE_{r}(\bfrp,\bfrp) & = & 
-\frac{p_0^2}{4\pi\epsilon_0\epsilon_1}\int_{0}^{\infty}dq\;\frac{q^3}{v'^2}\alpha_{1}(q)K_1^2(qv'),\;\;\;\;\;\;(\bfp0\perp\hat{z}) 
\nonumber \\ \bfp0\cdot\bfE_{r}(\bfrp,\bfrp) & = & 
-\frac{p_0^2}{2\pi\epsilon_0\epsilon_1}\int_{0}^{\infty}dq\;\frac{q^3}{v'^2}\alpha_{0}(q)K_1^2(qv'),\;\;\;\;\;\;(\bfp0\parallel\hat{z})\label{eq:pEtipappendix} 
\eea
which were given in Eq.~(\ref{eq:pEtip}).  Examining further the solutions to 
$\alpha_{0,1}$, it can easily be shown that their asymptotic expansions in the 
limit $qv_0\gg 1$ take the form
\be 
\alpha_{0,1}(q){\approx}\frac{1}{\pi}\frac{\epsilon_1-\epsilon_2}{\epsilon_1+\epsilon_2}e^{2qv_0}.\;\;\;\;\;\;(qv_0{\gg}1) 
\ee
At the same time, in the limit $qv'\gg 1$ the behavior of $K_1^2$ is given by 
$K_1^2(qv')\approx(\pi/2qv')e^{-2qv'}$, and thus as $v'{\rightarrow}v_0$ the 
integrands of Eq.~(\ref{eq:pEtipappendix}) exhibit very long tails due to the 
presence of terms ${\sim}e^{-2q(v'-v_0)}$ at large $q$.  The tail is the origin 
of the divergence that we expected on physical grounds.  Using these expansions 
as well as the fact that the decay rate is proportional to 
$\textrm{Im}\;(\bfp0\cdot\bfE)$, the integrals can be evaluated exactly and 
yield the non-radiative decay rates given in Eq.~(\ref{eq:tipnonrad}).

\section{Boundary element method}\label{sec:BEM}

Our numerical implementation of the boundary element method~(BEM) closely 
follows the method derived in~\cite{abajo02}.  Here we briefly outline the main 
ideas of BEM while referring the reader to~\cite{abajo02} for more details, and 
we discuss the key elements of our implementation.

We assume that our system contains a set of known, time-harmonic source charges 
and currents $\rho_{\footnotesize\textrm{ext}}$,${\bf 
j}_{\footnotesize\textrm{ext}}$ in the presence of some scattering dielectric 
body whose surface is denoted $S$~(although we discuss one body here, BEM is 
easily generalizable to treat multiple scatterers).  In the case of interest, 
$S$ represents the surface of a metallic nanotip, while the external source 
corresponds to an oscillating point dipole ${\bfp0}e^{-i\omega t}$ at some 
location $\bfrp$.  For simplicity we also assume that we are working with 
non-magnetic media, and we denote by $\epsilon_j$~($j=1,2$) the dimensionless 
electric permittivities outside and inside $S$, respectively.  The underlying 
principle behind BEM is that the scalar and vector potentials $\phi_{j}(\bfr)$ 
and ${\bf A}_{j}(\bfr)$ in each region can be written~(in the Lorenz gauge) in 
the form
\bea \phi_{j}(\bfr) & = & 
\frac{1}{4\pi\epsilon_{0}\epsilon_{j}}\int\,d\bfrp\,G_{j}(\bfr-\bfrp)\rho_{\footnotesize\textrm{ext}}(\bfrp) 
+\frac{1}{4\pi\epsilon_{0}\epsilon_{j}}\int_{S}\,d\bfs\,G_{j}(\bfr-\bfs)\sigma_{j}(\bfs), 
\\ {\bf A}_{j}(\bfr) & = & \frac{\mu_0}{4\pi}\int\,d\bfrp\,G_{j}(\bfr-\bfrp){\bf j}_{\footnotesize\textrm{ext}}(\bfrp) 
+\frac{\mu}{4\pi}\int_{S}\,d\bfs\,G_{j}(\bfr-\bfs){\bf h}_{j}(\bfs), \\ 
G_{j}(\bfr) & = & \frac{e^{ik_{j}r}}{r}, \eea
where $G_j$ is the Green's function in a medium of uniform $\epsilon_j$, and 
$k_{j}=\sqrt{\epsilon_j}(\omega/c)$.  Physically, the equations above state that 
the fields in region $j$ can be described as a result of the combination of the 
external sources and some effective surface charge and current distributions 
$\sigma_{j},{\bf h}_{j}$ on $S$.  In general, these effective distributions do 
not have physical significance; for example, they do not correspond to actual 
charges and currents, and the distributions in region $1$ and region $2$ are not 
necessarily equal~(\textit{e.g.}, $\sigma_{1}(\bfs)\neq\sigma_{2}(\bfs)$).  The 
values of $\sigma_{j},{\bf h}_{j}$ are not known initially, but a set of linear 
integral equations for these distributions results from enforcing various 
boundary conditions for the scalar and vector potentials at $S$.  In particular, 
$\phi$, ${\bf A}$, $\bfD_{\perp}$, and $\bfH_{\parallel}$ must be continuous at 
the boundary.  To calculate the distributions numerically, if the boundary $S$ 
is finite, one can mesh up the surface into a finite number of grid points.  
Assuming that $\sigma_j,{\bf h}_j$ are constant over each grid point, the linear 
integral equations become a set of linear equations in the values of 
$\sigma_{j},{\bf h}_j$ that can be solved straightforwardly.  Once these 
distributions are known, the potentials and then the fields $\bfE,\bfH$ can be 
calculated.

In our problem of interest, we assume that the dipole is located on the $z$-axis 
and oriented along $\hat{z}$, while the nanotip is described by a paraboloid of 
revolution around the $z$-axis.  Due to the axial symmetry of the system, BEM 
simulations are advantageous because one only needs to calculate the unknown 
distributions along one dimension instead of over the entire two-dimensional 
surface $S$.  At the same time, the source is a dipole oscillating at constant 
frequency, and thus the external charges and currents are calculated quite 
easily.  In BEM~(at least in the current formulation), it is necessary that the 
nanotip surface $S$ be finite, and we implement this numerically by tapering and 
rounding off the nanotip far from the region of interest.  In general, any 
termination can result in some back-reflection of the guided plasmon, and this 
results in some small oscillations of the fields due to interference with the 
forward-propagating plasmon, as barely seen, \textit{e.g.}, in 
Fig.~\ref{fig:BEMplot}.  In our simulations, the reflected amplitude is kept to 
within a few percent.  Very fine meshes were used to ensure accuracy; in most of 
our simulations, for example, the spacing between points in the regions of 
constant $R$ was approximately $\lambda_{\footnotesize\textrm{pl}}/400$.

\section{Derivation of coupled-mode equations}\label{sec:coupledmodeeqns}

In this section we derive the equations of evolution for two electromagnetically 
coupled systems based on Lorentz reciprocity.

First we derive the Lorentz reciprocity equation generally.  Assuming 
non-magnetic media, suppose that 
$\{\bfEdu{1}{}(\bfr),\bfHdu{1}{}(\bfr),\epsilondu{1}{}(\bfr)\}$ and 
$\{\bfEdu{2}{}(\bfr),\bfHdu{2}{}(\bfr),\epsilondu{2}{}(\bfr)\}$ separately 
satisfy Maxwell's Equations.  At this point the systems $1,2$ and their field 
solutions are not necessarily related to each other at all.  In the following we 
assume that all fields $\bfE(\bfr,t)=\bfE(\bfr) e^{-i\omega 
t}$,$\bfH(\bfr,t)=\bfH(\bfr) e^{-i\omega t}$ have harmonic time dependence.  
Using the vector identity
\be \nabla\cdot({\bf a}\times{\bf b})={\bf b}\cdot(\curl{\bf a})-{\bf 
a}\cdot(\curl{\bf b}), \ee
and the curl relations of Maxwell's Equations we can write
\bea \nabla\cdot(\bfEdu{1}{}\times\bfHdu{2}{\ast}) & = & 
\bfHdu{2}{\ast}\cdot(\curl\bfEdu{1}{})-\bfEdu{1}{}\cdot(\curl\bfHdu{2}{\ast}) 
\nonumber
\\ & = &
\bfHdu{2}{\ast}\cdot(i\omega\mu_{0}\bfHdu{1}{})-\bfEdu{1}{}\cdot(i\omega\epsilon_0\epsilondu{2}{\ast}\bfEdu{2}{\ast}),\label{eq:1} 
\eea
and similarly
\bea \nabla\cdot(\bfEdu{2}{\ast}\times\bfHdu{1}{}) & = & 
\bfHdu{1}{}\cdot(\curl\bfEdu{2}{\ast})-\bfEdu{2}{\ast}\cdot(\curl\bfHdu{1}{}) 
\nonumber
\\ & = &
\bfHdu{1}{}\cdot(-i\omega\mu_{0}\bfHdu{2}{\ast})-\bfEdu{2}{\ast}\cdot(-i\omega\epsilon_0\epsilondu{1}{}\bfEdu{1}{}).\label{eq:2} 
\eea
Adding up Eqs.~(\ref{eq:1}) and~(\ref{eq:2}) yields the equation for Lorentz 
reciprocity,
\be 
\nabla\cdot(\bfEdu{1}{}\times\bfHdu{2}{\ast}+\bfEdu{2}{\ast}\times\bfHdu{1}{})=i\omega\epsilon_0\bfEdu{1}{}\cdot\bfEdu{2}{\ast}(\epsilondu{1}{}(\bfr)-\epsilondu{2}{\ast}(\bfr)).\label{eq:lorentzreciprocity} 
\ee

We now derive coupled-mode equations for two waveguides based on the Lorentz 
reciprocity equation above.  This derivation closely follows that 
of~\cite{barclay03}.  We emphasize that the nature of the waveguides can be 
quite general, \textit{e.g.}, they can be any type of normal dielectric or 
plasmon waveguide.  We let the indices $\mu,\nu=a,b$ refer to the system 
consisting of waveguide $a$ without the presence of system $b$, and $b$ the 
system consisting of waveguide $b$ without the presence of $a$.  We also assume 
that the surrounding dielectrics for systems $a,b$ are the same, \textit{i.e.}, 
$\epsilondu{a}{}(r=\infty)=\epsilondu{b}{}(r=\infty)$, and that the waveguides 
are co-propagating along the $z$-direction.  It is assumed that the total 
electric field for the system consisting of waveguides $a$ and $b$ together can 
be written as
\be 
\bfEdu{T}{}(\bfr)=\sum_{\nu=a,b}C_{\nu}(z)\bfEdu{\nu}{}(\bfr),\label{eq:solutionansatz} 
\ee
with a similar expression for $\bfH$.  That is, we assume that the total field 
can be written as a linear superposition of the unperturbed modes of systems 
$a,b$.  For the case where systems $a,b$ each have one allowed mode, the index 
$\nu$ refers to these unperturbed modes. In general, when $a,b$ have $N_{a,b}$ 
allowed unperturbed modes, $\nu$ is understood to be an index that covers all of 
these modes.  We can derive exact equations of motion for $C_{\nu}(z)$ by using 
Eq.~(\ref{eq:lorentzreciprocity}). Specifically, we will let the index $1=T$ in 
Eq.~(\ref{eq:lorentzreciprocity}) refer to the total fields 
$\bfEdu{T}{}(\bfr),\bfHdu{T}{}(\bfr)$ and the dielectric profile of the combined 
system $\epsilon_T(\bfr)$, while we will let the index $2=\mu$ refer to any one 
of the allowed, unperturbed modes of systems $a,b$.  Substituting this into 
Eq.~(\ref{eq:lorentzreciprocity}) yields
\be 
\nabla\cdot(\bfEdu{T}{}\times\bfHdu{\mu}{\ast}+\bfEdu{\mu}{\ast}\times\bfHdu{T}{})=i\omega\epsilon_0\bfEdu{T}{}\cdot\bfEdu{\mu}{\ast}(\epsilondu{T}{}(\bfr)-\epsilondu{\mu}{\ast}(\bfr)), 
\ee
or
\be 
\sum_{\nu=a,b}\nabla\cdot\left(C_{\nu}(z)\bfEdu{\nu}{}\times\bfHdu{\mu}{\ast}+C_{\nu}(z)\bfEdu{\mu}{\ast}\times\bfHdu{\nu}{}\right)=i\omega\epsilon_0\sum_{\nu=a,b}C_{\nu}(z)\bfEdu{\nu}{}\cdot\bfEdu{\mu}{\ast}\left(\epsilondu{T}{}(\bfr)-\epsilondu{\mu}{\ast}(\bfr)\right). 
\ee
Applying Stokes' Theorem to this result gives
\bea 
\frac{\partial}{{\partial}z}\sum_{\nu=a,b}\int\;d^2\bfrho\left(C_{\nu}(z)\bfEdu{\nu}{}\times\bfHdu{\mu}{\ast}+C_{\nu}(z)\bfEdu{\mu}{\ast}\times\bfHdu{\nu}{}\right)\cdot\hat{z}
 & = &
i\omega\epsilon_0\sum_{\nu=a,b}C_{\nu}(z)\int\;d^2\bfrho\;\bfEdu{\nu}{}\cdot\bfEdu{\mu}{\ast}\left(\epsilondu{T}{}(\bfr)-\epsilondu{\mu}{\ast}(\bfr)\right).\label{eq:unsimplified} 
\eea
The left-hand side can be further simplified,
\bea LHS & = & 
\frac{\partial}{{\partial}z}\sum_{\nu=a,b}\int\;d^2\bfrho\left(C_{\nu}(z)\bfEdu{\nu}{}\times\bfHdu{\mu}{\ast}+C_{\nu}(z)\bfEdu{\mu}{\ast}\times\bfHdu{\nu}{}\right)\cdot\hat{z}
\\ & = &
\sum_{\nu=a,b}\frac{dC_{\nu}}{dz}\left(\int\;d^2\bfrho\;(\bfEdu{\nu}{}\times\bfHdu{\mu}{\ast}+\bfEdu{\mu}{\ast}\times\bfHdu{\nu}{})\cdot\hat{z}\right) 
+C_{\nu}(z)\left(\frac{\partial}{{\partial}z}\int\;d^2\bfrho\;(\bfEdu{\nu}{}\times\bfHdu{\mu}{\ast}+\bfEdu{\mu}{\ast}\times\bfHdu{\nu}{})\cdot\hat{z}\right)
\\ & = &
\sum_{\nu=a,b}\frac{dC_{\nu}}{dz}\left(\int\;d^2\bfrho\;(\bfEdu{\nu}{}\times\bfHdu{\mu}{\ast}+\bfEdu{\mu}{\ast}\times\bfHdu{\nu}{})\cdot\hat{z}\right) 
+C_{\nu}(z)\left(i\omega\epsilon_0\int\;d^2\bfrho\;\bfEdu{\nu}{}\cdot\bfEdu{\mu}{\ast}(\epsilondu{\nu}{}(\bfr)-\epsilondu{\mu}{\ast}(\bfr))\right),\label{eq:simplelhs} 
\eea
where we have applied Stokes' Theorem on Eq.~(\ref{eq:lorentzreciprocity}) to 
get the last line.

Substituting Eq.~(\ref{eq:simplelhs}) back into Eq.~(\ref{eq:unsimplified}) 
yields a set of $N_a+N_b$ coupled, first-order differential equations:
\bea \sum_{\nu}\frac{dC_{\nu}}{dz}P_{\nu\mu}(z) & = &
-i\omega\epsilon_0\sum_{\nu}C_{\nu}(z)K_{\nu\mu}(z),\label{eq:couplemodeeqns} \\
P_{\nu\mu}(z) & \equiv & 
\int\;d^2\bfrho\;(\bfEdu{\nu}{}\times\bfHdu{\mu}{\ast}+\bfEdu{\mu}{\ast}\times\bfHdu{\nu}{})\cdot\hat{z},
\\ K_{\nu\mu}(z) & \equiv &
\int\;d^2\bfrho\;\bfEdu{\nu}{}\cdot\bfEdu{\mu}{\ast}(\epsilondu{\nu}{}(\bfr)-\epsilondu{T}{}(\bfr)). 
\eea
We emphasize that these coupled-mode equations are exact within the ansatz given 
by Eq.~(\ref{eq:solutionansatz}).

\section{Radiative scattering}\label{sec:radiativelossesappendix}

Suppose that in the presence of roughness, the first-order scattered field $E_z$ 
is given by Eq.~(\ref{eq:Eexpansion}).  Using the expressions derived in 
Eq.~(\ref{eq:tm0mode}) for the unperturbed, incident plasmon field, and letting 
$\kp$ denote the unperturbed plasmon wavevector, the total~(incident plus 
scattered) fields to first order in $p$ are given by
\bea \bfE_{1}^{\footnotesize\textrm{total}} & = &
\left(\frac{i\kp\koneperp}{k_1^2}b_{1}H^{\prime}_{0}(\koneperp\rho)e^{i{\kp}z}+p\int_{-\infty}^{\infty}d{\hp}\,\frac{i\hp\honeperp}{k_1^2}H^{\prime}_{0}(\honeperp\rho)e^{i{\hp}z}A(\hp)\right)\hat{\rho} \nonumber \\
& & 
+\left(\frac{\koneperp^2}{k_1^2}b_{1}H_{0}(\koneperp\rho)e^{i{\kp}z}+p\int_{-\infty}^{\infty}d{\hp}\,\frac{\honeperp^2}{k_1^2}H_{0}(\honeperp\rho)e^{i{\hp}z}A(\hp)\right)\hat{z}
\nonumber \\
\bfE_{2}^{\footnotesize\textrm{total}} & = & 
\left(\frac{i\kp\ktwoperp}{k_2^2}b_{2}J^{\prime}_{0}(\ktwoperp\rho)e^{i{\kp}z}+p\int_{-\infty}^{\infty}d{\hp}\,\frac{i\hp\htwoperp}{k_2^2}J^{\prime}_{0}(\htwoperp\rho)e^{i{\hp}z}B(\hp)\right)\hat{\rho}
\nonumber \\
& & 
+\left(\frac{\ktwoperp^2}{k_2^2}b_{2}J_{0}(\ktwoperp\rho)e^{i{\kp}z}+p\int_{-\infty}^{\infty}d{\hp}\,\frac{\htwoperp^2}{k_2^2}J_{0}(\htwoperp\rho)e^{i{\hp}z}B(\hp)\right)\hat{z}
\nonumber \\
H_{1,\phi}^{\footnotesize\textrm{total}} & = & 
\frac{1}{\omega\mu_0}\left[i{\koneperp}b_{1}H^{\prime}_{0}(\koneperp\rho)e^{i{\kp}z}+p\int_{-\infty}^{\infty}d{\hp}\,i{\honeperp}H^{\prime}_{0}(\honeperp\rho)e^{i{\hp}z}A(\hp)\right]
\nonumber \\
H_{2,\phi}^{\footnotesize\textrm{total}} & = & 
\frac{1}{\omega\mu_{0}}\left[i{\ktwoperp}J^{\prime}_{0}(\ktwoperp\rho)b_{2}e^{i{\kp}z}+p\int_{-\infty}^{\infty}d{\hp}\,i{\htwoperp}J^{\prime}_{0}(\htwoperp\rho)e^{i{\hp}z}B(\hp)\right].\label{eq:fields} 
\eea

The boundary condition equations in Eq.~(\ref{eq:boundarycond}) can be solved by 
plugging in the fields above, carefully expanding the equations as a power 
series in $p$, and then solving for each order of $p$, utilizing the expansions
\bea \hat{t} & = & \hat{z}+p\frac{d\zeta}{dz}\hat{\rho}+\mathcal{O}(p^2), \nonumber \\ F_{i,m}(k_{i\perp}\rho_0) & = & F_{i,m}({\kiperp}R)+p\zeta{\kiperp}F^{\prime}_{i,m}({\kiperp}R)+\mathcal{O}(p^2), \nonumber \\
F_{i,m}\left(\hiperp\rho_0\right) & = & 
F_{i,m}({\hiperp}R)+p\zeta{\hiperp}F^{\prime}_{i,m}({\hiperp}R)+\mathcal{O}(p^2),\label{eq:pexpansions} 
\eea
where $F_{1,m}(x)=H_{m}(x)$ and $F_{2,m}(x)=J_{m}(x)$.  The resulting 
$\mathcal{O}(p^0)$ equations are trivially satisfied by the plasmon fields of a 
smooth nanowire, while the $\mathcal{O}(p)$ equations are found to be
\bea & & 
\int_{-\infty}^{\infty}d\hp\,\left[\frac{\honeperp^2}{k_1^2}H_{0}({\honeperp}R)A(\hp)-\frac{\htwoperp^2}{k_2^2}J_{0}({\htwoperp}R)B(\hp)\right]e^{i{\hp}z}=
\nonumber \\
& & 
\;\;\;\;\;\;\;\left[\frac{\ktwoperp^3}{k_2^2}b_{2}\zeta(z)J^{\prime}_{0}({\ktwoperp}R)-\frac{\koneperp^3}{k_1^2}b_{1}\zeta(z)H^{\prime}_{0}({\koneperp}R) 
+\frac{i\kp\ktwoperp}{k_2^2}b_{2}\frac{d\zeta}{dz}J^{\prime}_{0}({\ktwoperp}R)-\frac{i\kp\koneperp}{k_1^2}b_{1}\frac{d\zeta}{dz}H^{\prime}_{0}({\koneperp}R)\right]e^{i{\kp}z} 
\nonumber
\\
& & 
\int_{-\infty}^{\infty}d\hp\,\left[{\honeperp}H^{\prime}_{0}({\honeperp}R)A(\hp)-{\htwoperp}J^{\prime}_{0}({\htwoperp}R)B(\hp)\right]e^{i{\hp}z}= 
\nonumber \\ & & 
\;\;\;\;\;\;\;\left[\ktwoperp^{2}b_{2}\zeta(z)J^{\prime\prime}_{0}({\ktwoperp}R)-\koneperp^{2}b_{1}\zeta(z)H^{\prime\prime}_{0}({\koneperp}R)\right]e^{i{\kp}z}.\label{eq:firstorderp} 
\eea
Here we assume that the metal inherently has no losses, \textit{i.e.}, 
$\textrm{Im}\,\epsilon_2=0$, such that $\kp$ is purely real.  Then, by plugging 
in the Fourier transform of $\zeta(z)$ given in 
Eq.~(\ref{eq:roughnesstransform}), the equations above become purely algebraic.  
It is tedious but straightforward to show that the solutions are given by 
Eq.~(\ref{eq:AB}), with the coefficients $f(\hp)$ and $g(\hp)$ defined via
\bea f(\hp) & = & 
\frac{{\htwoperp}NJ_{0}({\htwoperp}R)-k_2^{2}M(\hp)J^{\prime}_{0}({\htwoperp}R)}{H_{0}({\honeperp}R)J^{\prime}_{0}({\htwoperp}R){\honeperp}k_2^2-H^{\prime}_{0}({\honeperp}R)J_{0}({\htwoperp}R){\htwoperp}k_1^2}
\nonumber \\
g(\hp) & = & 
\frac{{\honeperp}NH_{0}({\honeperp}R)-k_1^{2}M(\hp)H^{\prime}_{0}({\honeperp}R)}{H_{0}({\honeperp}R)J^{\prime}_{0}({\htwoperp}R){\honeperp}k_2^2-H^{\prime}_{0}({\honeperp}R)J_{0}({\htwoperp}R){\htwoperp}k_1^2}
\nonumber \\
M(\hp) & = & 
\left(\frac{\koneperp^3}{k_1^2}-\frac{\kp\koneperp(\hp-\kp)}{k_1^2}\right)b_{1}H^{\prime}_{0}({\koneperp}R) 
-\left(\frac{\ktwoperp^3}{k_2^2}-\frac{\kp\ktwoperp(\hp-\kp)}{k_2^2}\right)b_{2}J^{\prime}_{0}({\ktwoperp}R) 
\nonumber \\ N & = & 
b_{1}k_{1\perp}^{2}H^{\prime\prime}_{0}({\koneperp}R)-b_{2}\ktwoperp^{2}J^{\prime\prime}_{0}({\ktwoperp}R).\label{eq:fgMN} 
\eea

To evaluate the expression for $\Gamma_{\footnotesize\textrm{rad,rough}}$ in 
Eq.~(\ref{eq:dissipationratio}), it is convenient to normalize the plasmon 
fields using Eq.~(\ref{eq:normalization}), in which case the denominator 
of~(\ref{eq:dissipationratio}) becomes $\hbar\omega/4L$ and the normalization 
coefficient for the field outside the wire is given by 
$b_{1}{\approx}\sqrt{\frac{\hbar\omega k_0^4 \epsilon_1^2 R^2}{\epsilon_0 
\tilde{V} C_{-1}^4 L}}$, as derived in Sec.~\ref{subsec:plasmondecay}.  Then, 
using the relationships $\kp{\approx}C_{-1}/R,\kiperp{\approx}iC_{-1}/R$ one can 
calculate the leading terms of $f(\hp)$ as $R{\rightarrow}0$,
\be f(\hp) \approx \frac{\htwoperp 
b_{1}(C_{-1}/R)^{2}\phi}{-2ik_1^2\htwoperp/\pi\honeperp R}, \ee
where we have defined 
$\phi{\equiv}(b_2/b_1)J_0^{\prime\prime}(iC_{-1})-H_{0}^{\prime\prime}(iC_{-1})$.  
In the equation above we have explicitly given the leading terms of the 
numerator and denominator of $f(\hp)$.  The ratio $b_1/b_2$ is given in
Eq.~(\ref{eq:b1b2}), which in the nanowire limit results in the simplification 
of $\phi$ given in Eq.~(\ref{eq:phidef}).

\section{Non-radiative scattering}\label{sec:matvec}

The elements of the matrices $M_i$ and vectors $\bfv_i$ appearing in the matrix 
integral equation~(\ref{eq:roughnessmatrix}) in the presence of surface 
roughness are given by
\bea M_{0}(h) & = & \left(\begin{array}{cc} \tilde{K}_{m}(hR) & 
-\tilde{I}_{m}(hR) \\ h\epsilon_{1}\tilde{K}^{\prime}_{m}(hR) &
-h\epsilon_{2}\tilde{I}^{\prime}_{m}(hR) \end{array}\right), \\
M_{1}(h,q) & = & \left(\begin{array}{cc}
q\tilde{K}^{\prime}_{m}(qR) & -q\tilde{I}^{\prime}_{m}(qR) \\
\epsilon_{1}(q^2\tilde{K}^{\prime\prime}_{m}(qR)+q(h-q)\tilde{K}_{m}(qR)) & 
-\epsilon_{2}(q^2\tilde{I}^{\prime\prime}_{m}(qR)+q(h-q)\tilde{I}_{m}(qR))
\end{array}\right)e^{i(h-q)z'}, \\
M_{2}(h,q,q^{\prime}) & = & \left(\begin{array}{cc} 
\frac{q^2}{2}\tilde{K}^{\prime\prime}_{m}(qR) &
-\frac{q^2}{2}\tilde{I}^{\prime\prime}_{m}(qR) \\
M_{2}^{21}(h,q,q') & M_{2}^{22}(h,q,q')
\end{array}\right)e^{i(h-q)z'}, \\
M_{2}^{21}(h,q,q') & = & 
\epsilon_{1}\left(\frac{q^3}{2}\tilde{K}^{\prime\prime\prime}_{m}(qR)+q^{2}(h-q-q^{\prime})\tilde{K}^{\prime}_{m}(qR)+\frac{1}{2}qq^{\prime}(h-q-q^{\prime})\tilde{K}^{\prime}_{m}(qR)\right), 
\\
M_{2}^{22}(h,q,q') & = & 
-\epsilon_{2}\left(\frac{q^3}{2}\tilde{I}^{\prime\prime\prime}_{m}(qR)+q^{2}(h-q-q^{\prime})\tilde{I}^{\prime}_{m}(qR)+\frac{1}{2}qq^{\prime}(h-q-q^{\prime})\tilde{I}^{\prime}_{m}(qR)\right), 
\\
\bfv_{0}(h) & = & -\left(\begin{array}{c} \tilde{I}_{m}(hR) \\
h\tilde{I}^{\prime}_{m}(hR)\end{array}\right)\tilde{K}_{m}(h\rho^{\prime}), \\
\bfv_{1}(h,q) & = & -\left(\begin{array}{c}
q\tilde{I}^{\prime}_{m}(qR) \\
q^{2}\tilde{I}^{\prime\prime}_{m}(qR)+q(h-q)\tilde{I}_{m}(qR)\end{array}\right)\tilde{K}_{m}(q\rho^{\prime})e^{i(h-q)z^{\prime}},
\\
\bfv_{2}(h,q,q^{\prime}) & = & -\left(\begin{array}{c}
\frac{q^2}{2}\tilde{I}^{\prime\prime}_{m}(qR) \\
\frac{q^3}{2}\tilde{I}^{\prime\prime\prime}_{m}(qR)+q^{2}(h-q-q^{\prime})\tilde{I}^{\prime}_{m}(qR)+\frac{1}{2}qq^{\prime}(h-q-q^{\prime})\tilde{I}^{\prime}_{m}(qR)\end{array}\right)\tilde{K}_{m}(q\rho^{\prime})e^{i(h-q)z^{\prime}}. 
\eea

%

\begin{table*}[p]
\begin{center}
\begin{tabular}{|c|c|c|c|} \hline
Roughness parameters  & $k_{0}R=0.1$~($R{\approx}16$~nm) & 
$k_{0}R=0.2$~($R{\approx}32$~nm) & $k_{0}R=0.3$~($R{\approx}48$~nm) \\ \hline

$a=0.1R$, $\delta=0.05R$~($s=0.5$) & $0.09\%$ & $0.5\%$ & $1.4\%$ \\ \hline

$a=0.1R$, $\delta=0.1R$~($s=1$) & 0.4\% & 1.9\% & 5.6\% \\ \hline

$a=R$, $\delta=0.05R$~($s=0.05$) & 0.9\% & 4.5\% & 12\% \\ \hline

$a=5R$, $\delta=0.05R$~($s=0.01$) & 2.8\% & 8.0\% & 10\% \\ \hline

$a=10R$, $\delta=0.1R$~($s=0.01$) & $7.0\%$ & $14\%$ & $16\%$ \\ \hline

$a=20R$, $\delta=0.1R$~($s=0.005$) & 0.9\% & 2.9\% & 3.8\% \\ \hline

$a=25R$, $\delta=0.1R$~($s=0.004$) & 0.3\% & 1.3\% & 1.8\%
\\ \hline \end{tabular}
\end{center}
\caption{Losses due to radiative scattering off of surface roughness for 
nanowires of varying sizes and roughness parameters.  The scattering rates are 
given in terms of the percentage increase in $\textrm{Im}\;\kp$ that one would 
expect over the values for a smooth nanowire.  \label{table:roughness}}
\end{table*}

\begin{table*}[p]
\begin{center}
\begin{tabular}{|c|c|c|} \hline
Roughness parameters  & $\Delta(\textrm{Re}\,\tilde{C}_{-1})$ 
 & $\Delta(\textrm{Im}\,\tilde{C}_{-1}/\textrm{Re}\,\tilde{C}_{-1})$ \\ \hline

$a=0.1R$, $\delta=0.01R$~($s=0.1$) & 0.2\% & 0.2\% \\ \hline

$a=0.1R$, $\delta=0.05R$~($s=0.5$) & 7.5\% & 6.8\%  \\ \hline

$a=R$, $\delta=0.01R$~($s=0.01$) & 0.03\% & 1.0\% \\ \hline

$a=R$, $\delta=0.05R$~($s=0.05$) & 0.9\% & 26\% \\ \hline

$a=R$, $\delta=0.1R$~($s=0.1$) & 3.5\% & 110\% \\ \hline

$a=10R$, $\delta=0.01R$~($s=0.001$) & $>0.01\%$ & 2.7\% \\ \hline

$a=10R$, $\delta=0.05R$~($s=0.005$) & 0.2\% & 67\% \\ \hline

$a=10R$, $\delta=0.1R$~($s=0.01$) & 0.8\% & 270\% \\ \hline

\end{tabular}
\end{center}
\caption{Losses and wavevector shifts due to non-radiative scattering off of 
surface roughness for nanowires with varying roughness parameters.  The shifts 
in $\textrm{Re}\,\tilde{C}_{-1}$ and changes in loss parameters
$\textrm{Im}\,\tilde{C}_{-1}/\textrm{Re}\,\tilde{C}_{-1}$ are given in terms of
percentage increase over the corresponding values for a smooth nanowire.  
\label{table:roughness2}}
\end{table*}


\begin{thebibliography}{99}
\bibitem{note1} It can be verified that the boundary condition equations for
the perpendicular field components $E_{\rho},H_{\rho}$ are redundant with those 
for the parallel field components.
\end{thebibliography}
\end{document}